\documentclass[12pt]{article}
\usepackage{a4wide}
\usepackage{amsmath}
\usepackage{amsfonts}
\usepackage{amssymb}
\usepackage{graphicx}
\usepackage{subfigure}

\usepackage{epsf}
\usepackage{hyperref}

\newcommand{\md}{{\rm d}}
\newcommand{\vt}{\vartheta}
\newcommand{\vp}{\varphi}
\newcommand{\lP}{\ell_{\rm P}}

\begin{document}
{\renewcommand{\thefootnote}{\fnsymbol{footnote}}
\hfill  IGC--08/6--3\\
\medskip
\begin{center}
{\LARGE  Lemaitre--Tolman--Bondi collapse from the perspective of loop quantum gravity}\\
\vspace{1.5em}
Martin Bojowald\footnote{e-mail address: {\tt bojowald@gravity.psu.edu}}
\\
\vspace{0.5em}
Institute for Gravitation and the Cosmos,
The Pennsylvania State
University,\\
104 Davey Lab, University Park, PA 16802, USA\\
\vspace{0.5em}
Tomohiro Harada\footnote{e-mail address: {\tt harada@rikkyo.ac.jp}}\\
\vspace{0.5em}
Department of Physics, Rikkyo University, Toshima, 
Tokyo  171-8501, Japan\\
\vspace{0.5em}
and Rakesh Tibrewala\footnote{e-mail address: {\tt rtibs@mailhost.tifr.res.in}}\\
\vspace{0.5em}
Tata Institute of Fundamental Research,\\
Homi Bhabha Road, Mumbai 400 005, India
\vspace{1.5em}
\end{center}
}

\setcounter{footnote}{0}

\begin{abstract}
 Lemaitre--Tolman--Bondi models as specific spherically symmetric
 solutions of general relativity simplify in their reduced form some
 of the mathematical ingredients of black hole or cosmological
 applications. The conditions imposed in addition to spherical
 symmetry turn out to take a simple form at the kinematical level of
 loop quantum gravity, which allows a discussion of their implications
 at the quantum level. Moreover, the spherically symmetric setting of
 inhomogeneity illustrates several non-trivial properties of lattice
 refinements of discrete quantum gravity. Nevertheless, the situation
 at the dynamical level is quite non-trivial and thus provides
 insights to the anomaly problem. At an effective level, consistent
 versions of the dynamics are presented which implement the conditions
 together with the dynamical constraints of gravity in an anomaly-free
 manner. These are then used for analytical as well as numerical
 investigations of the fate of classical singularities, including
 non-spacelike ones, as they generically develop in these models. None
 of the corrections used here resolve those singularities by regular
 effective geometries. However, there are numerical indications that
 the collapse ends in a tamer shell-crossing singularity prior to the
 formation of central singularities for mass functions giving a
 regular conserved mass density.  Moreover, we find quantum
 gravitational obstructions to the existence of exactly homogeneous
 solutions within this class of models. This indicates that
 homogeneous models must be seen in a wider context of inhomogeneous
 solutions and their reduction in order to provide reliable dynamical
 conclusions.
\end{abstract}

\section{Introduction}

Black holes provide one of the most active areas of gravitational
research, and a prime testing ground for quantum gravity. Many aspects
can already be analyzed under the assumption of spherical symmetry,
which is sufficient to describe non-rotating black holes.  In vacuum,
for instance, a reduced quantization is possible
\cite{SphKl1,SphKl2,Kuchar}.  This reduction has thus often been used
in diverse approaches to quantum gravity, and also loop quantum
gravity \cite{Rov,ALRev,ThomasRev} has provided its own formulation
\cite{SymmRed,SphSymm,SphSymmHam,BlackHoles,SphSymmUniform}. This
allows one to use the loop representation constructed in the full
theory in a simpler setting in which, as one hopes, one can find and
analyze physical solutions. The result is a set of many coupled
partial difference equations for a wave function on midisuperspace,
which reflects the discreteness of spatial geometry realized in a loop
quantization. As a consequence, spacelike classical singularities as
they occur e.g.\ in the best-known case of the Schwarzschild solution
are removed because their high curvature regimes no longer present
boundaries to quantum evolution \cite{SphSymmSing,BSCG}.  So far,
these present the only inhomogeneous singularities analyzed by the
methods of loop quantum gravity, and the situation appears much more
non-trivial than in homogeneous models. Due to this complexity,
several questions remain open. In particular, inhomogeneous models
allow not only spacelike singularities but also null or timelike ones,
which may even be naked; see e.g.\
\cite{LTBNakedJoshi,NakedSingForm1,NakedSingForm2} and references
therein.  Their fate presents additional problems of high interest
which can be studied in spherical symmetry.

Unfortunately, even the spherically symmetric equations in loop
quantum gravity are difficult to tackle, and so it becomes interesting
to look at further reductions which can preserve the physical setting
but provide mathematical simplifications. Some of the possible effects
have been introduced in a more ad-hoc manner in the models of
\cite{HWBlackHole,HWHorizon,HWCollapse,DilatonBohr}, and were used
mainly to understand implications on the horizon dynamics. A drawback
of such an approach is that quantum corrections are not clearly linked
to key ingredients of a loop quantization, or any quantization one
could apply to the full, unrestricted theory. We thus intend to
introduce new models which, starting from the spherically symmetric
reduction along the general lines of \cite{SphSymm}, provide further
simplifications. Moreover, we will pay special attention to the
anomaly problem which is required for full consistency. The
possibility followed here is a reformulation of the spherically
symmetric constrained system as it is used in the canonical definition
of Lemaitre--Tolman--Bondi (LTB) models
\cite{Lemaitre,TolmanSol,Bondi}.  In this classical reduction, one
solves the diffeomorphism constraint and, inserting the solutions into
the Hamiltonian constraint, simplifies the set of equations. This
eliminates some of the equations and removes several gauge issues. The
remaining equations are then easier to quantize and solve.

The imposed conditions make the constrained system partially second
class, and thus are not straightforward to implement at the quantum
level. Symmetry reduction, which is also partially second class and
can be done at the kinematical quantum level
\cite{SymmRed,SymmQFT,Reduction,Rieffel,SymmStatesInt}, provides some
guidelines, but the gauge fixing conditions required for LTB models
are more complicated. While these issues are discussed here, we
postpone a direct implementation at the dynamical quantum level and
rather perform in this article an analysis of the quantum correction
terms one can expect for the classical equations. Nevertheless, we
will be able to highlight several important issues at the quantum
level, such as the role of dynamical lattice refinements of discrete
states \cite{InhomLattice,CosConst}. In particular, we will see that
not all refinement schemes discussed so far in anisotropic homogeneous
models can be embedded in spherically symmetric ones. But a large
class still remains because only the direction dependence, not the
size dependence of refinements is determined.

What we provide in the main part of this paper is an analysis of
consistent sets of quantum corrected equations of motion for a type of
non-perturbative inhomogeneities, which has not yet been available in
other models studied in the framework of loop quantum
gravity. Previous equations either refer to homogeneous models
\cite{BouncePert,BouncePot,QuantumBounce,BounceSqueezed}, where strict
effective equations can be derived as analyzed in a canonical setting
in \cite{EffAc,Karpacz,EffCons}, but the consistency issue
trivializes, or to perturbative inhomogeneities as in
\cite{InhomEvolve,HamPerturb,ScaleInvLQC,CCPowerLoop,SuperInflLQC,Vector,TensorBackground,Tensor}. In
our case, inhomogeneities can be non-perturbative but their equations
are arrived at only after performing a partial gauge fixing.

We will thus be able to restrict the form of possible quantum
corrections by the condition that they provide consistent deformations
of the classical equations. Thus, we will discuss how quantum
corrections due to loop quantum gravity should occur in order to
guarantee consistent LTB-like solutions. We will not be deriving
strict effective equations, but we will determine consistent
anomaly-free consequences of different types of quantum
corrections. For several of the consistent choices we study
consequences for inhomogeneous spherically symmetric systems based on
analytical as well as numerical considerations. Our main interest
here, as often in this context, is the fate of classical
singularities. Compared to other models which by now have been studied
extensively in loop quantum cosmology \cite{LivRev}, a new feature is
the possibility of non-spacelike singularities in LTB
systems. Interestingly, none of the corrections studied here resolve
such singularities directly (nor spacelike ones in these models), in
contrast to what their analogs do so easily in homogeneous models.
However, numerically we find indications that shell-crossing
singularities generically occur in the presence of the corrections,
where only shell-focusing singularities would do in classical theory.
The shell-crossing singularities are expected to be weakly extendable,
or to be avoidable by realistic matter.  This is one example for new
effects in inhomogeneous models, which not surprisingly prove
themselves much more non-trivial than homogeneous ones.

\section{Spherically symmetric spacetimes and LTB variables}

The line element in a general spherically symmetric spacetime with
polar coordinates $(x,\vartheta,\varphi)$ can be written as
\begin{equation} \label{dsss}
\md s^{2}=-e^{2\nu}\md t^{2}+e^{2\lambda}\md x^{2}+R^{2}
(\md\vartheta^{2}+\sin^{2}\vartheta \md\varphi^{2}),
\end{equation}
where $\nu=\nu(t,x)$, $\lambda=\lambda(t,x)$ and $R=R(t,x)$ are
functions of the time and radial coordinates $t$ and $x$.
Alternatively, the form suitable for a canonical analysis, as done
below, is
\begin{equation} \label{CanonMetric}
  \md s^2= -N(x,t)^2\md t^2+ L(x,t)^2(\md x+N^x(x,t)\md t)^2 
+R(x,t)^2(\md\vartheta^{2}+\sin^{2}\vartheta \md\varphi^{2})
\end{equation}
where $N$, $N^x$, $L$ and $R$ are again free functions of the radial
coordinate $x$ and time $t$ only. In a canonical quantization, the
lapse function $N$ and the shift vector $N^x$ appear as Lagrange
multipliers of the Hamiltonian and diffeomorphism constraints,
respectively. They are thus not dynamical, unlike the remaining
functions $L$ and $R$ which have non-trivial momenta $P_L$ and $P_R$,
together forming the kinematical phase space of spherically symmetric
gravity. In (\ref{dsss}), the shift vector has already been chosen to
vanish, while $\nu$ and $\lambda$ are related to $N$ and $L$ in
obvious ways.

\subsection{Matter source}

We start with a spherically symmetric matter source whose
stress-energy tensor $T^{\mu}_{\nu}$ is diagonalized in the form
\begin{equation}
T^{\mu}_{\nu}=\left(
\begin{array}{cccc}
-\epsilon& 0 & 0 & 0 \\
0 & \Sigma & 0 & 0 \\
0 & 0 & \Pi & 0\\
0 & 0 & 0 & \Pi
\end{array}
\right),
\end{equation}
where $\epsilon=\epsilon(t,x)$, $\Sigma=\Sigma(t,x)$ and
$\Pi=\Pi(t,x)$ are the energy density, the radial pressure and the
tangential pressure, respectively.  Einstein's equation and the
conservation law then reduce to a set of partial differential
equations given by
\begin{eqnarray}
&& m'= 4\pi G \epsilon R^{2}R'
\label{eq:m'}\\
&&\dot{m}= - 4\pi G\Sigma R^{2}\dot{R}
\label{eq:mdot}\\
&&\dot{R}'= \dot{R}\nu'+R' \dot{\lambda}
\label{eq:rdot'}\\
&&\Sigma'= -(\epsilon+\Sigma )\nu'-2 (\Sigma -\Pi)\frac{R'}{R} 
\label{eq:sigma'}\\
&& e^{-2\nu}\left(\ddot{R}-\dot{\nu}\dot{R}\right)-e^{-2\lambda}
\nu' R'=-\frac{m}{R^{2}}-\frac{4}{3}\pi (\Sigma+2\Pi) R
\label{eq:rddot}
\end{eqnarray}
where we have introduced
\begin{equation}
 m=\frac{R}{2}\left(1-R'^{2}e^{-2\lambda}+\dot{R}^{2}e^{-2\nu}\right),
\label{eq:ms_mass}
\end{equation}
called the Misner--Sharp mass.

The Misner--Sharp mass $m$ is generally defined as
\begin{equation}
m=\frac{1}{2}R(1-\nabla_{A}R\nabla^{A}R)\,,
\end{equation}
where $A$ takes values 0 and 1 corresponding to the 2-manifold
coordinatized by $t$ and $x$. At spherical trapping horizons
\cite{trapping} we have $R=2m$. (To see this, one can use the simple
criterion for the spherically symmetric trapping horizon as the place
where constant area radius surfaces become null \cite{ClosedHor}. This
is the boundary of the region of spherical trapped surfaces; there may
be non-spherical trapped surfaces outside that region
\cite{TrappedVaidya,OuterTrappedVaidya}.)  In an asymptotically flat
spacetime, the asymptotic value of the Misner--Sharp mass at spatial
infinity equals the ADM mass $M_{\rm ADM}$.  We can define a conserved
current from the Misner--Sharp mass, which is called the Kodama
current and given by
\begin{equation}
j^{A}=\epsilon^{AB}\nabla_Bm\,,
\end{equation}
where $\epsilon_{AB}$ is the antisymmetric tensor satisfying
$\epsilon_{AB}\epsilon^{B}_{~~C}=g_{AC}$.  By definition, the current
$j^{A}$ satisfies the conservation law $\nabla_Aj^{A}=0$. The Kodama
current provides the energy density
\begin{equation}
\epsilon=\frac{m'}{4\pi G R^{2}R'}
\label{eq:effective_energy_density}
\end{equation}
in agreement with (\ref{eq:m'}).

\subsection{Classical LTB spacetimes}

In classical Einstein gravity, there is an exact solution which
describes a spherically symmetric collapse system sourced by
inhomogeneous dust: the Lemaitre-Tolman-Bondi solution
\cite{Lemaitre,TolmanSol,Bondi}.  For dust, we have $\Sigma=\Pi=0$
such that Eq.~(\ref{eq:mdot}) implies that
\begin{equation}
m= \frac{1}{2}F(x)
\end{equation}
is an arbitrary function of the radial coordinate only. From
Eq.~(\ref{eq:m'}), we then have
\begin{equation} \label{eps}
\epsilon=\frac{F'}{8 \pi G R^{2}R'}
\end{equation}
for the dust density. Then, Eq.~(\ref{eq:sigma'}) implies that
$\nu=\nu(t)$ can be made to vanish by rescaling the time coordinate
$t$ such that it becomes the dust proper time. Eq.~(\ref{eq:rdot'})
can be integrated to give
\begin{equation}
(R'e^{-\lambda})^{2}=1+\kappa(x)\,,
\end{equation}
where $\kappa(x)>-1$ is another arbitrary function. 
The resulting line element can be written as 
\begin{equation}
\md s^{2}=-\md t^{2}+\frac{R'^{2}}{1+\kappa(x)}\md x^{2}+
R^{2}(\md\vartheta^{2}+\sin^{2}\vartheta \md\varphi^{2})
\end{equation}
where the only dynamical function left satisfies, from Eq.~(\ref{eq:ms_mass}),
\begin{equation}
\dot{R}^{2} =\kappa(x)+ \frac{F(x)}{R}\,.
\label{eq:rdot}
\end{equation}

Here we concentrate on the special case where $\kappa(x)=0$, called
marginally bound LTB solution, such that
\begin{equation}\label{ltbmet}
 \md s^{2}=-\md t^{2}+R'^{2}\md x^{2}+
R^{2}(\md\vartheta^{2}+\sin^{2}\vartheta \md\varphi^{2})
\end{equation}
and Eq.~(\ref{eq:rdot}) becomes
\begin{equation}
\dot{R}^{2} =\frac{F(x)}{R}\,.
\label{eq:energy_conservation}
\end{equation}
At a trapping horizon, this expression equals one.  We can integrate
the equation of motion to
\begin{equation} \label{ts}
 t-t_{s}(x)=\pm \frac{2R^{3/2}}{3F(x)^{1/2}},
\end{equation}
where $t_{s}(x)$ is an arbitrary function specifying initial values at
$t=0$. Its freedom can be absorbed by defining the function $R$ in
terms of the coordinate $x$ at $t=0$. For instance, requiring
$R|_{t=0}=x$ and choosing the lower sign in (\ref{ts}) for a
collapsing model, we have $t_s(x)= \frac{2}{3}x^{3/2}F(x)^{-1/2}$ and
thus
\begin{equation} \label{GenSol}
 R(x,t)= \left(x^{3/2}-\frac{3}{2}\sqrt{F(x)}t\right)^{2/3}\,.
\end{equation}
(An example is the self-similar solution $R=x(1-at/x)^{2/3}$ obtained
for $F(x)=\lambda x$ as studied in \cite{LTBSelfSim}.)  Of special
interest in those solutions is the small-$x$ behavior of $R$ which may
be regular or singular.  In particular, the energy density (\ref{eps})
can be divergent where $R'=0$, giving rise to shell-crossing
singularities. While solutions are no longer valid beyond this point,
shell-crossing singularities are deemed avoidable by more realistic
matter models; see e.g.\ \cite{RegularEinsteinVlasov}. Moreover,
solutions are extendable through shell-crossing singularities in a
distributional sense \cite{SingAnalysis,CrossingSing}.  In addition to
this, energy density as well as the Ricci scalar can diverge when
$R=0$, giving rise to a shell-focusing or central singularity which is
the main interest here. The properties of shell-focusing singularities
in LTB models have been extensively studied
\cite{LTBNumerical,Christodoulou,LTBCollapseFinal,LTBStructureSing}.
In particular, it was proven that naked singularities form generically
in this system \cite{Christodoulou,Newman,LTBSingII}.

This provides a new arena for effects of quantum gravity, which we
will come back to in more detail in Sec.~\ref{s:App}.  We will
be using loop quantum gravity which requires Hamiltonian techniques.
In the canonical formulation of spherically symmetric Einstein
gravity, the Hamiltonians of the gravitational sector, $H_{\rm grav}$,
and of the matter sector, $H_{\rm dust}$, for the LTB system provide
the expressions
\begin{equation}
H_{\rm grav}=-\frac{\dot{R}^{2}R'+R(\dot{R}^{2})'}{2G}
=-\frac{(\dot{R}^{2}R)'}{2G} \quad,\quad
H_{\rm dust}=\frac{F'}{2G}
\end{equation}
after replacing the momentum in terms of $\dot{R}$.  The total
Hamiltonian constraint $H_{\rm grav}+H_{\rm dust}=0$ then yields
Eq.~(\ref{eq:energy_conservation}). Below, we will provide a detailed
canonical analysis based on Ashtekar variables, which will allow us to
incorporate some corrections as they are expected from loop quantum
gravity. (See \cite{LTBADM1,LTBADM2,LTBADMnonmarg} for a canonical
analysis in ADM variables, and \cite{LTBd1,LTBd2} for an analysis in
arbitrary dimensions.)

\subsection{Connection variables}

For a canonical formulation only the spatial part of the metric
provides the dynamical degrees of freedom. Moreover, the most
successful canonical quantization of gravity, loop quantum gravity, is
based on a densitized triad $E^a_i$ rather than a spatial metric
$q_{ab}$, which are related by $E^a_iE^b_i=\det(q_{cd})q^{ab}$.
(We use tangent space indices $a,b,\ldots$ and internal gauge indices
$i,j,\ldots$.)  The spatial metric for a spherically symmetric system
in components of this variable is given by
\begin{equation} \label{ltbE}
\md s^{2}=\frac{E^{\varphi}(x)^{2}}{|E^x(x)|}\md x^{2}+|E^{x}(x)|
(\md\vartheta^2+\sin^2\vartheta\md\varphi^2)
\end{equation}
where instead of the spherically symmetric spatial metric components
$L$ (which equals $E^{\varphi}/\sqrt{|E^x|}$) and $R$ (which equals
$\sqrt{|E^x|}$) the spherically symmetric triad components $E^x$ and
$E^{\varphi}$ appear. Written as a densitized vector field taking
values in su(2) with basis $\tau_i$, these components define the
spherically symmetric densitized triad by
\begin{equation} \label{E}
 E=E^x(x)\tau_3\sin\vt\frac{\partial}{\partial x}+
(E^1(x)\tau_1+E^2(x)\tau_2)\sin\vt\frac{\partial}{\partial\vt}+
(E^1(x)\tau_2-E^2(x)\tau_1)\frac{\partial}{\partial\vp}\,.
\end{equation}
such that $(E^{\vp})^2=(E^1)^2+(E^2)^2$. (For more details on this
decomposition see \cite{SymmRed,SphSymm}. Notice that the sign of
$E^x$ is not restricted to be positive, while $E^{\vp}$ is defined to
be non-negative. The sign of $E^x$ thus determines the orientation of
the triad since ${\rm sgn}\det (E^a_i)={\rm sgn} E^x$; it plays an
important role for fundamental singularity resolution \cite{BSCG}, but
not for most of the analysis done in this article.) The remaining
freedom of the three functions $E^x$, $E^1$ and $E^2$ not contained in
the components $E^x$ and $E^{\vp}$ is pure gauge since it does not
affect the metric.  It can be parameterized by an angle
$\eta=\arctan(E^2/E^1)$ which is subject to U(1)-gauge
transformations. The corresponding Gauss constraint is $G[\lambda]=
\int\md x \lambda(x)((E^x)'+P^{\eta})$ where $P^{\eta}$ is the
momentum of $\eta$.

Fields canonically conjugate to the original triad components $E^I$
are Ashtekar connection components of
\begin{equation} \label{A}
 A=A_x(x)\tau_3\md x+(A_1(x)\tau_1+A_2(x)\tau_2)\md\vt+
(A_1(x)\tau_2-A_2(x)\tau_1)\sin\vt\md\vp+ \tau_3\cos\vt\md\vp\,.
\end{equation}
Also here we introduce the U(1)-gauge invariant quantity
$A_{\vp}^2=A_1^2+A_2^2$ and the gauge angle $\beta=\arctan(A_2/A_1)$.
Only the difference $\alpha:=\eta-\beta$ of the angles is gauge
invariant, while each of them can be changed by the same amount with a
gauge transformation. However, the new parameter $A_{\vp}$ is not
canonically conjugate to $E^{\vp}$; see \cite{SphSymm} for details. In
terms of the U(1)-gauge invariant triad component $E^{\vp}$, its
conjugate variable is rather given by $A_{\vp}\cos\alpha=\gamma
K_{\vp}$ which happens to be proportional to an extrinsic curvature
component \cite{SphSymmHam}. (The constant of proportionality is given
by the Barbero--Immirzi parameter $\gamma$
\cite{AshVarReell,Immirzi}.)  The canonical pairs we will be dealing
with are thus $(A_x,E^x)$, $(\gamma K_{\vp},E^{\vp})$ for which
\begin{equation} \label{Poisson}
 \{A_x(x),E^x(y)\} = 2\gamma G\delta(x,y)\quad,\quad \{\gamma
 K_{\varphi}(x), E^{\varphi}(y)\}= \gamma G\delta(x,y)
\end{equation}
and another pair $(\eta,P^{\eta})$ for the gauge angle. The relation
between the remaining Ashtekar connection component $A_x$ and
extrinsic curvature is $A_x=\gamma K_x-\eta'$ where the spatial
derivative $\eta'$ of the angle is the (negative) $x$-component of the
spin connection. Its angular components are given in terms of
\begin{equation} \label{Gammaphi}
 \Gamma_{\vp}=-\frac{(E^x)'}{2E^{\vp}}
\end{equation}
which determines the relation between $A_{\vp}$ and $K_{\vp}$ as
$A_{\vp}^2 = \Gamma_{\vp}^2+\gamma^2K_{\vp}^2$ without reference to
$\alpha$.

Solving the Gauss constraint removes the gauge angle and its momentum,
and implies that invariant objects can depend on the $x$-component of
the Ashtekar connection only through the extrinsic curvature quantity
$A_x+\eta'=\gamma K_x$. We are then left with only two canonical pairs
$(K_x,E^x)$ and $(K_{\varphi},E^{\varphi})$. A further reduction can
be implemented by using the LTB form of the variables corresponding to
the metric (\ref{ltbmet}).  Comparing the spatial parts of the metrics
(\ref{ltbmet}) and (\ref{ltbE}) one obtains the LTB relation
\begin{equation}
E^{\varphi}(x)=\frac{1}{2}|E^{x}(x)|'\,.
\label{ephi}
\end{equation}
In particular, we have $\Gamma_{\varphi}=-{\rm sgn}E^x$
everywhere. For a complete reduction of canonical degrees of freedom,
this is to be accompanied by a condition between $K_x$ and
$K_{\varphi}$. We will determine this by consistency with the
constraints, such that (\ref{ephi}) is preserved in time.

\subsection{Constraints}

Constraints in real Ashtekar variables as used here have been derived
in \cite{SphSymm,SphSymmHam}.  The diffeomorphism constraint is given
by
\begin{equation}
D[N^{x}]=\int\md xN^{x}(2A'_{1}E^{1}+2A'_{2}E^{2}-A_{x}(E^{x})')
\end{equation}
where $A_{1}=A_{\varphi}\sin\beta$, $A_{2}=-A_{\varphi}\cos\beta$,
$E^{1}=E^{\varphi}\sin\eta$, $E^{2}=-E^{\varphi}\cos\eta$. Using these
definitions in the above equation one obtains
\begin{equation}
  D[N^{x}]=\int\md xN^{x}\left(2(A_{\varphi}\cos\alpha)'E^{\varphi}+ 
2(\alpha'+\beta')E^{\varphi}A_{\varphi}\sin\alpha-A_{x}(E^{x})'\right)
\end{equation}
Using the relations $A_{\varphi}\cos\alpha=\gamma K_{\varphi}$,
$A_{\varphi}\sin\alpha=\Gamma_{\varphi}$, $\alpha+\beta=\eta$,
$A_{x}=-\eta'+\gamma K_{x}$ and
$\Gamma_{\varphi}=-(E^{x})'/2E^{\varphi}$, this can be expressed
solely in terms of U(1)-gauge invariant quantities as
\begin{equation}
D[N^{x}]=\int\md xN^{x}\left(2\gamma K_{\varphi}'E^{\varphi}-
\gamma K_{x}(E^{x})'\right)\,.
\end{equation}

We can consider the marginally bound LTB condition, as given by
(\ref{ephi}), as a gauge-fixing condition of the diffeomorphism
constraint to form a second class pair. On the gauge fixing
surface, the diffeomorphism constraint can be replaced by
\begin{equation}
D[N^{x}]=\int\md xN^{x}\gamma (E^{x})'({\rm sgn}(E^x)K_{\varphi}'-K_{x})\,.
\end{equation}
Thus, for the diffeomorphism constraint to be satisfied
one must have
\begin{equation} \label{kphi}
 K_{\varphi}'=K_{x} {\rm sgn}E^x\,.
\end{equation}
This provides the LTB condition for the conjugate variables to $E^I$
analogous to (\ref{ephi}). Having solved the second class constraints
resulting from gauge-fixing $D[N^x]$, we proceed to the Hamiltonian
constraint and insert the solutions.

In spherical symmetry, the gravitational part of the Hamiltonian
constraint is
\begin{equation} \label{Hamclass}
  H_{\rm grav}[N]=-\frac{1}{2G}\int\md xN(x)|E^{x}|^{-1/2} 
\left((1-\Gamma_{\vp}^{2}+K_{\varphi}^{2})E^{\vp}+
2\gamma^{-1}K_{\vp}E^{x}(A_{x}+\eta')+2E^{x}\Gamma'_{\vp}\right)\,.
\end{equation}
With $\Gamma_{\varphi}=-1$ under the LTB condition in (\ref{ephi}), we
have
\begin{equation}
H_{\rm grav}[N]=-\frac{1}{2G}\int\md xN(x)|E^{x}|^{-1/2}
\left(K_{\vp}^{2}E^{\vp}+2K_{\varphi}E^{x}K_{x}\right)
\end{equation}
or, in unsmeared form,
\begin{equation}
H_{\rm grav}=-\frac{1}{2G}\left(\frac{K_{\vp}^{2}E^{\vp}}{\sqrt{|E^{x}|}}+
2K_{\vp}K_{x}\sqrt{|E^{x}|}\right)
\label{hg}
\end{equation}
where $E^{\varphi}$ and $K_{\varphi}$ are to be understood as
functionals of $E^x$ and $K_x$, respectively, via the LTB conditions,
or as functions of $R$ and its momentum $P_R$. Thus, $H_{\rm grav}$ is
left as the sole constraint to restrict initial values and generate
equations of motion for LTB models. In fact, the conditions
(\ref{ephi}) and (\ref{kphi}) can be seen to be consistent with the
classical equations of motion for $E^x$, $E^{\varphi}$, $K_x$ and
$K_{\varphi}$ generated by (\ref{hg}) (in addition to solving the
diffeomorphism constraint identically). Thus, the reduction to LTB
form is dynamically consistent and provides correct space-time
solutions. Maintaining this consistency will be our main guideline in
the analysis of quantum effects.

\section{Effects of a loop quantization}

Like any quantization of an interacting system, quantum gravity
implies corrections to classical equations of motion and thus forces
us to re-address consistency issues of classical reductions. As we
will see explicitly, for a loop quantization this provides valuable
feedback on the consistency or possible anomalies of the overall
framework. In LTB systems, once a consistent reduction has been found,
it can be used for applications to black hole singularity issues. We
focus on loop specific issues which do not arise in Wheeler--DeWitt
type quantizations of \cite{LTBADM1,LTBADM2,LTBADMnonmarg} which have
for instance been applied to Hawking radiation
\cite{LTBADMHawking,LTBADMHawking2,LTBd2}.

\subsection{Basics of spherically symmetric loop quantum gravity}

A loop quantization \cite{LoopRep} of spherically symmetric gravity is
based on holonomies
\begin{equation}\label{hx}
h_e(A_x) = \exp\left(\tfrac{1}{2}i \smallint_e A_x\md x\right)
\quad,\quad
 h_v(K_{\vp}) =\exp(i \gamma K_{\vp}(v))\quad,\quad
h_v(\eta)=\exp(i\eta(v))
\end{equation}
for the configuration variables instead of linear expressions in
connection or extrinsic curvature components. Here, we have used edges
$e$ and vertices $v$ in the radial line. The use of these variables is
strongly motivated by general results of loop quantum gravity
\cite{Rov,ALRev,ThomasRev}: Background independence, i.e.\
quantization in the absence of a metric other than the physical one
determined by $E^a_i$, is realized in a well-defined representation of
smeared basic variables for holonomies together with fluxes as
2-dimensionally integrated densitized triads. But operators for
connection or extrinsic curvature components themselves do not exist.

An orthonormal basis of gauge invariant states in the connection
representation is given by \cite{SphSymm}
\begin{equation} \label{GaugeInvSpinNetwork}
 T_{g,k,\mu}=\prod_{e\in g} \exp\left(\tfrac{1}{2}i k_e
\smallint_e(A_x+\eta')\md x\right)  \prod_{v\in g}
\exp(i\mu_v \gamma K_{\vp}(v))
\end{equation}
with integer labels $k_e$ and positive real labels $\mu_v$ on edges
$e$ and vertices $v$, respectively, forming a finite graph $g$ in the
1-dimensional radial line. The labels determine the connection
dependence by irreducible representations of the groups spanned by the
holonomies. (These groups are U(1) for $A_x$-holonomies and the Bohr
compactification $\bar{\mathbb R}_{\rm Bohr}$ of the real line for
$K_{\vp}$-holonomies; see \cite{SphSymm} for details.)  The densitized
triad, i.e.\ momenta conjugate to the connection components, will be
derivative operators which are quantized in the full theory in the
form of fluxes $\int_S E^a_i\tau^in_a \md^2y$ as integrals over
surfaces $S$ with co-normals $n_a$. Their explicit action in spherical
symmetry is
\begin{eqnarray}
 \hat{E}^x(x) T_{g,k,\mu} &=& \gamma\lP^2
\frac{k_{e^+(x)}+k_{e^-(x)}}{2} T_{g,k,\mu} \label{Exspec}\\
 \int_{\cal I}\hat{E}^{\vp}T_{g,k,\mu} &=& \gamma\lP^2
\sum_{v\in{\cal I}} \mu_v T_{g,k,\mu}\label{Epspec}
\end{eqnarray}
where $\ell_{\rm P}^2=G\hbar$ is the Planck length squared and
$e^{\pm}(x)$ denote the neighboring edges to a point $x$,
distinguished from each other using a given orientation of the radial
line. (We have $k_{e^+(x)}=k_{e^-(x)}$ if $x$ is not a vertex of the
graph.) The $\hat{E}^{\varphi}$-operators only exist in smeared form
after integrating over arbitrary radial intervals ${\cal I}$.
All flux operators have discrete spectra: eigenstates as seen
in (\ref{Exspec}) and (\ref{Epspec}) are normalizable. But only
$\hat{E}^x$ has a discrete set of eigenvalues, while
$\hat{E}^{\varphi}$-eigenvalues fill the real line. (Their eigenstates
are elements of the non-separable Hilbert space of square integrable
functions on the Bohr compactification of the real line.)

These basic operators can be used for composite operators as well,
providing well-defined but rather complicated constraint operators.
Instead of dealing directly with these operators, we will extract some
typical effects as phenomenological corrections to the classical
equations and analyze their potential implications with more
ease. This has been done in quite some detail in cosmological
applications
\cite{PowerLoop,ScaleInvLQC,CCPowerLoop,SuperInflLQC,InhomEvolve,Vector,TensorBackground,Tensor,RefinementInflation,RefinementMatter,TensorHalf},
and we start the same in this paper for inhomogeneous models in the
spherically symmetric context. In this way, we provide the first
examples where full inhomogeneities, rather than perturbative ones as
in cosmology, are studied in this way. But we emphasize that we do not
derive complete effective equations as per
\cite{EffAc,Karpacz,EffCons} in this paper, which is rather an
exploration of possible effects. Nevertheless, restrictions by
consistency already provide interesting lessons. The general
consistency issue is similar to that studied after partial gauge
fixings in spherical symmetry in \cite{SphSymmUniform}, where it was
analyzed based on the general ideas of \cite{ConsistDisc,UniformDisc},
and in \cite{HolBH}. Our analysis here provides complementary results
in a different setting, where we make sure that potential anomalies
arising from different correction terms cancel each other. Moreover,
the effective treatment allows us to arrive more directly at
properties of physical solutions.

\subsection{Implementing the LTB conditions}

The LTB conditions (\ref{ephi}) and (\ref{kphi}) in terms of
densitized triad and extrinsic curvature variables are well-suited to
an implementation at the level of spin networks. They refer directly
to basic expressions of the quantization (provided that one just
exponentiates the relation (\ref{kphi}) to result in holonomies) and
can thus easily be formulated as conditions for kinematical states. In
this way, the LTB reduction can be performed at the quantum
level. However, consistency issues of the dynamics are not easy to
deal with at the complete quantum level. We will therefore describe
the kinematical constructions here, proceed to a phenomenological
effective level in Sec.~\ref{s:CorrLTB} and then study its
consistency. By the link to the initial loop quantization, this
indirect route will nevertheless provide feedback on the full theory.

{}From the triad relation (\ref{ephi}) we derive a condition for fluxes
simply by integrating over arbitrary radial intervals ${\cal I}$:
\begin{equation}
 \int_{\cal I} E^{\vp} = \frac{1}{2} |E^x|_{\partial {\cal I}}
\end{equation}
where $\partial{\cal I}$ is the boundary of ${\cal I}$ at which $E^x$
is evaluated, taking into account orientation to have the correct
signs. This relation can be imposed on triad eigenstates
(\ref{GaugeInvSpinNetwork}), where (\ref{Exspec}) and (\ref{Epspec})
imply
\begin{equation} \label{LTBQuant}
 \mu_v = \frac{1}{2} (|k_{e^+(v)}|-|k_{e^-(v)}|)
\end{equation}
for any vertex $v$. This directly eliminates all vertex labels in
favor of the edge labels which remain free, analogously to the
function $|E^x|=R^2$ which classically determines an LTB metric
completely.

On these reduced states, it turns out, the LTB condition for holonomy
operators is already implemented. Upon integration and exponentiation,
we have
\begin{equation} \label{kphiOp}
 \exp\left({\textstyle\frac{1}{2}}i{\rm sgn}(E^x) 
\smallint_{v_1}^{v_2} (A_x+\eta')
\md x\right)=
 \exp\left({\textstyle\frac{1}{2}}i\gamma K_{\vp}(v_1)\right)
\exp\left(-{\textstyle\frac{1}{2}}i\gamma K_{\vp}(v_2)\right)
\end{equation}
expressed solely in terms of elementary holonomy operators. This
condition is realized in the sense that the left and right hand sides,
as multiplication operators, have the same action on solutions to the
LTB condition satisfying (\ref{LTBQuant}). In fact, the left hand side
simply increases the label of the edge between $v_1$ and $v_2$ (which
we assume to be two adjacent vertices) by one. Thus, it changes both
$k_{e^+(v_1)}$ and $k_{e^-(v_2)}$ by $\pm1$ depending on their sign.
The two operators on the right hand side, on the other hand, change
the vertex label $\mu_{v_1}$ by $\frac{1}{2}$ and $\mu_{v_2}$ by
$-\frac{1}{2}$ in the right way to respect the condition
(\ref{LTBQuant}) if it was realized for the original state. (If there
are vertices $v$ between $v_1$ and $v_2$, $k_{e^+(v)}$ and
$k_{e^-(v)}$ change by the same value such that (\ref{LTBQuant})
remains implemented without changing $\mu_v$.)

Notice that, unlike conditions for a symmetry reduction, the two LTB
conditions for densitized triads and extrinsic curvature have
vanishing Poisson brackets with each other (but not with the
constraints).  Thus, the curvature condition can indeed be implemented
on the solution space of the triad condition. It does not add further
conditions for states because they are written in a specific
polarization. LTB states are thus simply represented by a chain of
integer labels $k_n$ for $n=0,1,\ldots$ which represents spatial
discreteness (a 1-dimensional lattice of independent sites) as well as
the discreteness of quantum geometry (integer $k_n$ as eigenvalues of
the area radius squared). In a connection representation, they can be
written as $ T_{\vec{k}}(z_0,z_1,\ldots)= \prod_n z_n^{k_n}$ where the
assignment $n\mapsto z_n:= \exp({\textstyle\frac{1}{2}}i\int_{e_n}
\gamma K_x\md x)$ is a generalized LTB connection.

While states can be reduced immediately to implement the LTB
conditions, further conditions do result for composite operators
because (\ref{kphiOp}) must be used if the action of any operator is
to be written on the LTB states where (\ref{LTBQuant}) has eliminated
vertex labels. This provides reductions, e.g.\ of constraint
operators, such that characteristic quantum gravity effects in loop
operators can be carried over to constraints for an LTB model.

\subsection{Inverse triad effects}
\label{s:InvTriad}

The first effect we turn to arises from the required quantization of
inverse powers of the densitized triad, such as $E^x$ in (\ref{hg}).
There is no direct quantization of such an inverse since $\hat{E}^x$
in a loop quantization (\ref{Exspec}) has a discrete spectrum
containing zero, and thus lacks an inverse operator. Nevertheless, one
can use general techniques to arrive at a well-defined operator which
reproduces $(E^x)^{-1}$ in a classical limit.  This quantization is
based on the Poisson relation between a holonomy and the volume, which
one can identify with an expression for
$\frac{1}{\sqrt{|E^{x}|}}$. More precisely, we have $4\pi\gamma G{\rm
sgn}(E^x)E^{\vp}/\sqrt{|E^x|}= \{A_x,V\}$ for the combination of triad
components appearing in the first term of (\ref{hg}), where
$V=4\pi\int\md x \sqrt{|E^x|}E^{\vp}$ is the classical expression for
volume in spherically symmetric setting. In these expressions, we
follow general constructions of the full theory \cite{QSDI}.

When quantized, the connection component is expressed through a
holonomy, and the Poisson bracket becomes a commutator. In order to
stay as close to the full theory as possible, we use SU(2)-holonomies
\begin{equation}
 h_x(A) = \exp(\tau_3\smallint A_x)=
 \cos({\textstyle\frac{1}{2}}\smallint A_x)+ 2\tau_3 
\sin({\textstyle\frac{1}{2}}\smallint A_x)
\end{equation}
which in their matrix elements provide the basic quantities
(\ref{hx}). (Generators of SU(2) are $\tau_j=-\frac{1}{2}i\sigma_j$ in
terms of Pauli matrices $\sigma_j$; path ordering is not necessary for
radial holonomies thanks to the symmetry reduction, which reduces the
gauge group to an Abelian one \cite{SymmRed}.) The commutator for these
holonomies becomes
\begin{eqnarray}
 h_x[h_x^{-1},\hat{V}] &=& \hat{V}-\cos(\tfrac{1}{2}\smallint A_x)
 \hat{V}\cos(\tfrac{1}{2}\smallint A_x)- \sin(\tfrac{1}{2}\smallint A_x )
 \hat{V}\sin(\tfrac{1}{2}\smallint A_x)\\
 &&+ 2\tau_3 \left(\cos(\tfrac{1}{2}\smallint A_x)
 \hat{V}\sin(\tfrac{1}{2}\smallint A_x)-\sin(\tfrac{1}{2}\smallint A_x )
 \hat{V}\cos(\tfrac{1}{2}\smallint A_x)\right) \nonumber
\end{eqnarray}
and appears in the constraint in the form ${\rm
tr}(\tau_3h_x[h_x^{-1},\hat{V}])$. This can be used in a quantization of
\[
{\rm
tr}(\tau_3h_x\{h_x^{-1},V\})= -{\rm tr}(\tau_3^2\{\smallint
A_x,V\})
=\frac{1}{2} \int_e\{A_x,V\}\sim \frac{1}{2}\ell_0 \{A_x,V\}
\]
where $\ell_0$ is the coordinate length of the edge used. When
inserted in the Hamiltonian constraint, $\ell_0$ for all edges
discretizes the integration measure $\md x$.  Eigenvalues can be
computed easily from the basic action of holonomies and fluxes: for
the operator
\begin{eqnarray}
 \widehat{\int_{\cal I} \frac{E^{\varphi}{\rm sgn}(E^x)}{\sqrt{|E^x|}}} &=&
\frac{-i}{2\pi\gamma G\hbar} {\rm
tr}(\tau_3h_x[h_x^{-1},\hat{V}])\label{InverseEx}\\
 &=& \frac{-i}{2\pi \gamma G\hbar} \left(\sin(\tfrac{1}{2}\smallint A_x)
 \hat{V}\cos(\tfrac{1}{2}\smallint A_x)-\cos(\tfrac{1}{2}\smallint A_x )
 \hat{V}\sin(\tfrac{1}{2}\smallint A_x)\right)\nonumber
\end{eqnarray}
we have eigenvalues
\begin{equation}
 \left(\widehat{\int_{\cal I}
 \frac{E^{\varphi}{\rm sgn}(E^x)}{\sqrt{|E^x|}}}\right)_{k,\mu} =
 2\sqrt{\gamma}\ell_{\rm P}
 |\mu_v|\left(\sqrt{|k_{e^+(v)}+k_{e^-(v)}+1|}-
\sqrt{|k_{e^+(v)}+k_{e^-(v)}-1|}\right)
\end{equation}
where $v$ is the starting point of the interval ${\cal I}$ used as the
edge in the holonomy.

Looked at for all values of $E^x$, eigenvalues of the resulting
operator do not agree exactly with the classical function
$E^{\varphi}/\sqrt{|E^x|}$ but show deviations especially at small
$E^x$. We can parameterize this by a correction function
$\alpha(E^{x})$ as
\begin{equation} \label{alpha}
\alpha({E^{x}}):=
\left(\widehat{\frac{1}{\sqrt{|E^x|}}}\right)_{k(E^x)}
\left(\sqrt{|\hat{E}^x|}\right)_{k(E^x)} =
2\frac{\sqrt{|E^{x}+\gamma \lP^{2}/2|}-
\sqrt{|E^{x}-\gamma \lP^{2}/2|}}{\gamma \lP^{2}}\sqrt{|E^{x}|}
\end{equation}
where the subscript $k(E^x)$ means that the eigenvalue of the operator
is taken at label $k_{e^+(v)}+k_{e^-(v)}=2E^x/\gamma\lP^2$ as it
follows from (\ref{Exspec}). Classically, i.e.\ for $\ell_{\rm
P}\to0$, we have $\alpha(E^x)=1$, and this limit is approached by
$\alpha$ for large $E^x$. With the correction, the expression for the
inverse power of the triad component $E^x$ is finite, just as in the
isotropic case \cite{InvScale}. The general behavior of the correction
function is illustrated in Fig.~\ref{LTBalpha} below. Similar
constructions have been used before in spherical symmetry, see e.g.\
\cite{HWBlackHole}. (While finiteness of inverse triad operators is
realized in isotropic and spherically symmetric models, this is not
expected to be a general property
\cite{BoundFull,DegFull}. Nevertheless, inverse triad operators are
well-defined in general situations of loop quantum gravity
\cite{QSDV}.)

One should note that $E^x$ refers to the total area of a whole orbit
at radius $x$, which can have macroscopic values. In this case,
$\alpha$ only slightly differs from one. However, a fully
inhomogeneous quantization would refer to flux values of individual
microscopic patches, where $E^x=\sum_np_n$ is a large sum of
microscopic contributions $p_n$, together giving the whole orbit
area. The single fluxes $p_n$ are much smaller and closer to the
Planck scale, which makes $\alpha$ differ from one more noticeably if
these fundamental fluxes are used. This is an illustration of the fact
that symmetric models often artificially suppress corrections from
inverse triad operators, as first noted in \cite{SchwarzN}. Inverse
triad corrections analogous to $\alpha$ have thus occasionally been
underestimated, especially in isotropic models of macroscopic
universes with large matter content. One can model the enhancements of
corrections in fully inhomogeneous states even in symmetric models by
using higher SU(2)-representations of holonomies in operators, not
just the fundamental one as understood in (\ref{InverseEx}). Then, the
expression for $\alpha$ changes essentially by replacing $\gamma
\lP^{2}$ in (\ref{alpha}) by $j\gamma \lP^{2}$ if $j$ is the spin of
the representation. (See \cite{Ambig,QuantCorrPert} for precise
formulas in those cases.) For our qualitative analysis here we can
focus on the expression (\ref{alpha}). In fact, later applications
mainly use the small-$E^x$ behavior near a center or a central
singularity where corrections are strong for any $j$.

Because classically, for $\lP\to0$, the function $\alpha(E^x)$
approaches one, the classical limit is correct if this function is
inserted as a multiplier of $1/\sqrt{E^x}$ in the Hamiltonian
constraint.  We can thus write (\ref{hg}) as
\begin{eqnarray}
H^{(I)}_{\rm grav}&=&-\frac{1}{G}\left(\frac{\sqrt{|E^{x}+\gamma \lP^{2}/2|}-
\sqrt{|E^{x}-\gamma \lP^{2}/2|}}{\gamma \lP^{2}}
K_{\varphi}^{2}E^{\varphi}+K_{\varphi}K_{x}\sqrt{|E^{x}|}\right)
\nonumber\\
&=& -\frac{1}{2G}\left(\frac{\alpha(E^x)}{\sqrt{|E^x|}} 
K_{\varphi}^2E^{\varphi} +2K_{\varphi}K_{x}
\sqrt{|E^{x}|}\right)\,. \label{Ham1}
\end{eqnarray}

The form of corrections is usually not fully unique due to the
presence of quantization ambiguities. Sometimes, however, they can be
restricted more strongly by relating, whenever possible, reduced
expressions to what one expects in the full theory. This provides
different options for specific corrections which can be analyzed for
robustness and the phenomenology they imply. In the present case, the
full theory may suggest an alternative corrected Hamiltonian
\begin{equation} \label{Ham2}
H^{(II)}_{\rm grav}= -\frac{1}{2G}\frac{\alpha(E^x)}{\sqrt{|E^x|}} 
(K_{\varphi}^2E^{\varphi} +2K_{\varphi}K_{x}E^{x})\,.
\end{equation}
Here, also the second term carries a correction function, which is
motivated if one takes into account that the $\sqrt{|E^x|}$ in the
second term of (\ref{hg}) arises from a cancellation in
$E^x/\sqrt{|E^x|}$ after inserting spherically symmetric variables
into the full constraint. Thus, the second term could also be expected
to have a correction by $\alpha$. We will analyze both cases below and
describe the differences they imply.

\subsection{Holonomy effects and lattice refinements}
\label{s:HolRef}

Another characteristic, and in fact eponymous, feature of loop
quantum gravity is that not components of the connection but rather
its holonomies are represented as operators. Since these are
non-linear objects, additional corrections by higher order terms of
the connection (or extrinsic curvature) will be present which again
can be evaluated by including them as correction terms in
phenomenological equations. This may appear as higher curvature
corrections as they involve higher powers of extrinsic curvature, but
we emphasize that this procedure will not give a complete picture
since higher derivative terms are missing. These can be computed at an
effective level \cite{EffAc,Karpacz,EffCons}, which would require much
more work not pursued here.

We can correct for the holonomy effects in the Hamiltonian constraint
(\ref{hg}) as follows. We assume $K_{x}$ to be fairly constant over a
given edge (or part of an edge for graph-changing operators) of the
graph whose holonomies appear as basic loop variables, so that
$\int_{v_{-}}^{v_{+}}K_{x}\approx \ell_0K_{x}$ where $\ell_0$ is the
coordinate length of the edge lying between the vertices $v_{-}$ and
$v_{+}$. Rather than using a precise loop quantization of (\ref{hg})
and computing its expectation value in terms of holonomies, we make
the following replacements in the Hamiltonian constraint:
$K_{\varphi}\rightarrow (\gamma\delta)^{-1}\sin(\gamma\delta
K_{\varphi})$ and $K_{x}\rightarrow (\gamma \ell_0)^{-1}\sin(\gamma
K_{x}\ell_0)$.  In addition to $\ell_0$, $\delta$ is a dimensionless
parameter whose role is discussed below.  With these corrections, the
constraint becomes
\begin{equation}
H^{(III)}_{\rm grav}=-\frac{1}{2G}\left(\frac{\sin^{2}(\gamma\delta
  K_{\varphi})}{\gamma^{2}\delta^{2}}
\frac{E^{\varphi}}{\sqrt{|E^{x}|}}+
2\frac{\sin(\gamma\delta K_{\varphi})}{\gamma\delta}
\frac{\sin(\gamma K_{x}\ell_0)}{\gamma \ell_0}\sqrt{|E^{x}|}\right)\,.
\label{hgnew}
\end{equation}

While this may not be the precise result from a complete effective
constraint, it captures the main effects of using holonomies as
periodic, rather than linear, functions in the curvature
components. Moreover, this simplest choice guarantees that also here
the classical limit, which involves a continuum limit $\delta\to0$ and
$\ell_0\to0$, is satisfied. Note, however, that one should not take
the full functional form too seriously but rather view the sine
functions as a convenient place-holder for a perturbative expansion in
higher powers of the $K$-components. Since there will be other
corrections as mentioned above, they could easily dominate most of the
expansion terms.

Although the parameters $\delta$ and $ \ell_0$ appear in similar
forms, their origins and roles are quite different. The parameter $
\ell_0$ arises as the coordinate length of a radial edge along which
we compute a holonomy. Its size is determined by the embedded graph we
act on, as well as the precise form of the Hamiltonian constraint
operator understood in the construction. While a Hamiltonian operator
does not depend on the embedding and thus $\ell_0$, that dependence
would arise in the cause of computing an effective Hamiltonian as the
expectation value in a state peaked on a classical
geometry. Specifying the classical geometry requires one to partially
fix the diffeomorphism gauge; the size of $\ell_0$ is then a direct
measure for the discreteness of the state in this setting. Also
$\delta$ measures the discreteness, but it does not refer to the
length of any edge. It is associated with curvature components
$K_{\varphi}$ along spherical symmetry orbits, and there is no room
for orbital edges in this reduced model. To understand the meaning of
$\delta$, we again have to look at what it should correspond to in a
full, unreduced setting as we did in Sec.~\ref{s:InvTriad} to discuss
the size of inverse triad corrections $\alpha$. In a fully
inhomogeneous setting, there would now be edges along spherical orbits
whose lengths correspond to $\delta$. For a configuration which is
nearly spherical, there should be a regular distribution of edges
forming a lattice on each symmetry orbit. The length of each edge, and
thus $\delta$, would decrease with an increasing number of lattice
plaquettes ${\cal N}$: $\delta\propto {\cal N}^{-1/2}$. In particular,
for finer lattices we have $\delta\to 0$ just as $\ell_0\to 0$,
approaching the continuum limit.

The precise form of $\delta$ depends on the exact state which is
approximated by a spherically symmetric one. In particular, the
argument shows that $\delta$, unlike $\ell_0$, may be phase-space
dependent if geometrical growth is accompanied by a refinement of the
lattice such that ${\cal N}(E^x)$ depends on $E^x$, e.g.\ by a
power-law form ${\cal N}(E^x)\propto |E^x|^k$. (Note that $4\pi
|E^x(x)|$ is the area of a sphere at radius $x$, and thus coordinate
independent. In fact, a densitized triad $E^x$ in one dimension
behaves like a scalar.)

In this way, we are naturally led to a refinement scheme of
phase-space dependent holonomies where point holonomies associated
with $K_{\varphi}$ depend on $E^x$, while $K_x$-holonomies along the
inhomogeneous direction are triad independent. In a reduction to
anisotropic but homogeneous models as in \cite{SchwarzN}, this
specific form has been shown to imply a dynamical law given by a
fundamental difference equation which cannot be implemented
equidistantly in minisuperspace variables. Equidistant versions of the
difference equation, which would result if holonomies for a connection
or extrinsic curvature component depend only on its conjugate triad
component, i.e.\ $E^{\varphi}$ for $K_{\varphi}$-holonomies, are not
embeddable in a spherically symmetric model. In this way,
inhomogeneous models can reduce some of the freedom involved in
choosing a refinement scheme for a homogeneous model. Nevertheless, at
least a 1-parameter freedom of ${\cal N}(E^x)\propto |E^x|^k$, or even
a different functional behavior, is left. It is only the direction
dependence of ${\cal N}$ which is restricted, not the
size-dependence. In particular, it is impossible to restrict the
corresponding freedom in isotropic models. What we can also see is the
fact that spherically symmetric refinement schemes which do not depend
on any auxiliary scales can give rise to apparently scale dependent
equations in a homogeneous reduction: when reduced to isotropy, a
non-trivial refinement scheme can always be expressed by a function
${\cal N}(a)$ of the scale factor $a$. In contrast to $E^x$, $a$ is
coordinate dependent since it rescales if spatial coordinates are
multiplied by a constant. The reduction from spherical symmetry thus
must automatically introduce a scaling dependent parameter $f_0$ such
that ${\cal N}(a)$ depends only on the coordinate independent
combination $f_0a$. This results in a well-defined way of non-trivial
refinement schemes with all the freedom as it is realized in spherical
symmetry.  Moreover, since the scaling dependence arises only in the
reduction to homogeneity, it cannot be used as a reliable criterion to
rule out refinement schemes if it is applied purely in homogeneous
situations.

The lattice refinement behavior is to be expected in any model on
general grounds \cite{InhomLattice,CosConst}; while a direct
derivation of the behavior of ${\cal N}(E^x)$ from a full Hamiltonian
operator is difficult, one can arrive at some properties and analyze
their consequences phenomenologically. For cosmology, such work has
been done in \cite{RefinementInflation,RefinementMatter,Vector,Tensor}
and is initiated here for black hole physics. (The interior of the
Schwarzschild black hole, which can be formulated as a homogeneous
model, has been studied from this perspective in
\cite{SchwarzN,BHIntHol,KSmubar}.)

\section{Corrected LTB models}
\label{s:CorrLTB}

An LTB reduction at the dynamical quantum level of spherically
symmetric systems is difficult because the combined algebra of LTB
conditions and constraints, when seen as an extended constrained
system, is not purely first class. Although, as mentioned, the
classical LTB conditions are preserved by the equations of motion
generated by the spherically symmetric Hamiltonian constraint, there
is no simple off-shell algebra between these functionals which one
could represent on the Hilbert space generated by spherically
symmetric spin network states. The various versions of quantum
corrected Hamiltonians we have provided so far are thus not yet LTB
reduced, although we have already removed the spin connection terms as
they drop out in the classical LTB reduction. Moreover, although we
did see that the classical LTB conditions can directly be taken over
to the kinematical quantum level, such a step is much more complicated
when combined with the quantum constraint algebra. If overall
consistency with the constraints is required, the LTB conditions
themselves may well require quantum corrections, too. In this section,
we will be exploring the possibility of LTB-like solutions at a
phenomenological effective level, allowing for corrections to
constraints as well as the classical LTB conditions.

At this stage, we still have two canonical pairs and two smeared
constraints (the diffeomorphism and corrected Hamiltonian
constraint). The corrected constraints in this form are automatically
consistent (i.e.\ first class) since the absence of the spin
connection terms implies the absence of spatial derivatives in $H_{\rm
grav}$; the Hamiltonian constraints thus commute with themselves. Even
if we add the non-dynamical dust contribution $H_{\rm dust}= F'/2G$ in
terms of the mass function $F(x)$, the system remains consistent. We
have dropped the spin connection terms in anticipation of the
imposition of LTB conditions, which solve the diffeomorphism
constraint identically and thus show that the Abelian Poisson bracket
$\{H[N],H[M]\}=0$, realized even in the quantum corrected case, is
correct. With such an algebra, the LTB form allows us to discuss the
anomaly issue more easily. However, there is still a potential anomaly
problem, whose solution allows us to draw feedback for quantizations
of the Hamiltonian constraint: the reduction to constraints of LTB
form is consistent only if there are LTB-like conditions, relating
$E^{\varphi}$ to $(E^x)'$ and $K_{\varphi}'$ to $K_x$, which are
preserved by the quantum corrected spherically symmetric
constraints. The requirement of preserved LTB conditions will, as we
will see, restrict the form of quantum corrections in different terms
of the constraints.

\subsection{Consistent LTB reductions}

To derive equations of motion for the metric component $R$ left as the
only degree of freedom in an LTB metric, we eliminate $K_{\varphi}$
and $K_{x}$ in favor of $\dot{E}^{\varphi}$ and $\dot{E}^{x}$ using
the equations of motion $\dot{E}^{I}=\{E^{I},H\}$ where
$I\in\{\varphi,x\}$. To evaluate these expressions we make use of the
canonical Poisson bracket relations
$\{K_{x}(x),E^{x}(y)\}=2G\delta(x-y)$ and
$\{K_{\varphi}(x),E^{\varphi}(y)\}=G\delta(x-y)$ as they follow from
(\ref{Poisson}), such that
\begin{equation}
K_{\varphi}=\frac{\dot{E}^{x}}{2\sqrt{|E^{x}|}}
\quad,\quad
K_{x}=\frac{1}{\sqrt{|E^{x}|}}\left(\dot{E}^{\varphi}-
\frac{\dot{E}^{x}E^{\varphi}}{2E^{x}}\right)
\end{equation}
for the classical constraint.

The key problem now is that (\ref{ephi}) combined with (\ref{kphi}) is
no longer preserved by the evolution equations generated by
(\ref{Ham1}) or (\ref{Ham2}) for $\alpha\not=1$, as can directly be
checked. If these equations were consistent, one could eliminate the
variable $E^{\varphi}$ in favor of $E^x$ in all equations of motion,
which is then expressed as $E^x=R^2$. (From now on we assume $E^x>0$
without loss of generality for the applications we are interested in.)
Thus, a complete set of equations given by the Hamiltonian constraint
for $\dot{R}$ and the second order evolution equation for $R$ would be
obtained. When the LTB conditions are not preserved, however, the
Hamiltonian constraint equation for $R$ will not be preserved by the
evolution equation.

Before deriving quantum corrected equations for $R$, we thus have to
find LTB conditions suitable for the quantum corrected dynamics. The
main conditions are (i) that they reduce to the classical LTB
conditions when quantum corrections vanish, (ii) that they still solve
the uncorrected diffeomorphism constraint identically as this is
necessary for a consistent constraint algebra, and (iii) that they be
preserved by the quantum corrected evolution equations. Condition (ii)
is also motivated by the fact that finite diffeomorphisms are
represented in loop quantum gravity directly by the action they
generate on phase space functions without requiring any quantum
corrections.

\subsubsection{Inverse triad corrections: first version}
\label{s:Corr1}

In agreement with the diffeomorphism constraint, we make the ansatz
\begin{equation} \label{LTB1}
(E^{x})'=2f(E^{x})E^{\varphi} \quad,\quad K_{\varphi}'=f(E^{x})K_{x}
\end{equation}
where, as indicated, $f(E^{x})$ is assumed to depend only on $E^{x}$
in algebraic form.  This function will be determined by demanding that
the new LTB conditions are preserved in time by (\ref{Ham1}). Writing
the first constraint in the smeared form $C_{\rm LTB}=\int \md
x\mu(x)(2f(E^{x})E^{\varphi}-(E^{x})')$, we require that the Poisson
bracket $\{C_{\rm LTB},\int \md yH^{(I)}_{\rm grav}\}$, which evaluates to
\begin{equation}
 \int \md z\mu(z)\left(-4K_{\varphi}E^{\varphi}\sqrt{E^{x}}\frac{\md
 f}{\md E^{x}}+ 2K_{\varphi}'\sqrt{E^{x}}+
 \frac{K_{\varphi}(E^{x})'}{\sqrt{E^{x}}}- \frac{2f\alpha
 K_{\varphi}E^{\varphi}}{\sqrt{E^{x}}}- 2fK_{x}\sqrt{E^{x}}\right)\,,
\end{equation}
vanishes for all $\mu(x)$. Using (\ref{LTB1}) to remove the derivative
terms we get
\begin{equation}
 \int \md
 z\mu(z)\left(-4K_{\varphi}E^{\varphi}\sqrt{E^{x}}\frac{\md f}{\md E^{x}}+
 \frac{2fK_{\varphi}E^{\varphi}}{\sqrt{E^{x}}}- \frac{2f\alpha
 K_{\varphi}E^{\varphi}}{\sqrt{E^{x}}}\right)\,.
\end{equation}
For this to vanish for all $\mu$, the integrand must vanish which
therefore gives the differential equation
\begin{equation} \label{falpha}
f(1-\alpha)=2E^{x}\frac{\md f}{\md E^{x}}
\end{equation}
for $f(E^{x})$ which, for $\alpha$ as in (\ref{alpha}), is solved by
\begin{equation}
 f(E^{x})=\frac{c_1\sqrt{E^{x}}e^{-\alpha/2}}{\left(\sqrt{E^{x}}+
 \sqrt{E^{x}-\gamma \lP^{2}/2}\right)^{1/2} \left(\sqrt{E^{x}}+
 \sqrt{E^{x}+\gamma \lP^{2}/2}\right)^{1/2}}\,.
\end{equation}
for $E^x>\gamma\ell_{\rm P}^2/2$ and
\begin{equation}
 f(E^{x})= \frac{c_2\sqrt{E^x} \exp\left(-\frac{1}{2}\alpha+\frac{1}{2}\arctan
 \left(\sqrt{E^x/(\gamma\ell_{\rm P}^2/2-E^x)}\right)\right)}{
 \sqrt{\sqrt{E^x}+\sqrt{E^x+\gamma\ell_{\rm P}^2/2}}}
\end{equation}
for $E^x<\gamma\ell_{\rm P}^2/2$.  The constant $c_1$ is fixed by
demanding that in the limit $E^{x}\rightarrow\infty$ the corrected LTB
conditions should go to their classical form. To ensure
$f(E^{x})\rightarrow1$ in the classical limit we have
$c_1=2\sqrt{e}$. The functions $f$ and $\alpha$ then have similar
fall-off behaviors at large $E^x$: $f(E^x)\sim
1+2^{-7}\gamma^2\ell_{\rm P}^4(E^x)^{-2}+\cdots$ while
$\alpha(E^x)\sim 1+2^{-5}\gamma^2\ell_{\rm P}^4(E^x)^{-2}+\cdots$. The
second constant $c_2$ is determined by continuity at
$E^x=\gamma\ell_{\rm P}^2/2$, which gives $c_2=
2^{5/4}e^{1/2-\pi/4}\gamma^{-1/4}\ell_{\rm P}^{-1/2}$.  It is easy to
check that with this form for $f(E^{x})$ the other LTB condition, $
K_{\varphi}'=f(E^{x})K_{x}$, is also preserved in time.

We now eliminate the connection components from the Hamiltonian in
favor of the triad components $E^{x}$ and $E^{\varphi}$ using the
equations of motion which give
\begin{equation} \label{kpx}
 K_{\varphi}=\frac{\dot{E}^{x}}{2\sqrt{E^{x}}}
\quad,\quad
 K_{x}=\frac{1}{\sqrt{E^{x}}}\left(\dot{E}^{\varphi}-
 \alpha\frac{\dot{E}^{x}E^{\varphi}}{2E^{x}}\right)\,.
\end{equation} 
We then further eliminate $E^{\varphi}$ using the new LTB conditions
to obtain
\begin{equation}
H^{(I)}_{\rm grav}=-\frac{1}{2G}\left(\frac{\alpha
(\dot{E}^x)^2(E^{x})'}{8f(E^{x})^{3/2}}+
\frac{\dot{E}^{x}(\dot{E}^{x})'}{2f\sqrt{E^{x}}}-
\frac{(\dot{E}^x)^2(E^{x})'}{4f(E^{x})^{3/2}}\right)\,.
\end{equation}
The total Hamiltonian constraint $H^{(I)}_{\rm grav}+H_{\rm dust}=0$
then becomes
\begin{equation} \label{ConsHam1E}
\frac{\alpha (\dot{E}^{x})^2(E^{x})'}{8f(E^{x})^{3/2}}+
\frac{\dot{E}^{x}(\dot{E}^{x})'}{2f\sqrt{E^{x}}}-
\frac{(\dot{E}^{x})^2(E^{x})'}{4f(E^{x})^{3/2}}-F'=0
\end{equation}
with the given $E^x$-dependence of $f$ and $\alpha$.

To obtain the evolution equation we take a time derivative of the
first equation in (\ref{kpx}) to obtain
\begin{equation} \label{kphidot one}
 \dot{K}_{\varphi}=\frac{2E^{x}\ddot{E^{x}}-
 (\dot{E}^{x})^2}{4|E^{x}|^{3/2}}\,.
\end{equation}
Since $\dot{K}_{\varphi}$ is also determined by
$\dot{K}_{\varphi}=\{K_{\varphi}, H^{(I)}_{\rm grav}\}=
-\alpha K_{\varphi}^{2}/2\sqrt{E^{x}}$
we have
\begin{equation} \label{evolution eqE}
4E^{x}\ddot{E}^{x}-(2-\alpha)(\dot{E}^x)^2=0
\end{equation}
We note that in this derivation of the evolution equation, $f(E^{x})$
does not appear anywhere. Nevertheless, its form is important for the
mutual consistency of the evolution equation (\ref{evolution eqE}) and the
Hamiltonian constraint equation (\ref{ConsHam1E}).

\begin{figure}
\begin{center}
\includegraphics[width=13cm]{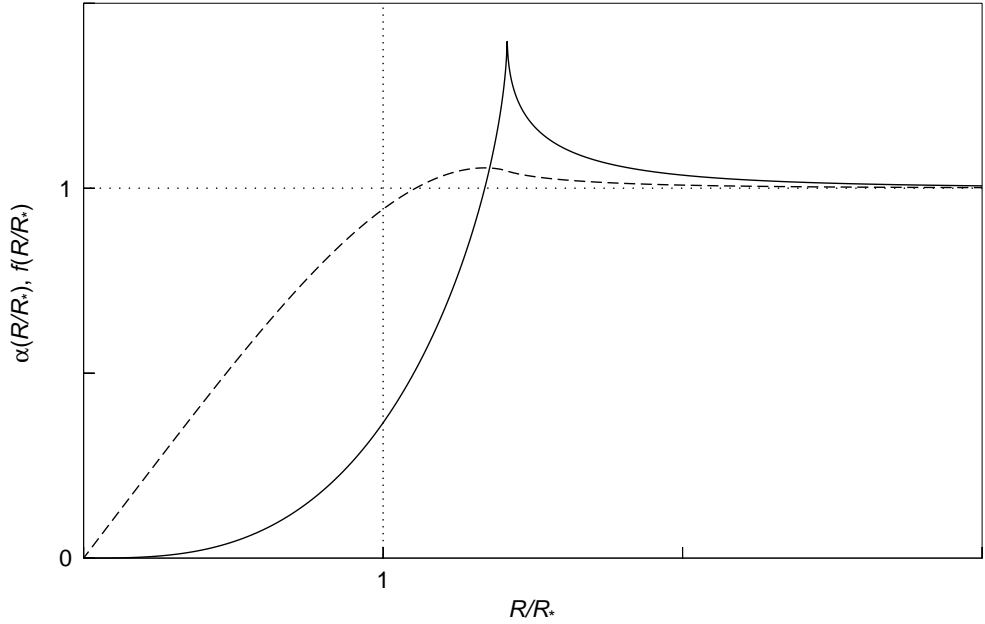}
\caption{\label{LTBalpha} The correction functions $\alpha(R)$ (solid) and
  $f(R)$ (dashed) where $R$ is taken relative to $R_*:=\sqrt{\gamma/2}\ell_{\rm
  P}$.}
\end{center}
\end{figure}

We can now write everything in terms of $R$ using the relation
$E^{x}=R^{2}$:
\begin{eqnarray} \label{alpha r}
 \alpha(R)&=&2\frac{\sqrt{|R^{2}+\gamma \lP^{2}/2|}-
 \sqrt{|R^{2}-\gamma \lP^{2}/2|}}{\gamma \lP^{2}}R\\
\label{f of r}
f(R)&=&\left\{\begin{array}{cl}
\frac{2R\exp(\frac{1}{2}(1-\alpha(R)))}{(R+\sqrt{R^{2}-\gamma \lP^{2}/2})^{1/2}
(R+\sqrt{R^{2}+\gamma \lP^{2}/2})^{1/2}} & \mbox{ for
} R>\sqrt{\gamma/2}\ell_{\rm P}\\
\frac{2^{5/4} R\exp\left(\frac{1}{2}(1-\alpha(R))+\frac{1}{2}\arctan
 \left(\sqrt{R^2/(\gamma\ell_{\rm P}^2/2-R^2)}\right)-\pi/4\right)}{
 \gamma^{1/4}\sqrt{\ell_{\rm P}}
\sqrt{R+\sqrt{R^2+\gamma\ell_{\rm P}^2/2}}} & \mbox{ for
} R<\sqrt{\gamma/2}\ell_{\rm P} \end{array}\right.\,.
\end{eqnarray}
These functions are shown in Fig.~\ref{LTBalpha}.
The first order equation (\ref{ConsHam1E}) can be written as 
\begin{equation} \label{ConsHam1}
 \dot{R}^{2}R'(\alpha(R)-1)+2R\dot{R}\dot{R}'+\dot{R}^{2}R'=f(R)F'
\end{equation}
and the evolution equation is
\begin{equation} \label{evolution eq}
 2R\ddot{R}+\dot{R}^{2}+(\alpha(R)-1)\dot{R}^{2}=0\,.
\end{equation}  
These equations can explicitly be seen to be consistent upon using the
differential relation (\ref{falpha}) between $f$ and $\alpha$.

\subsubsection{Inverse triad corrections: second version}
\label{s:Corr2}

Starting from (\ref{Ham2}), which gives equations of motion
\begin{eqnarray}
 \dot{E}^x &=& 2\alpha K_{\varphi}\sqrt{E^x}\quad,\quad
 \dot{E}^{\varphi} = \alpha K_{\varphi} \frac{E^{\varphi}}{\sqrt{E^x}}
 +\alpha K_x\sqrt{E^x}\\
 \dot{K}_x &=& \alpha K_{\varphi}^2 \frac{E^{\varphi}}{2(E^x)^{3/2}}-
 \alpha K_{\varphi}\frac{K_x}{\sqrt{E^x}}- \frac{\md\alpha}{\md E^x}
 K_{\varphi}^2\frac{E^{\varphi}}{\sqrt{E^x}}- 2\frac{\md\alpha}{\md E^x}
 K_{\varphi}K_x\sqrt{E^x}\\
 \dot{K}_{\varphi} &=& -\alpha \frac{K_{\varphi}^2}{2\sqrt{E^x}}
\end{eqnarray}
we can proceed similarly.  Also with these equations, the classical
LTB conditions will not be preserved such that the reduction would be
inconsistent. However, one can verify that the corrected LTB
conditions
\begin{equation}
 E^{\varphi}=\frac{1}{2\alpha}(E^x)' \quad,\quad \alpha K_x=K_{\varphi}'
\end{equation}
are preserved as before. 

Proceeding with these equations of motion and LTB conditions, we obtain
\begin{equation}
 \left(\frac{\dot{R}^2R}{\alpha^2}\right)'-F'=0
\end{equation}
as the Hamiltonian constraint equation in the presence of dust. Thus,
the corrected equation for $R$ is
\begin{equation} \label{ConsHam2}
 \dot{R}^2R=\alpha^2(F(r)+c(t))\,.
\end{equation}
The evolution equation derived via $\dot{K}_{\varphi}$ is
\begin{equation} \label{EOM2}
 2R\ddot{R}+\dot{R}^2= 2\frac{\md\log\alpha}{\md\log R} \dot{R}^2
\end{equation}
with a quantum correction on the right hand side. Taking a time
derivative of (\ref{ConsHam2}), using $\dot{F}=0$ and eliminating $F$
from the resulting equation via (\ref{ConsHam2}) indeed produces
(\ref{EOM2}) provided $c(t)=c={\rm const}$. Thus, the system is
consistent, and the freedom of $c(t)$ in (\ref{ConsHam2}) is reduced
to a constant which can be absorbed in the mass function.

\subsubsection{Holonomy corrections}
\label{s:CorrHol}

We now look for a consistent formulation with corrections due to
holonomy effects. It turns out that the Hamiltonian (\ref{hgnew})
leads to equations which are algebraically complicated to handle.  To
avoid technical difficulties, we first look at a Hamiltonian where
$K_{\varphi}$ appears via the function
$(\gamma\delta)^{-1}\sin(\gamma\delta K_{\varphi})$ but $K_x$ has its
classical appearance (i.e.\ the continuum limit $\ell_0\to0$ has been
taken):
\begin{equation}
 H^{(IIIa)}_{\rm grav}=-\frac{1}{2G}\left(\frac{\sin^{2}\left(\gamma\delta
 K_{\varphi}\right)}{\gamma^{2}\delta^{2}}\frac{E^{\varphi}}{\sqrt{E^{x}}}+
 2\frac{\sin\left(\gamma\delta
 K_{\varphi}\right)}{\gamma\delta}K_{x}\sqrt{E^{x}}\right)\,.
 \label{kphi corrected hamiltonian}
\end{equation}
This describes holonomy corrections in regions where $K_{\varphi}$ is
large but $K_x$ remains small, or for states with a dense radial
lattice such that $\ell_0$ is small.  Again, the classical LTB
conditions would not be preserved, and therefore we look for an
alternative of the form
\begin{equation}
 (E^{x})'=2g(K_{\varphi})E^{\varphi} \quad,\quad K_{\varphi}'=
g(K_{\varphi})K_{x}
\end{equation}
compatible with the diffeomorphism constraint, where $g(K_{\varphi})$
is assumed to depend only on $K_{\varphi}$. As before, this dependence
will be self-consistently verified by demanding that the new LTB
conditions are preserved in time: $\{\int \md
x\mu(x)(2gE^{\varphi}-(E^{x})'),\int \md yH^{(IIIa)}_{\rm grav}\}=0$
for all $\mu(x)$. This Poisson bracket evaluates to
\begin{eqnarray}
 \int \md z\mu(z)\left(\frac{\sin^{2}(\gamma\delta
   K_{\varphi})}{\gamma^{2}\delta^{2}}\frac{E^{\varphi}}{\sqrt{E^{x}}}
 \frac{\md g}{\md K_{\varphi}}-2g\frac{\sin(\gamma\delta
   K_{\varphi})\cos(\gamma\delta
   K_{\varphi})}{\gamma\delta}\frac{E^{\varphi}}{\sqrt{E^{x}}}\right. 
\nonumber \\
 -\left.2g\cos(\gamma\delta K_{\varphi})K_{x}\sqrt{E^{x}}+2\cos(\gamma\delta
 K_{\varphi})K'_{\varphi}\sqrt{E^{x}}+\frac{\sin(\gamma\delta
   K_{\varphi})}{\gamma\delta}\frac{(E^{x})'}{\sqrt{E^{x}}}\right)\,.
\end{eqnarray}
Substituting for $(E^{x})'$ and $K'_{\varphi}$ from the corrected LTB
conditions, we have
\begin{eqnarray}
\int \md z\mu(z)\left( \frac{\sin^{2}(\gamma\delta
K_{\varphi})}{\gamma^{2}\delta^{2}}
\frac{E^{\varphi}}{\sqrt{E^{x}}}\frac{\md g}{\md K_{\varphi}}-
2g\frac{\sin(\gamma\delta K_{\varphi})\cos(\gamma\delta
K_{\varphi})}{\gamma\delta}\frac{E^{\varphi}}{\sqrt{E^{x}}}\right. \nonumber \\
+\left.\frac{2g\sin(\gamma\delta
K_{\varphi})}{\gamma\delta}\frac{E^{\varphi}}{\sqrt{E^{x}}}\right)\,.
\end{eqnarray}
For this to be zero for all $\mu$, the integrand must vanish which
implies the differential equation
\begin{equation}
 \frac{\sin(\gamma\delta K_{\varphi})}{\gamma\delta}\frac{\md g}{\md
 K_{\varphi}}=2(\cos(\gamma\delta K_{\varphi})-1)g
\end{equation}
solved by
\begin{equation}
g(K_{\varphi})=c\cos^{4}(\gamma\delta K_{\varphi}/2)\,.
\end{equation}
The classical limit $g\to1$ for $\delta\to0$ fixes the constant of
integration $c=1$. One can check that the other LTB
condition is consistent with this choice for $g(K_{\varphi})$.

As before we now eliminate the connection components in favor of the triad
components. The equations of motion give
\begin{eqnarray} \label{exdot}
\dot{E}^{x}&=&\frac{2\sin\left(\gamma\delta
K_{\varphi}\right)}{\gamma\delta}\sqrt{E^{x}}\\
\label{ephidot}
\dot{E}^{\varphi}&=&\frac{\sin\left(\gamma\delta
K_{\varphi}\right)\cos\left(\gamma\delta
K_{\varphi}\right)}{\gamma\delta}\frac{E^{\varphi}}{\sqrt{E^{x}}}+
\cos\left(\gamma\delta
K_{\varphi}\right)K_{x}\sqrt{E^{x}}\,.
\end{eqnarray}
We use (\ref{exdot}) to express the (co)sine function in terms of $E^{x}$:
\begin{equation} \label{sine and cos in terms of ex}
 \sin\left(\gamma\delta K_{\varphi}\right) = \frac{\gamma\delta}{2}
 \frac{\dot{E}^{x}}{\sqrt{E^{x}}} \quad,\quad
 \cos\left(\gamma\delta K_{\varphi}\right) = \pm\sqrt{1-\frac{\gamma^{2}
 \delta^{2}}{4}\frac{(\dot{E}^x)^2}{E^{x}}}
\end{equation}
and solve (\ref{ephidot}) for
\begin{equation}
K_{x}=\frac{1}{\cos(\gamma\delta K_{\varphi})\sqrt{E^{x}}}
\left(\dot{E}^{\varphi}+\frac{\sin\left(\gamma\delta K_{\varphi}\right)
\cos\left(\gamma\delta K_{\varphi}\right)}{\gamma\delta}
\frac{E^{\varphi}}{\sqrt{E^{x}}}\right)\,.
\end{equation}
After substituting for the sine and the cosine (choosing the plus sign
in the cosine), this can be written as
\begin{equation} \label{kx}
 K_{x}=\left(\dot{E}^{\varphi}-\frac{\dot{E}^{x}E^{\varphi}}{2E^{x}}
 \sqrt{1-\frac{\gamma^{2}\delta^{2}}{4}\frac{(\dot{E}^x)^2}{E^{x}}}\right)
 \left(E^{x}-\frac{\gamma^{2}\delta^{2}}{4}(\dot{E}^x)^2\right)^{-1/2}\,.
\end{equation} 
Substituting for $K_{x}$ and $\sin\left(\gamma\delta K_{\varphi}\right)$
back in the expression for the Hamiltonian (\ref{kphi corrected
hamiltonian}) we have
\begin{equation} \label{hamiltonian with kphi corrections} 
H^{(IIIa)}_{\rm grav}=-\frac{1}{2G}\left(
\frac{\dot{E}^{x}\dot{E}^{\varphi}}{\sqrt{E^{x}-
\frac{1}{4}\gamma^{2}\delta^{2}(\dot{E}^x)^2}}-
\frac{(\dot{E}^x)^2E^{\varphi}}{4(E^{x})^{3/2}}\right)\,.
\end{equation}
The new LTB condition
$E^{\varphi}=(E^{x})'/2g=(E^{x})'/2\cos^{4}(\gamma\delta
K_{\varphi}/2)$ allows us to eliminate $E^{\varphi}$ using (\ref{sine
and cos in terms of ex}):
\begin{equation}
 \cos^{4}(\gamma\delta K_{\varphi}/2)=\frac{1}{4}\left(1+\sqrt{1-
\frac{\gamma^{2}
 \delta^{2}}{4}\frac{(\dot{E}^x)^2}{E^{x}}}\right)^{2}
\end{equation}
which gives
\begin{equation}
E^{\varphi}=\frac{2(E^{x})'}{\left(1+\sqrt{1-
\frac{1}{4}\gamma^{2}\delta^{2}\frac{(\dot{E}^x)^2}{E^{x}}}\right)^2}\,.
\end{equation}
Substituting for $E^{\varphi}$ and its time derivative in
(\ref{hamiltonian with kphi corrections}) implies
\begin{eqnarray}
-2GH^{(IIIa)}_{\rm grav}&=&\frac{2\dot{E}^{x}
(\dot{E}^{x})'}{(E^{x})^{\frac{1}{2}}\left(1+
\sqrt{1-\frac{1}{4}\gamma^{2}\delta^{2}
\frac{(\dot{E}^x)^2}{E^{x}}}\right)^2\sqrt{1-
\frac{1}{4}\gamma^{2}\delta^{2}\frac{(\dot{E}^x)^2}{E^{x}}}}
-\frac{(\dot{E}^x)^2(E^{x})'}{2(E^{x})^{\frac{3}{2}}
\left(1+\sqrt{1-\frac{1}{4}\gamma^{2}\delta^{2}
\frac{\dot{E}^{x^{2}}}{E^{x}}}\right)^2}\nonumber\\
&&+\frac{\gamma^{2}\delta^{2}\dot{E}^{x}(E^{x})'(2E^{x}\dot{E}^{x}\ddot{E}^{x}-
(\dot{E}^x)^3)}{2(E^{x})^{\frac{5}{2}}\left(1+\sqrt{1-
\frac{1}{4}\gamma^{2}\delta^{2}\frac{(\dot{E}^x)^2}{E^{x}}}\right)^3
\left(1-\frac{1}{4}\gamma^{2}\delta^{2}\frac{(\dot{E}^x)^2}{E^{x}}\right)}\,.
\nonumber
\end{eqnarray}

We now derive the evolution equation consistent with 
(\ref{hamiltonian with kphi corrections}). On the one hand, we have the
equation of motion
\begin{equation} \label{kphi dot from eq of motion}
\dot{K}_{\varphi}=-\{K_{\varphi},H^{(IIIa)}_{\rm grav}\}=
\frac{\sin^{2}(\gamma\delta K_{\varphi})}{2\gamma^{2}\delta^{2}}
\frac{1}{\sqrt{E^{x}}}
\end{equation}
and, on the other hand, differentiating (\ref{exdot}) with respect to
time gives
\begin{equation}
\dot{K}_{\varphi}=\frac{1}{2\cos(\gamma\delta K_{\varphi})}
\left(\frac{\ddot{E}^{x}}{\sqrt{E^{x}}}-
\frac{(\dot{E}^x)^2}{2(E^{x})^{3/2}}\right)\,.
\end{equation}
Combining these two equations and writing everything in terms of $R$,
we have
\begin{equation} \label{evolution eq for kphi correction}
 2R\ddot{R}+\dot{R}^{2}\sqrt{1-\gamma^{2}\delta^{2}\dot{R}^{2}}=0\,.
\end{equation}

The evolution equation can now be used to eliminate the second time
derivative of $E^x$ from $H_{\rm grav}^{(IIIa)}$, which together with
$E^{x}=R^{2}$ and in combination with the matter part provides the
Hamiltonian constraint equation
\begin{equation}  \label{ConsHam3}
4\dot{R}^{2}R'\sqrt{1-\gamma^{2}\delta^{2}\dot{R}^{2}}+
8R\dot{R}\dot{R}'=F'\left(1+\sqrt{1-\gamma^{2}\delta^{2}
\dot{R}^{2}}\right)^{2}\sqrt{1-\gamma^{2}\delta^{2}\dot{R}^{2}}\,.
\end{equation}
We note that in the limit $\delta\rightarrow0$ we recover the
classical equation which also justifies the choice of plus sign in
(\ref{sine and cos in terms of ex}).

Finally, we could use a Hamiltonian where only $K_x$ has been replaced
by periodic functions,
\begin{equation}
H^{(IIIb)}_{\rm grav}=-\frac{1}{2G}\left(\frac{K_{\varphi}^{2}
E^{\varphi}}{\sqrt{E^{x}}}+
2K_{\varphi}\frac{\sin\left(\gamma K_{x}\ell_0\right)}{\gamma \ell_0}
\sqrt{E^{x}}\right)\,.
\label{kx corrected hamiltonian}
\end{equation}
In this case, however, the corrected LTB conditions will take a more
complicated form because correction functions will have to depend on
all the phase space variables, as one can check by making an ansatz as
before.  We leave this complicated case open for future work and
proceed with a general discussion and applications of the consistent
versions found.

\subsection{Discussion}
\label{s:ConsDisc}

We have provided several cases of consistent equations of motion for
the variables of a metric of LTB form, but with dynamics carrying
corrections as they are expected from loop quantum gravity. While we
have discussed inverse triad and one form of holonomy corrections
separately, they can be seen to be combined consistently simply in a
multiplicative form of the correction functions in the LTB
conditions. For instance, the first version of inverse triad
corrections and the holonomy correction we used can be consistently
combined in this way to result in a Hamiltonian constraint equation
\begin{equation} \label{first order eq with combined effects 1}
\alpha\dot{R}^{2}R'+\frac{2R\dot{R}\dot{R}'}{\sqrt{1-
\gamma^{2}\delta^{2}\dot{R}^{2}}}=f_{\delta}F'
\end{equation}
where $f_{\delta}[R]= f(R)\left(1+\sqrt{1-
\gamma^{2}\delta^{2}\dot{R}^{2}}\right)^{2}$
together with the evolution equation
\begin{equation} \label{evolution eq for combined effect 1}
2R\ddot{R}=-\alpha\dot{R}^{2}\sqrt{1-\gamma^{2}\delta^{2}\dot{R}^{2}}\,.
\end{equation}
With the second version of inverse triad corrections, we have
\begin{eqnarray} \label{first order eq in time for combined effect 2}
&&-4\alpha^{2}\dot{R}^{2}R'\sqrt{1-\frac{\gamma^{2}\delta^{2}
\dot{R}^{2}}{\alpha^{2}}}-4\alpha^{2}\dot{R}^{2}R'-
4\gamma^{2}\delta^{2}\dot{R}^{4}R'+8\alpha^{2}R\dot{R}\dot{R}'
\left(1+\sqrt{1-\frac{\gamma^{2}\delta^{2}\dot{R}^{2}}{\alpha^{2}}}\right)
 \nonumber \\
&&+\frac{8\alpha^{2}R^{2}\dot{R}^{2}R'}{\sqrt{R^{4}-
(\gamma l_{p}^{2}/2)^{2}}}\left(1+\sqrt{1-\frac{\gamma^{2}\delta^{2}
\dot{R}^{2}}{\alpha^{2}}}\right)=F'\alpha^{4}\left(1+\sqrt{1-
\frac{\gamma^{2}\delta^{2}\dot{R}^{2}}{\alpha^{2}}}\right)^{3}\sqrt{1-
\frac{\gamma^{2}\delta^{2}\dot{R}^{2}}{\alpha^{2}}} \nonumber \\
\end{eqnarray}
and
\begin{equation} \label{evolution eq for combined effect 2}
2R\ddot{R}=2\dot{R}^{2}-\frac{2R^{2}\dot{R}^{2}}{\sqrt{R^{4}-
(\gamma \ell_{\rm P}^{2}/2)^{2}}}-\dot{R}^{2}\sqrt{1-
\frac{\gamma^{2}\delta^{2}\dot{R}^{2}}{\alpha^{2}}}\,.
\end{equation}

While general properties of an LTB reduction allowed us to keep the
constraints consistent without severe limitations on quantum
correction functions, consistency conditions did remain. The remaining
constraints automatically form a first class system provided that the
Hamiltonian constraint is free of spatial derivatives, which is
realized if spin connection terms (or the 3-dimensional Ricci
curvature) drop out as it happens under the classical LTB
conditions. The consistency conditions arose at the level of
formulating the LTB conditions, because the classical ones are no
longer preserved under evolution corresponding to quantum corrected
constraints. We thus corrected the LTB conditions, too, such that in
their new form they were preserved under the quantum corrected
equations of motion as they are generated by a Hamiltonian whose spin
connection contribution vanishes.

With these conditions we are still identically satisfying the
classical diffeomorphism constraint and thus no new anomalies in the
constraint algebra can arise. However, the corrected LTB conditions do
not make the classical spin connection $\Gamma_{\varphi}$ equal $-1$,
which was assumed in the simplified classical Hamiltonians such as
(\ref{Ham1}). Thus, to be fully consistent we must assume that the
expressions containing the spin connection themselves carry quantum
corrections and read $f(E^x)\Gamma^{(I)}_{\varphi}=
-(E^x)'/2E^{\varphi}$ for the equations in Sec.~\ref{s:Corr1},
$\alpha(E^x)\Gamma^{(II)}_{\varphi}=- (E^x)'/2E^{\varphi}$ for
Sec.~\ref{s:Corr2} and $g(K_{\varphi})\Gamma^{(IIIa)}_{\varphi}=
-(E^x)'/2E^{\varphi}$ in Sec.~\ref{s:CorrHol}.  Thus, for consistency
additional corrections of this form must arise in the terms of the
Hamiltonian constraint containing the spin connection in such a way
that they vanish after imposing the quantum corrected LTB conditions.

That the spin connection terms carry their own corrections is a
reasonable expectation: There are inverse triad components and, in a
loop quantization, the spin connection is rather indirectly expressed
via $A_a^i$ and $\gamma K_a^i=A_a^i-\Gamma_a^i\propto
\{A_a^i,\{H^{({\rm E})},V\}\}$ using the Euclidean part $H^{({\rm
E})}$ of the Hamiltonian constraint \cite{QSDI}. Corrections are thus
expected from the inverse triad as well as from holonomies. The
specific form is difficult to determine because the full theory does
not provide operators for the non-covariant spin connection
components, but as demonstrated here it can be derived and justified
by the production of a consistent set of equations. In fact, if such
corrections occur, our equations provide a fully consistent LTB
system. In this way, consistency determines what further quantum
corrections must be entailed by a primary correction such as
$\alpha$. Since not all corrections in a Hamiltonian can equally
easily be computed, independent consistency considerations provide
useful relations between different terms. For instance, the spin
connection is more difficult to quantize than $1/E^x$, and its
corrections can thus more easily be found via consistency.

An important physical implication is that this suggests additional
effects because the space-time metric (\ref{ltbmet}) is no longer just
corrected by different solutions for $R(t)$ solving the corrected
constraint and evolution equations, but also by an additional
pre-factor in terms of $\alpha$, $f$ or $g$ in front of $L^2\propto
(R')^2$ which is no longer exactly $(R')^2$. This would, for instance,
affect the appearance of horizons.

\section{Applications}
\label{s:App}

Our focus in this paper for applications of the above equations is the
fate of the classical singularity which appears at $R=0$. There are
two possibilities for how such a singularity could be avoided in
effective equations. Dynamically, $R(t,x)$ may be bounded away from
zero for all $x$, in which case the behavior shown would be comparable
to a cosmological bounce. This can sometimes occur if isotropic
cosmological models exhibiting a bounce are matched to a spherically
symmetric outside region in a generalized Oppenheimer--Snyder manner
\cite{Collapse}. The outright spherically symmetric situation studied
here is, however, subject to different corrected equations and so one
has to provide a new analysis.

The second possibility is that the value $R=0$ is assumed, but that
this does not result in a singular space-time just as Minkowski space
in polar coordinates has $R=0$ at $x=0$. If $R=0$ occurs, one thus has
to proceed with a more detailed analysis to understand the space-time
neighborhood of the region where $R=0$.

Compared to homogeneous equations, this problem is of a new
quality. As we have seen, there is a non-trivial anomaly problem which
we were able to resolve in different versions of quantum corrected LTB
models. The presence of consistency conditions, which do not arise in
homogeneous models because they are subject to just a single
constraint, makes the form of quantum corrections more
restricted. Thus, several different terms in the constraints must
receive quantum corrections in a way closely related to each
other. Still, we have explicitly shown that non-trivial quantum
corrections are allowed.

In addition to the anomaly issue, spatial inhomogeneity allows
different types of singularities in classical general relativity. In
particular, not just spacelike singularities can occur as in
homogeneous models, but also null \cite{Christodoulou,Newman} or
timelike ones \cite{TimelikeBirth}. This has interesting general
ramifications concerning the consistency of quantum gravity in the
sense of allowing stable ground states, as discussed in
\cite{SingValue}, and it underlines the interest in inhomogeneous
models. In what follows, we present an initial analysis based on
analytical as well as numerical methods.

\subsection{Analytical properties}
\label{s:Analytical}

If there is a ``bounce'' where the area radius $R$ attains a non-zero
minimum value, we have $\dot{R}=0$ which can be substituted in the
above equations to check the possibility for this to happen in quantum
gravity.  From Eq.~(\ref{ConsHam1}) (for the first version of inverse
triad corrections) or Eq.~(\ref{ConsHam3}) (for holonomy corrections)
as well as the two combinations (\ref{first order eq with combined
effects 1}) and (\ref{first order eq in time for combined effect 2})
we can immediately see that this is not possible unless we drop the
condition $F'>0$ which classically avoids shell-crossing
singularities. Thus, we either have to drop this condition, possibly
taking into account quantum geometry corrections in the matter sector,
or retain the non-bouncing behavior of the classical models. For the
second version of inverse triad corrections, Eq. (\ref{ConsHam2}), we
would require $F=0$ at the bounce, which looks difficult to achieve in
a generic collapse model. (Conditions on $F$ may be avoided if
$\dot{R}'$ diverges where $\dot{R}=0$, but this does not appear
generic.)

There does not appear to be a simple conclusion about bounces as they
occur, e.g., in homogeneous models. We are looking at specific regimes
and certain types of quantum corrections which, by themselves, may
make a bounce difficult to occur. Moreover, we have restricted the
analysis to marginal models, which classically includes spatially flat
Friedmann--Robertson--Walker models as the interior region of
Oppenheimer--Snyder collapse, but not isotropic models with positive
spatial curvature. The latter (or scalar matter with negative
potential \cite{Cyclic}) would be required for a bounce based on inverse
triad corrections \cite{BounceClosed}. There is thus no contradiction
with known matching results based on isotropic interiors
\cite{Collapse}, but the fact that a bounce does not follow
straightforwardly, compared to the relative ease by which this can be
obtained in isotropic models, may also be taken as a warning sign
concerning the robustness of homogeneous bounces.

Similarly, we have considered holonomy corrections only of a special
form which made the analysis more manageable. Holonomy corrections
give rise to bounces more generally than inverse triad corrections
\cite{GenericBounce,QuantumBigBang,BouncePert,BounceSqueezed}. One
could thus expect that a full treatment of holonomy corrections should
give rise to general bounces also in LTB models. However, even though
we did not do such an analysis, such a bounce cannot be generic for
the following reason: simply choosing a fine spatial graph and thus
small enough $\ell_0$ makes the holonomy corrections studied here the
relevant ones. Since these corrections do not provide an automatic
bounce, a bounce cannot be generic in this inhomogeneous
system. Finally, there is a third effect due to quantum variables
which provides corrections in effective equations. Also this has not
been included here, but it is unlikely to result in a general bounce
given that it does not do so in isotropic models (where it could even
prevent a bounce which would otherwise occur based on holonomy
corrections \cite{QuantumBounce,BounceSqueezed}). We thus conclude that
singularities in LTB systems do not appear to be resolved by bounces.

\subsubsection{Corrected LTB equations as cosmological models}

As the simplest case, we first consider a vacuum solution where, in
the absence of dust, $F'=0$ must be satisfied.  If $R'\neq 0$ holds,
$R=R(x)$, i.e.\ a static configuration, is a trivial solution to any
of the corrected equations (\ref{ConsHam1}), (\ref{ConsHam2}) and
(\ref{ConsHam3}). While this corresponds to the classical Minkowski
space solution, since $R(x)$ can then easily be introduced as a
coordinate instead of $x$, there are quantum corrections for small
$R$: Our corrected LTB metric, using $L=R'/f(R)$ for (\ref{ConsHam1})
to be specific, reads
\begin{equation}
 \md s^2 = -\md t^2+\frac{\md R^2}{f(R)^2}+R^2(\md\vartheta^2+
\sin^2\vartheta\md\varphi^2)
\end{equation}
which asymptotically presents Minkowski space. As we will discuss in
more detail below, the appearance of $f(R)$ shows that quantum effects
originating in the spatial discreteness of loop quantum gravity spoil
some of the exact symmetries such as spatial homogeneity known to
exist in classical solutions.

If there is dust, it is of interest to see whether we can have a
Friedmann solution in this system.  For this we choose $x$ such that
it coincides with the circumferential radius at $t=0$, and make an
ansatz of the form $R(t,x)=a(t)x$. If such a solution exists, as it
does in the classical case, the LTB metric reduces to a
Friedmann--Robertson--Walker one where $a$ is identified with the
scale factor. The dust density profile then becomes
\begin{equation}
f(x)F'=8\pi G \epsilon_{0} x^{2} 
\end{equation}
where $\epsilon_{0}$ is the initial uniform density. (According to our
general choice of $R(0,x)=x$, the scale factor is normalized to
$a_0=1$ at $t=0$ in the cosmological context.)
Substituting this into Eq.~(\ref{ConsHam1}), we have
\begin{equation}\label{eq:Friedmann}
\dot{a}^{2}a=\frac{8\pi G \epsilon_{0}}{3+(\alpha(ax)-1)}
\frac{f(ax)}{f(x)}\,.
\end{equation}
The left-hand side depends only on $t$ while the right-hand side
depends non-trivially on $ax$. Hence, the corrected LTB system does
not admit a solution of Friedmann form.

There is an additional effect which prevents Friedmann solutions for
the corrected equations, because our LTB form of the metric receives
quantum corrections, too, as a consequence of consistency. The
corrected LTB metrics have coefficient $L=R'/f(R)$ in the case of
$H^{(I)}$, and $L=R'/\alpha(R)$ in the case of $H^{(II)}$. This
changes the metric in addition to the corrected dynamics of the metric
component $R$. In particular, the metrics are no longer homogeneous
because of the non-trivial $R$-dependence. If we were interested in
spatial volumes of finite regions in constant $t$ slices, for an
approximate solution of the form $R=a(t)x$ they would become $V=4\pi
a^3\int\md x x^2/f(ax)$ and $V=4\pi a^3\int\md x x^2/\alpha(ax)$,
respectively. This illustrates an interesting difference between these
two cases which both come from inverse triad corrections: for $f(ax)$,
we have the small-$x$ expansion $f(ax)\propto ax+O(a^2x^2)$, while for
$\alpha$ it reads $\alpha(ax)\propto a^3x^3+O(a^4x^4)$. Thus, in the
first case the spatial volumes vanish at $a=0$ as in the classical
case, while the second case implies diverging volumes $V\sim \int
x^{-1}\md x$ even of finite regions near $a=0$. This suggests further
implications of the behavior near a classical singularity, which due
to the required inhomogeneity do not appear easy to discern.

\subsubsection{Effective densities}

As a further consequence of corrections, we note that the mass
function $F(x)$ is no longer directly related to the Misner--Sharp
mass.  We need to distinguish the latter from the dust mass which can
be defined as
\begin{equation}
M(x)=\frac{F(x)}{2G}\,.
\end{equation}
The asymptotic value $M_{\rm dust}=\lim_{x\to \infty}M(x)$ corresponds
to the total mass of dust. The Misner--Sharp mass, on the other hand,
now takes the form
\begin{equation}
 m= \frac{R}{2}\left(1-\left(\frac{R'}{L}\right)^2+\dot{R}^2\right)
\end{equation}
whose expression changes for the corrected LTB conditions because this
affects the relation between $R'$ and $L$. From the corrected mass, we
can then derive an effective density $\epsilon= m'/4\pi GR^2R'$. For
instance, for the condition consistent with $H^{(I)}_{\rm grav}$ we
have a Misner--Sharp mass
\begin{equation} \label{mI}
 m^{(I)}=\frac{1}{2}R(1-f^2+\dot{R}^2)= m_{\rm class}-\frac{1}{2}R(f^2-1)
\end{equation}
which, upon using (\ref{ConsHam1}), leads to an effective density
\begin{equation} \label{epsI}
 \epsilon^{(I)}= \frac{1}{8\pi GR^2} \left(\frac{f(R)F'}{R'}-
 (\alpha(R)-1)(\dot{R}^2-2f(R)^2)-
 (f(R)^2-1)\right)\,.
\end{equation}

Similarly, for the equations following from $H^{(II)}_{\rm grav}$, we
have $m=\frac{1}{2}R(1-\alpha^2+\dot{R}^2)$ and thus
\begin{equation}
 \epsilon^{(II)}= \frac{1}{8\pi G} \left(\frac{\alpha^2F'}{R^2R'}-
 \frac{\alpha^2-1}{R^2}+ \frac{2}{\alpha} \frac{\md\alpha}{\md R}
 \frac{\dot{R}^2}{R}- 2\frac{\alpha}{R}\frac{\md\alpha}{\md
 R}\right)\,.
\end{equation}
In particular, in this case the horizon condition $2m=R$ reads
$\dot{R}^2= \alpha^2$, which by the Hamiltonian constraint equation
agrees with $\alpha^2F/R$. Thus, in terms of $F$ the horizon condition
$F=R$ is uncorrected in this case, although $R$ as a function of time
is corrected compared to the classical behavior.

The correction terms to the effective densities may be nonzero even in
vacuum regions devoid of dust. Depending on the regime, they can be
positive or negative according to the signs of $\alpha(R)-1$, $f(R)-1$
and their derivatives involved.

\subsubsection{Existence of self-similar solutions?}

The classical equation can rather easily be analyzed using
self-similar solutions; see e.g.\ \cite{LTBSelfSim}. One can first
write the classical constraint equation as $ \dot{R}^2= F(x)/R$ and
then, for the special case of a linear mass function $F(x)=\lambda x$,
find an explicit solution for $R(x,t)$ of self-similar form which
depends on $t$ only via the function $1-at/x$ with a constant
$a=\frac{3}{2}\sqrt{\lambda}$. For such a self-similar solution, the
structure of the singularity has been analyzed in \cite{LTBSing}.

If we use the second version of inverse triad corrections, this
equation is simply changed by multiplying the mass function with
$\alpha(R)^2$. Thus, for a linear mass function there is no longer a
self-similar solution. One would have to incorporate the new factor by
changing the mass function, if a self-similar solution is to be
obtained. But this is not straightforward since $\alpha$ depends not
on $x$ but on the unknown function $R$ which is to be solved for.

For the first version of inverse triad corrections the equation
changes more radically. In this case, we can bring the constraint
equation to the form of an integro-differential equation for $R$,
\begin{equation}
 \left(\frac{\md R}{\md t}\right)^2 = \frac{1}{R} \int\md x f[R(x)]F'(x) - 
\frac{1}{R}  \int \md R (\alpha-1)\left(\frac{\md R}{\md t}\right)^2\,.
\end{equation}
If $\alpha-1$ is small, one can solve this iteratively by inserting the
equation for $(\md R/\md t)^2$ in the integral:
\[
 \left(\frac{\md R}{\md t}\right)^2 = \frac{F(x)}{R} +\frac{1}{R}
\int\md x (f-1)F'(x) - \frac{1}{R} 
\int \md R (\alpha-1)F(x)/R + \cdots \,.
\]
The difficulty in solving this is that $F(x)$ depends on $x$ rather
than $R$, so we have to know $R(x)$ as a solution and invert it before
doing the integration. But this equation already shows qualitatively
that the quantum correction makes the solution more non-local,
which may prevent the existence of self-similar solutions.
Moreover, there will be additional time-dependent effects which do not
occur classically. This is so because $R(t,x)$, which we need to know
in order to replace $x$ by $R$ in the integrand, also depends on
$t$. Thus, the integrals are really time dependent, which one can
understand as replacing the classical $F(x)$ by a new function
\begin{eqnarray}
 {\cal F}(x,t) &=& F(x)+ \int\md x(f-1)F'(x)
-\int \md R (\alpha-1)F(x(R,t))/R\\
 && -\int \md R(\alpha-1)/R \int\md x (f-1)F'(x)
+ \int \md R
(\alpha-1)/R \int^R \md \tilde{R} (\alpha-1)F(x(\tilde{R},t))/\tilde{R} 
+ \cdots \nonumber
\end{eqnarray}
appearing on the right hand side of $(\md R/\md t)^2 = {\cal F}(x,t)/R$.

This refers only to the case where $\alpha-1$ is small, i.e.\ we have
perturbative corrections to the inverse triad effects. It would not
allow one to analyze the deeper quantum regime where $\alpha$ differs
significantly from one. For this regime we would have to use other
techniques, such as the expansions of the following subsection.

However, generally speaking, since the characteritic length scale
$\ell_{\rm P}$ is explicitly introduced into the corrections, we
cannot expect this kind of self-similar solutions, which are called
complete self-similar solutions or self-similar solutions of the first
kind. With the characteristic length scale, we can only expect
incomplete self-similar solutions, e.g.  kinematic self-similar
solutions in this context; see
\cite{SelfSimilarity,KinSelfSim,KinSelfSimGR}.

\subsubsection{Small-$x$ expansion}

For $R\gg \sqrt{\gamma}l_{\rm P}$, $\alpha\to 1$ and the classical
limit is recovered. However, for $R\lesssim \sqrt{\gamma} l_{\rm P}$,
the deviation from classical theory becomes of order unity. Here, deep
quantum effects might be revealed by a closer analysis.  We point out
that such a deep quantum regime is less reliable if only one type of
quantum effect is considered. Nevertheless, an analysis of single
effects can provide various possibilities and guide further
developments. Moreover, the corrections studied here, based on inverse
triad and holonomy corrections, can be combined without changing the
conclusions.

To have a regular center in an inhomogeneous case we assume that $F$
and $R$ admit the following expansions at the center.
\begin{equation}
F(x)=F_{3}x^{3}+F_4x^4+\cdots \quad,\quad 
R(t,x)=R_{1}(t)x+R_2(t)x^2+\cdots 
\end{equation}
where the dots denote higher order terms with respect to $x$, and
$F_{i}$ are constants but $R_{i}$ may be $t$-dependent.
In this way, the
classical expression for energy density gives an expansion of the form
\begin{equation} \label{density profile}
\epsilon(t,x)=\epsilon_{0}(t)+\epsilon_{1}(t)x+\epsilon_{2}(t)x^{2}+\cdots
\end{equation}
from $F(x)$ and $R(t,x)$, where $\epsilon_0= 3 F_{3}/8\pi G R_1^3$.
Classically, the lowest order then gives $\dot{R}_{1}^{2}=F_{3}/R_{1}$
with solution
\begin{equation}
R_{1}=\left(C\pm \frac{3\sqrt{F_{3}}}{2}t\right)^{2/3},
\end{equation}
where $C$ is an arbitrary constant. Hence, for the collapsing case,
$R_{1}$ monotonically decreases and becomes zero in a finite proper
time --- the central singularity develops where $\epsilon_0\to\infty$.
When we choose the radial coordinate $x$ so that $R=x$ at $t=t_{0}$,
we find 
\begin{equation}
R_{1}=\left(1 \pm \frac{3\sqrt{F_{3}}}{2}(t-t_{0})\right)^{2/3},
\end{equation}
and hence, $R_{1}$ vanishes at $t=t_{s}$, where
\begin{equation}
t_{s}=t_{0}\mp \frac{2}{3\sqrt{F_{3}}}.
\end{equation}
This behavior can be checked also in the presence of quantum
corrections to see if anything of the singularity changes. For the
first version of inverse triad corrections, we use the small-$x$ behavior
\begin{equation} \label{alpha r near x=0}
\alpha=\left(\frac{2}{\gamma\lP^{2}}\right)^{3/2}R_{1}^{3}x^{3} \quad,\quad
f=\sqrt{\frac{8e^{1-\pi/2}}{\gamma \lP^{2}}}R_{1}x
\end{equation}
of the correction functions.  As a result, if $F(x)$ has a cubic term
$F_{3}x^{3}$ as the lowest order, we find $\dot{R}_{1}=0$. However,
the additional factor of $x$ in the dust energy density proportional
to $f(R)F'/R^2R'$ then shows that we can allow a quadratic term
$F_2x^2$ in the mass function to produce the desired regular expansion
for energy density, although the total effective energy density
is still diverging at the center because of the ``vacuum''
contributions in (\ref{epsI}).

Using the various series expansions in \eqref{ConsHam1}, to lowest
order in $x$ we get
\begin{equation}
\dot{R}_{1}^{2}=\frac{2^{3/2}e^{\frac{1}{2}-
\frac{\pi}{4}}F_{2}}{(\gamma\lP^{2})^{1/2}}
\end{equation}
which is solved by
\begin{equation} \label{solution for r1}
R_{1}(t)=1\pm\left[\frac{2^{3/2}e^{\frac{1}{2}-
\frac{\pi}{4}}F_{2}}{(\gamma\lP^{2})^{1/2}}\right]^{1/2}(t_{0}-t)
\end{equation}
where the plus sign corresponds to a collapsing dust cloud and where
we have chosen the initial condition $R_{1}(t_{0})=1$. From here we
see that the central singularity, corresponding to $R_{1}(t)=0$ is
formed at
\begin{equation} \label{t for singularity}
t=t_{0}+\left[\frac{2^{3/2}e^{\frac{1}{2}-
\frac{\pi}{4}}F_{2}}{(\gamma\lP^{2})^{1/2}}\right]^{-1}
\end{equation} 
For the sake of comparison we note that classically the central
singularity forms at $t=t_{0}+2/3\sqrt{F_{3}}$. In terms of
the initial density profile, $F_{3}=8\pi G\epsilon_{0}(0)/3$ (note that
$\epsilon_{0}(0)$ is not the complete dust density profile at the initial
time but a coefficient in the series expansion for the dust density)
and therefore in terms of the initial density the
time for singularity formation is $t=t_{0}+1/\sqrt{6\pi G\epsilon_{0}}$.
For the quantum corrected case a similar consideration gives $F_{2}=4\pi
G\epsilon_{0}(\gamma\lP^{2})^{1/2}/2^{3/2}e^{\frac{1}{2}-\frac{\pi}{4}}$
which implies that the central singularity forms at $t=t_{0}+1/\sqrt{4\pi
G\epsilon_{0}}$.

For the second version of inverse triad corrections, the equation to
be solved is $\dot{R}^2R=\alpha^{2}F$ from (\ref{ConsHam2}).  For
small $R$, (\ref{alpha r near x=0}) implies that the lowest order term
on the right hand side goes as $x^{9}$ if we use the same form of the
expansion of $F(x)$ as classically. In this case, we can show
$\dot{R}_{1}=0$ as in the first version.  However, now our density is
$\epsilon\propto (\alpha^2F')/R^2R'$ if the divergent
``vacuum'' contribution is subtracted, which means that $\alpha^2 F$
should be required to have a leading term cubic in $x$, such that $F$
itself can have lower order terms. In this case, the singularity 
is not prevented either. Nevertheless, a finite neighborhood may
look different from the classical space-time near a central
singularity which, however, would require a more detailed analysis.

For our version of holonomy corrections, the equation to be solved is
(\ref{ConsHam3}).  After expanding in powers of $x$ and equating the
coefficients on the two sides we obtain (to order $x^{2}$)
$R_{1}\dot{R}_{1}^{2}=F_{3}$ which is the same as in the classical
case and gives
\begin{equation}
R_{1}(t)=\left(1-\frac{3}{2}\sqrt{F_{3}}(t-t_0)\right)^{2/3}
\end{equation}
implying that the singularity occurs at time $t=t_0+2/3\sqrt{F_{3}}$
(the same as in the classical case). At order $x^{3}$ we have the
equation for $R_{2}$ which also does not have any quantum
corrections. Effects of the $\gamma^{2}\delta^{2}$ factor occur only
at order $x^{4}$ and higher. Thus the small $x$ behaviour with this
correction is the same as for the classical LTB model. The
combinations of inverse triad and holonomy corrections in (\ref{first
order eq with combined effects 1}) and (\ref{first order eq in time
for combined effect 2}) lead to the same conclusions.

We conclude that there is no indication that the corrections
implemented here prevent the LTB singularity from forming. In
particular, naked singularities as they appear in these models do not
seem resolved automatically by a loop quantization. Whether they are
indeed naked singularities in the presence of quantum effects requires
further analysis of the effective space-time: the surroundings of the
singularity may be sufficiently different from the classical naked
case such that the singularity becomes spacelike. However, we have
shown that a situation which gives rise to a naked singularity
classically also gives rise to a singularity (of some form) under the
quantum effects considered here. An analysis whether naked
singularities remain naked may be of interest in the context of cosmic
censorship, which we will come back to in the numerical analysis.

A correction not considered here is the effect of $K_x$-holonomies
which are computationally more complicated and may be crucial in some
regimes. In fact, taking the general form (\ref{GenSol}) of classical
solutions indicates that $K_{\varphi}=\dot{R}=\sqrt{F/R}$ is
subdominant to
\[
 K_x= \dot{R}'= \sqrt{F/R} \left(\frac{F'}{2F}-
 \frac{1}{2}\frac{\sqrt{x}-tF'/F}{R^{3/2}}\right)
\]
near $x=0$. This may present a chance for holonomy corrections to
remove the singularity, after all. However, the size of holonomy
corrections is also state-dependent: a small $\ell_0$ suppresses
$K_x$-corrections even if $K_x$ is large. Thus, as already noted these
corrections cannot result in generic avoidance of singularities. (Note
that also in \cite{HolBH} only $K_{\varphi}$-corrections were
included, although the conclusion was that the Schwarzschild
singularity could be resolved in this way.)  This clearly shows the
non-trivial behavior of inhomogeneous situations compared to
homogeneous ones.

\subsection{Numerical analysis}

Although some analytical results are available, the region of
intermediate values for $R$ not in the asymptotic regimes $R\ll
\sqrt{\gamma}\lP$ or $R\gg\sqrt{\gamma}\lP$ is difficult to
analyze. We thus complement the preceding analysis by numerical
studies of inverse triad corrections.

For the first version, we transform the constraint equation with
\begin{equation}
K\equiv \dot{R}^{2}-\frac{F}{R} \,,
\end{equation}
to
\begin{eqnarray}
 K'&=&-\frac{1}{2R^{2}}\left[K+\left(K+\frac{F}{R}\right)(\alpha(R)-1)\right]
(R^{2})'+(f-1)F', 
\label{eq:K'}\\
\dot{R}&=&\pm\sqrt{K+\frac{F}{R}}\,.
\label{eq:Rdot}
\end{eqnarray}
We spatially integrate the constraint equation at each time step and
evolve $R$ using (\ref{eq:Rdot}).  The constraint is always satisfied
within $10^{-4}$ accuracy.  The spatial integration is done using a
fourth-order scheme while the time evolution uses a second-order
scheme.

The second version is given by a set of equations which is first
integrated by $R\dot{R}^{2}=\alpha^{2}F$.  The numerical
implementation of this equation is easy because we no longer need
spatial integration.  Our numerical scheme is second-order, which is
sufficiently accurate and stable for the present purpose.

It should be noted that there is scale invariance in both sets of
equations which are invariant under scaling $R(t,x)$, $F(x)$ and $t$
as $R(t/\beta,x)$, $\beta^{2} F(x)$ and $t/\beta$, where $\beta$ is a
positive constant. This scale invariance greatly simplifies the
analysis. If the functional form of $F(x)$ is the same up to a
constant factor, the evolution is similar up to the scaling of
time. We thus do not need to investigate the full parameter
space. Through this scaling, the classical Misner-Sharp mass $m_{\rm
class}$ scales as $\beta^{2} m_{\rm class}$, while the corrected mass
$m$ does not.  For holonomy corrections, the scaling behavior is
lost. Here, a dedicated analysis of the whole parameter freedom is
required to draw reliable conclusions, which we postpone to future
work.

\subsubsection{Initial condition}

The function $F(x)$ corresponds to the conserved mass because it
is constant for a comoving observer.  Then,
\begin{equation}
\rho_{\rm cons}\equiv \frac{F'}{8\pi G R^{2}R'}
\end{equation}
is regarded as the conserved mass density. Note that this differs from
effective energy densities which incorporate quantum effects.

We choose the radial coordinate $x$ so that 
\begin{equation}
x=\frac{R_{0}(x)}{R_{0}(x_{\rm max})}\,,
\end{equation}
i.e.\ $0\le x\le 1$ for the region of computation,
where $R_{0}(x)\equiv R(0,x)$. As matter models,
it is useful to consider two different cases: (i) uniform models where
\begin{equation}
F(x)=\left\{\begin{array}{cc}
F_{0}x^{3} & \mbox{ for } 0\leq x \leq x_{\rm s} \\
F_{1} & \mbox{ for } x_{\rm s}< x
\end{array}
\right. 
\end{equation}
and with 
\begin{eqnarray}
x_{\rm s}=\frac{R_{\rm s}}{R_{0}(x_{\rm max})}\quad, \quad
F_{0}=\frac{2GM}{x_{\rm s}^{3}}\quad, \quad
F_{1}=2 GM
\end{eqnarray}
in terms of the physical parameters given by the initial radius
$R_{\rm s}$ and the total conserved mass $M$; and (ii) quadratic
models where
\begin{equation}
F(x)=\left\{\begin{array}{cl}
\displaystyle{F_{0}\left(\frac{x^{3}}{3}-\frac{x^{5}}{5x_{\rm s}^{2}}\right)} 
& \mbox{ for }0\leq x \leq x_{\rm s} \\
F_{1} & \mbox{ for }x_{\rm s}< x
\end{array}
\right. 
\end{equation}
and 
\begin{eqnarray}
x_{\rm s}=\frac{R_{\rm s}}{R_{0}(x_{\rm max})}\quad, \quad
F_{0}= 15 \frac{GM}{x_{\rm s}^{3}}\quad, \quad
F_{1}= 2 G M\,.
\end{eqnarray}

Although we restrict ourselves to these two density profiles, there
still are many possible sets of values for $R_{\rm s}$ and $M$.  We
choose $R_{\rm s}=1$ and $0.1$ and $M=0.01$, where the Planck length
$\ell_{\rm P}$ is chosen to be unity. In units used for the numerical
analysis, we have a critical radius 
$R_{*}= \sqrt{\gamma/2} \ell_{\rm P}
\sim0.25\ell_{\rm P}$.  So, $R_{\rm s}=1$ and $0.1$ represent the cases
where the initial size of the dust cloud is above and below the
critical one, respectively.  As we have already seen, we can recover
the general mass scale by rescaling $F(x)$ and $t$ as $\beta^{2} F(x)$
and $t /\beta$.  In other words, the dynamics of the dust cloud 
follows this scaling relation.
However, it should be noted that as the corrected Misner-Sharp
mass will not scale in such a simple way, 
the kinematics of null geodesics on the corrected spacetime
will not follow this scaling.
This means that the condition for horizon formation
may depend on the mass scale, i.e. $\beta$.
Due to our choice for the mass parameter $M=0.01$,
we can see the correction effect on the null expansions very clearly.
The outer boundary of the calculated region is chosen
to be $R_{0}(x_{\rm max})=2$. 
Our calculation region covers both the
classical and effective regimes.

\begin{figure}[htbp]
\begin{center}
\includegraphics[width=0.45\textwidth]{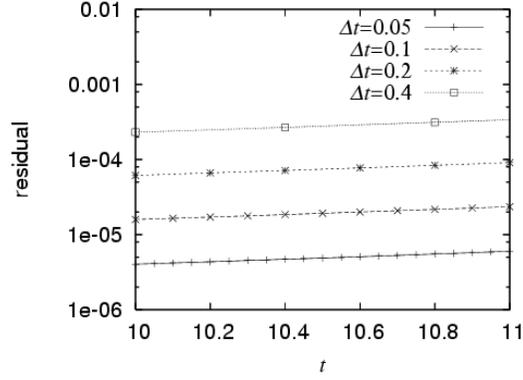}
\end{center}
\caption{\label{fg:convergence}Convergence test for time integration.}
\end{figure}

We have implemented a convergence test to the exact solution for
time integration:  For $\alpha=1$ and $f=1$, we have the marginally
bound Lemaitre-Tolman-Bondi solution
\begin{equation}
R_{\rm ex}(t,x)=\left(R(0,x)^{3/2}-\frac{3}{2}\sqrt{F(x)}t\right)^{2/3}.
\end{equation}
Fig.~\ref{fg:convergence} shows the residual 
of the numerical solution $R_{n,i}$ from the exact solution 
$R_{\rm ex}(t,x)$
\begin{equation}
|R_{n,i}-R_{\rm ex}(t_{n},x_{i})|
\end{equation}
for $R_{0,i}=10$ and $R_{\rm ex}(0,x_{i})=10$, where $n$ and $i$ label
the time step and the spatial grid point, respectively. The dust
parameters are set to $M=1$ and $R_{\rm s}=10$.  One can see that as we
decrease the time step $\Delta t$ while fixing $\Delta x=0.05$, the
residual decreases as $(\Delta t)^{2}$.

In the following we fix $\Delta x=10^{-4}$, where $x=1$ corresponds to
the outer boundary of the calculated region. We have confirmed that 
the numerical solution will not qualitatively change 
if we double $\Delta x$. The time step $\Delta t$ is chosen so that the 
physical quantities on each grid point will change their 
values within 1\% at each time step.

\subsubsection{Classical collapse}

Since there is no characteristic scale in classical theory, we have
two independent scalings of $R(t,x)$, $F(x)$ and $t$ to $\eta
R(t/\beta,x)$, $\beta^{2} \eta^{3} F(x)$ and $t/\beta$, where $\beta$
and $\eta$ are constants.  Hence, we can recover the results for
general radius and mass parameters from a simulation done with only
one set of parameters.

\begin{figure}[htbp]
\begin{center}
\begin{tabular}{cc}
\subfigure[Density profile]{\includegraphics[scale=.4]
{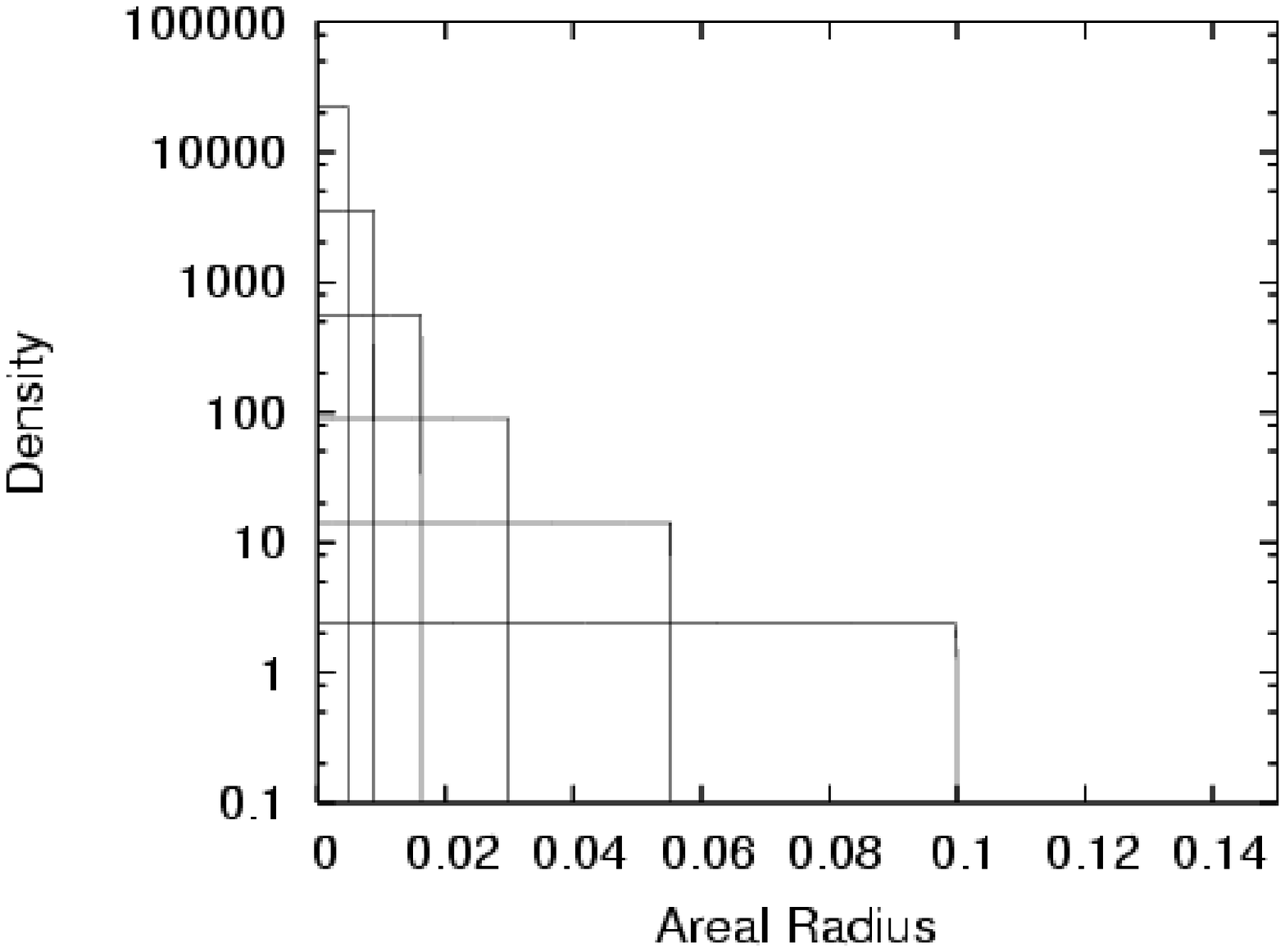}}&
\subfigure[Velocity profile]{\includegraphics[scale=.4]{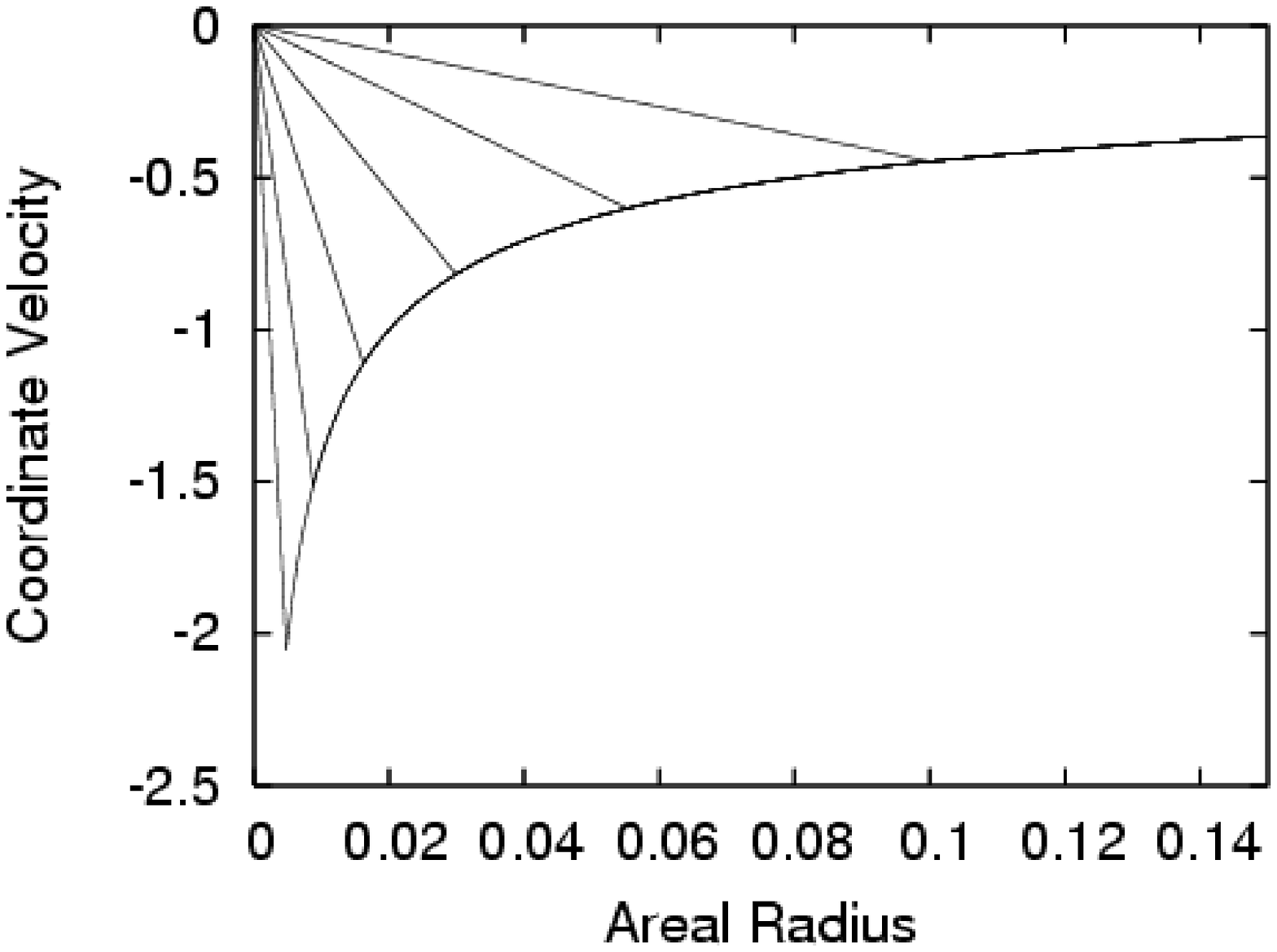}}\\
\subfigure[Quasi-local mass]
{\includegraphics[scale=.4]{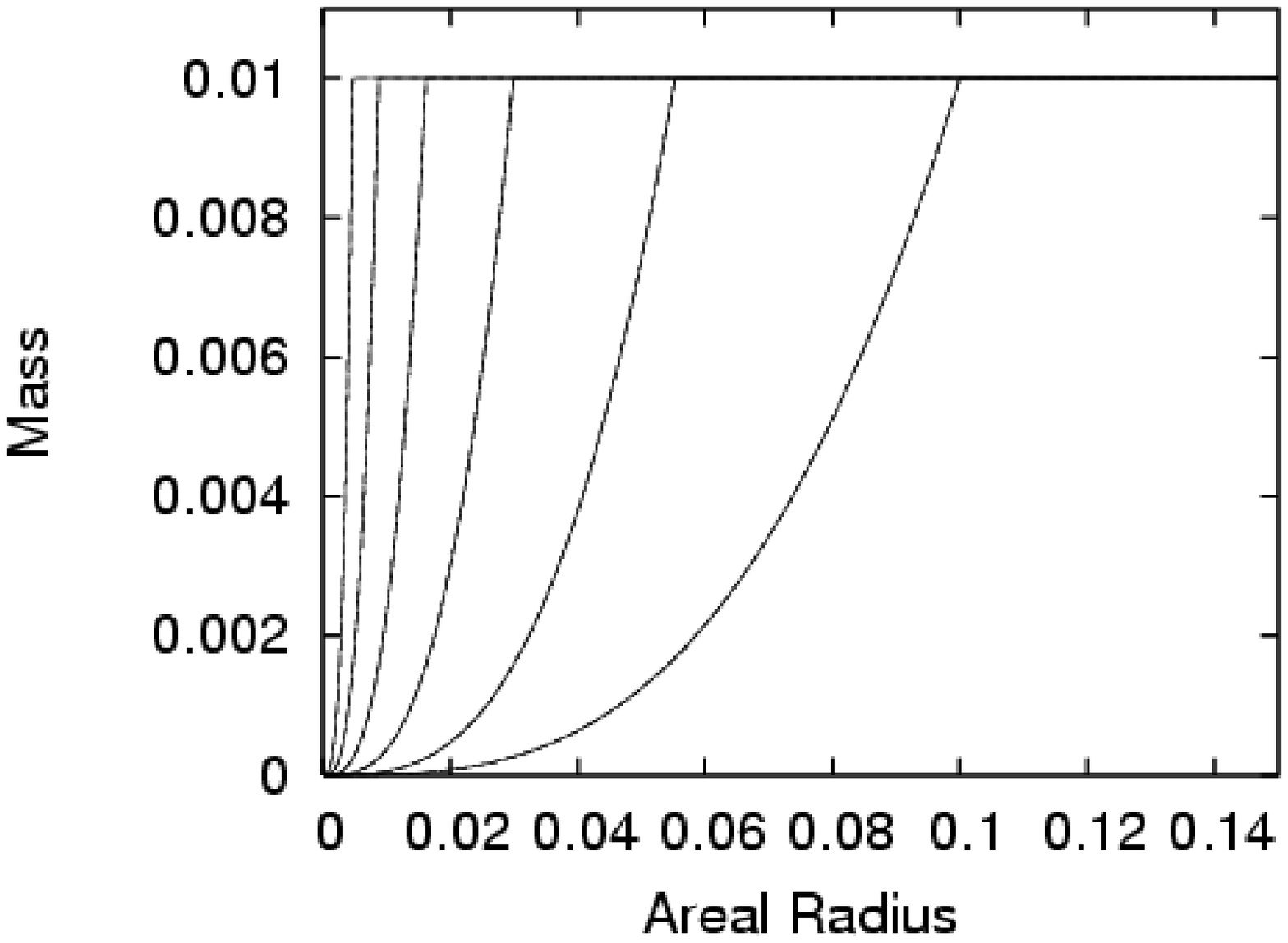}}&
\subfigure[Mass-radius ratio]
{\includegraphics[scale=.4]{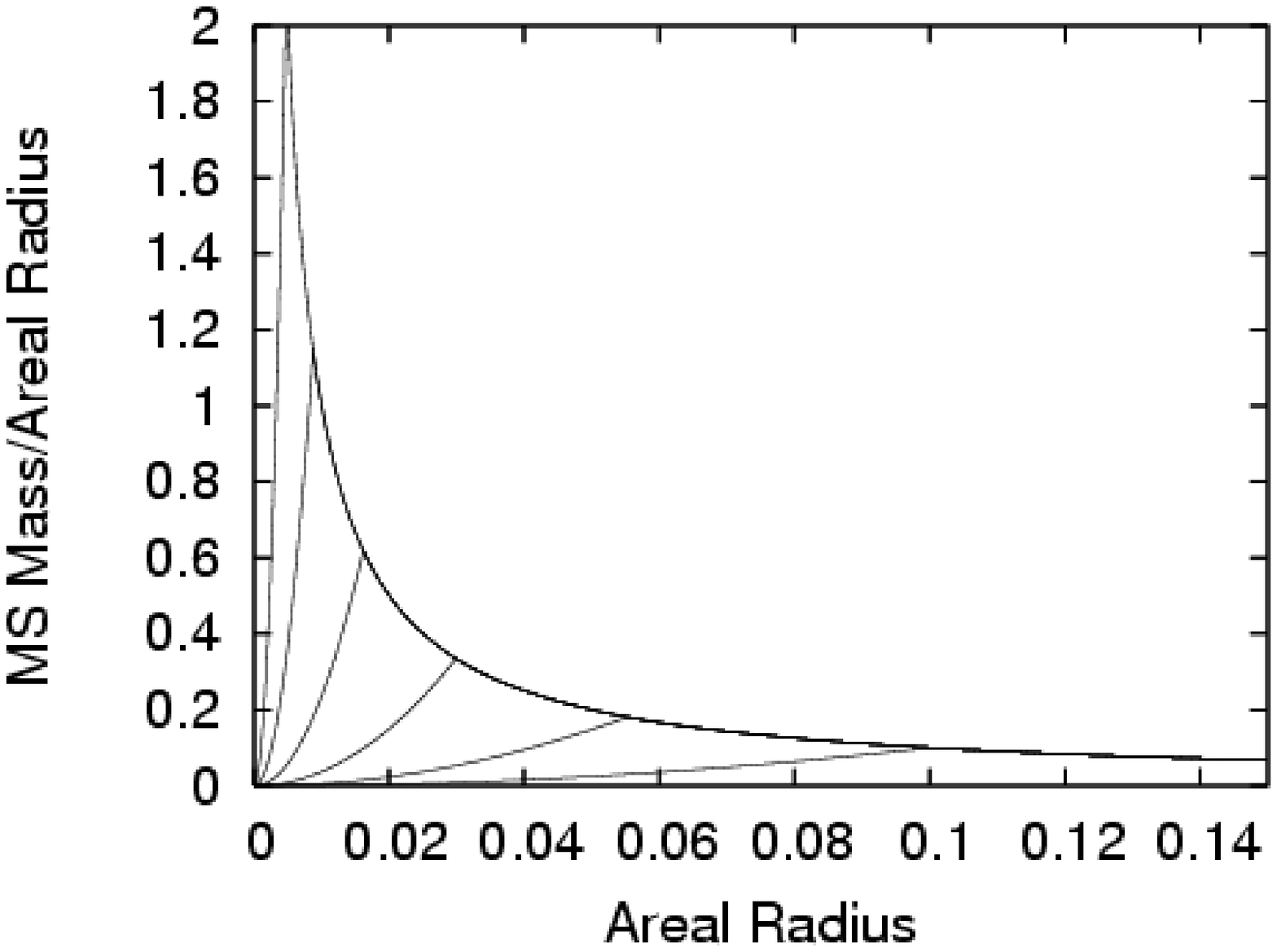}}
\end{tabular}

\end{center}
\caption{\label{fg:osr01}
The collapse of a homogeneous dust ball with $R_{\rm s}=0.1$ and $M=0.01$
in classical general relativity: the snapshots at $t=0$, 0.0887925729, 
0.125072481, 0.139516828, 0.145267646 and 0.147557255 are plotted.
}
\end{figure}

Fig.~\ref{fg:osr01} shows the collapse of an initially homogeneous
ball in classical general relativity.  The total conserved mass $M$ is
set to be 0.01 and the initial radius of the cloud surface $R_{\rm s}$
is set to be 0.1.  Fig.~\ref{fg:osr01} (a) shows the evolution of the
density profile, where the conserved mass density and the effective
density coincide with each other in the classical case.
Fig.~\ref{fg:osr01} (b) shows the evolution of the velocity profile.
Fig.~\ref{fg:osr01} (c) shows the evolution of the mass profile, where
also the conserved mass and the Misner-Sharp mass coincide with each
other in this classical case.  Fig.~\ref{fg:osr01} (d) shows the
evolution of the ratio between the Misner-Sharp mass $m$ and the area
radius $R$.  This ratio becomes a half at trapping horizons. {}From
Fig.~\ref{fg:osr01} (a), we can see that the density profile in the
ball remains uniform during the collapse. The solution is given by the
marginally bound Oppenheimer-Snyder solution~\cite{OppSny} where the
singularity is massive and spacelike.  In this exact solution, the
singularity appears at $t=\sqrt{2}/3(0.1^{3}/0.01)^{1/2}\simeq
0.1490712\cdots$. Not only an event horizon but also a trapping
horizon always appear in this solution. The singularity is always
hidden within the event horizon as well as the trapping horizon.
Cosmic censorship holds in this collapse model.

\begin{figure}[htbp]
\begin{center}
\begin{tabular}{cc}
\subfigure[Density profile]{\includegraphics[scale=.4]
{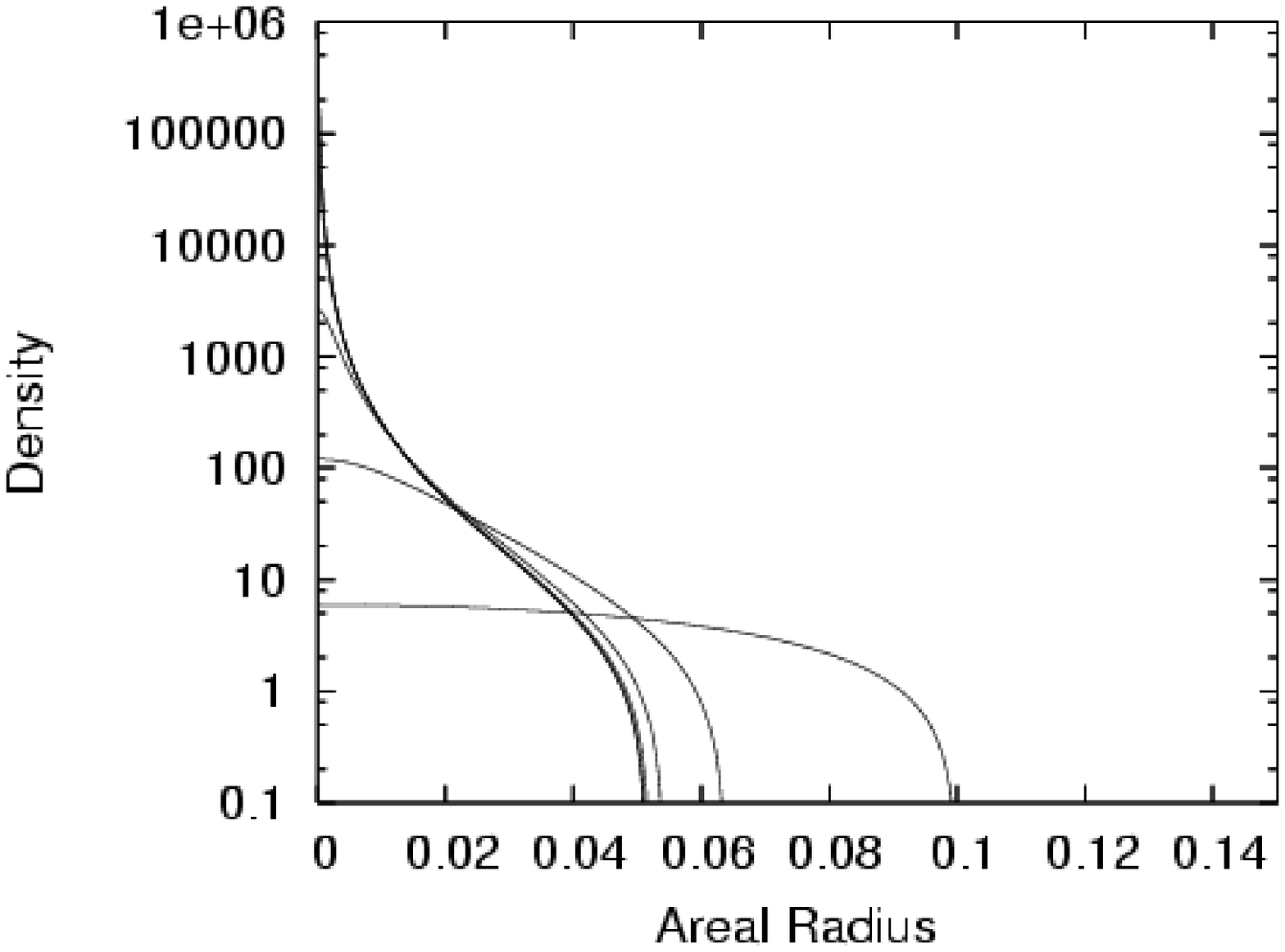}}&
\subfigure[Velocity profile]{\includegraphics[scale=.4]{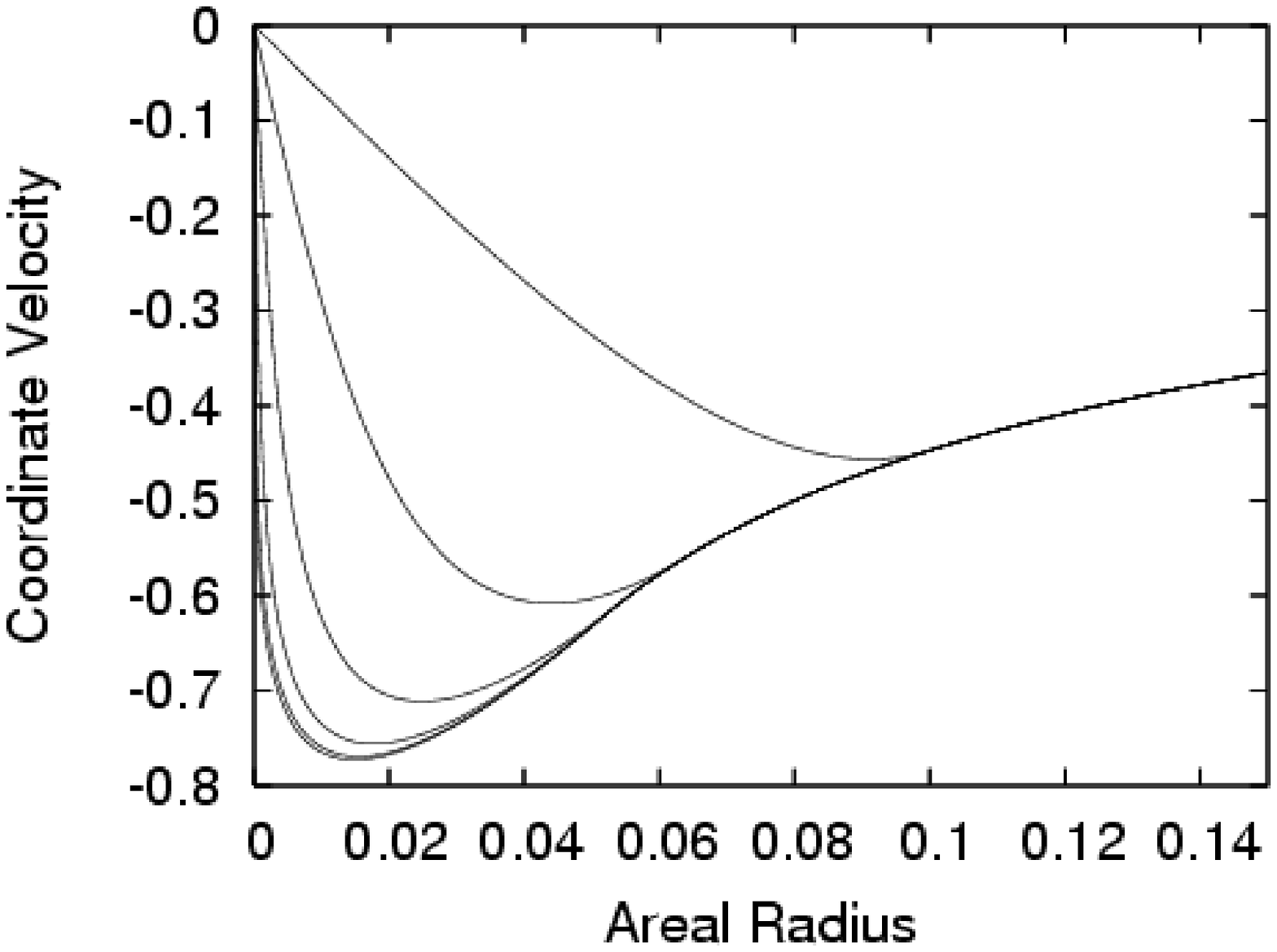}}\\
\subfigure[Quasi-local mass]
{\includegraphics[scale=.4]{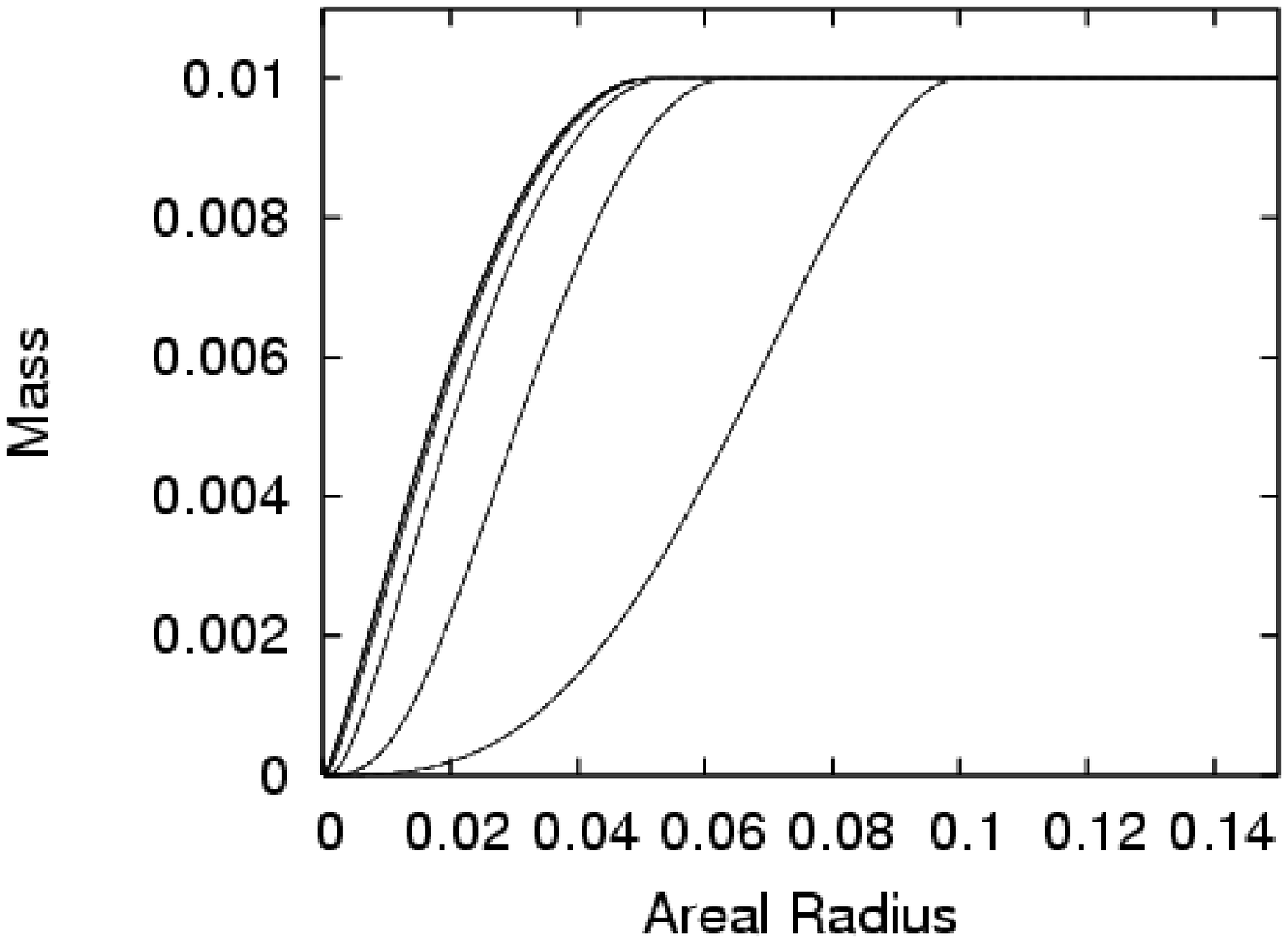}}&
\subfigure[Mass-radius ratio]
{\includegraphics[scale=.4]{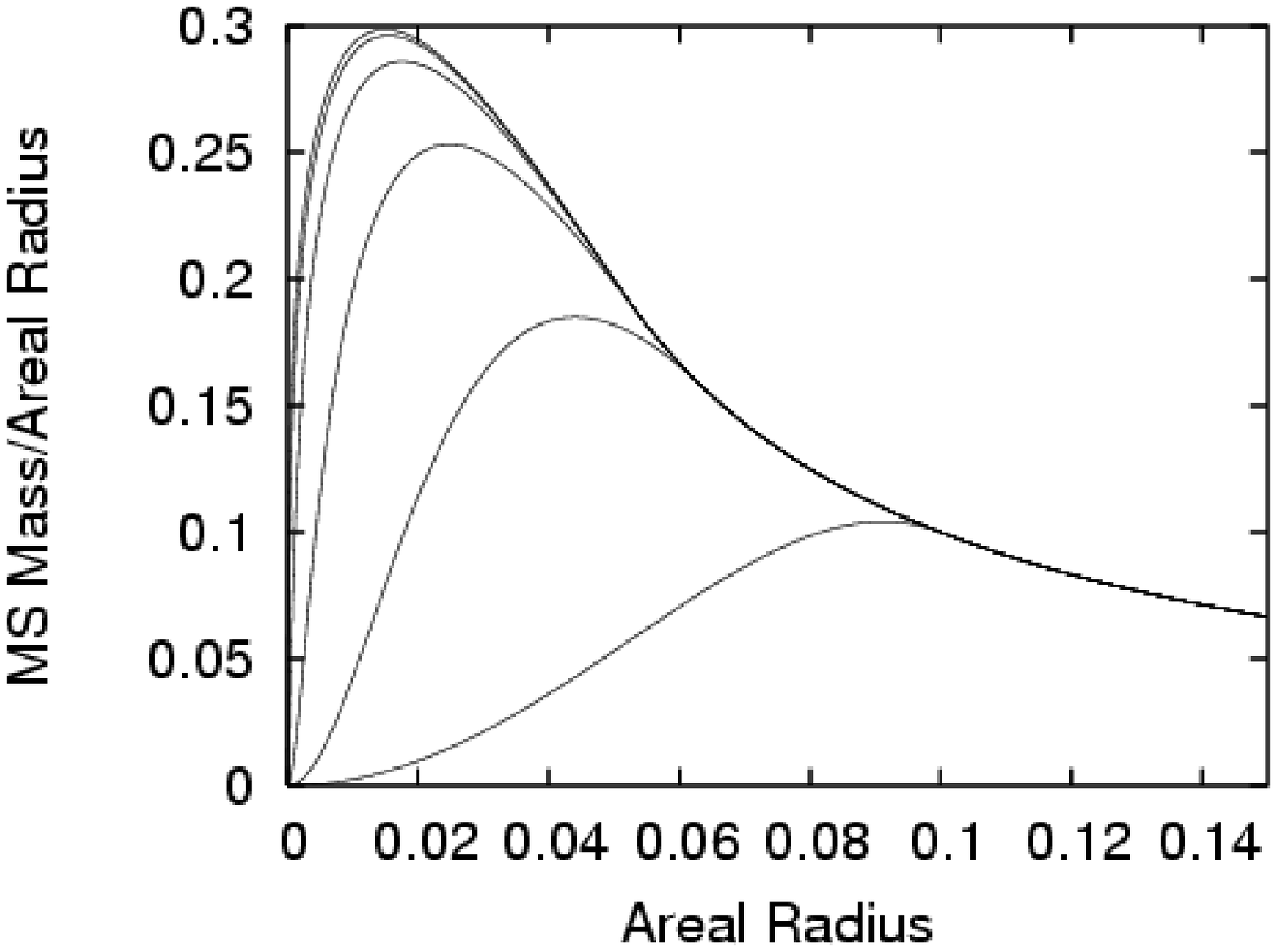}}
\end{tabular}
\caption{\label{fg:ltbr01}The collapse of an inhomogeneous dust ball
with $R_{\rm s}=0.1$ and $M=0.01$ in classical general relativity:
the snapshots at $t=0$,
0.0736495616, 0.0898356836, 0.0933232155, 0.0940745701, 
and 0.0942363423 are plotted.}
\end{center}
\end{figure}

Fig.~\ref{fg:ltbr01} shows the collapse of an initially inhomogeneous
ball with $R_{\rm s}=0.1$ and $M=0.01$ in classical general relativity.
{}From Fig.~\ref{fg:ltbr01} (a), we can see that the central density
grows very rapidly while the surrounding region falls into the central
region more slowly.  This induces strong inhomogeneity near the center
and finally the calculation breaks down soon after $t=0.0942363423$.
In fact, the solution is exactly given by the marginally bound
Lemaitre-Tolman-Bondi solution~\cite{Lemaitre,TolmanSol,Bondi}.  The
peculiar behavior at the center seen in the numerical solution
presents a shell-focusing singularity.  In the present class of the
Lemaitre-Tolman-Bondi solutions, the shell-focusing singularity has
been shown to be massless, generic and locally
naked~\cite{Christodoulou}, and moreover curvature
strong~\cite{Newman,LTBSingII}.  It can be globally naked depending on
the values for $R_{\rm s}$ and $M$.  We can determine whether a
trapping horizon forms or not before the singularity formation by
looking at the value of $m/R$ shown in Fig.~\ref{fg:ltbr01} (d), which
gives a maximum of about 0.30 achieved at $t=0.0942363423$.  Although this
might be slightly larger if we go closer to the singularity, the real
value is not so different from 0.30.  Since this ratio is a half at a
trapping horizon, the present result means that no trapping horizon is
formed in this case before the singularity is formed.
The existence of a trapping horizon
implies an event horizon outside or coinciding with it (but not vice
versa) in classical general relativity~\cite{HawkingEllis}.

\subsubsection{Inverse triad corrections: First version}

\begin{figure}[htbp]
\begin{center}
\begin{tabular}{cc}
\subfigure[Density profile]{\includegraphics[scale=.4]
{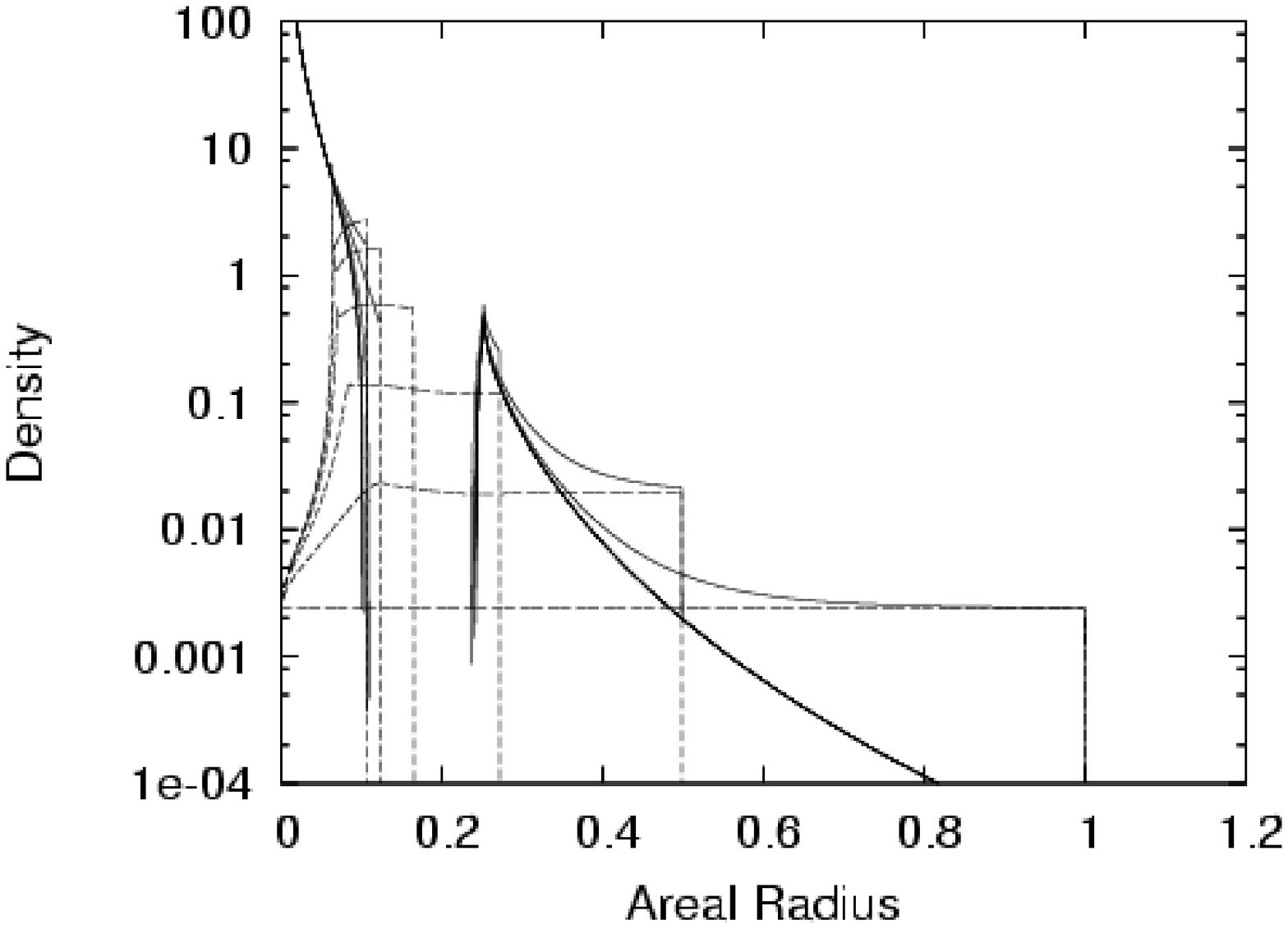}}&
\subfigure[Velocity profile]{\includegraphics[scale=.4]{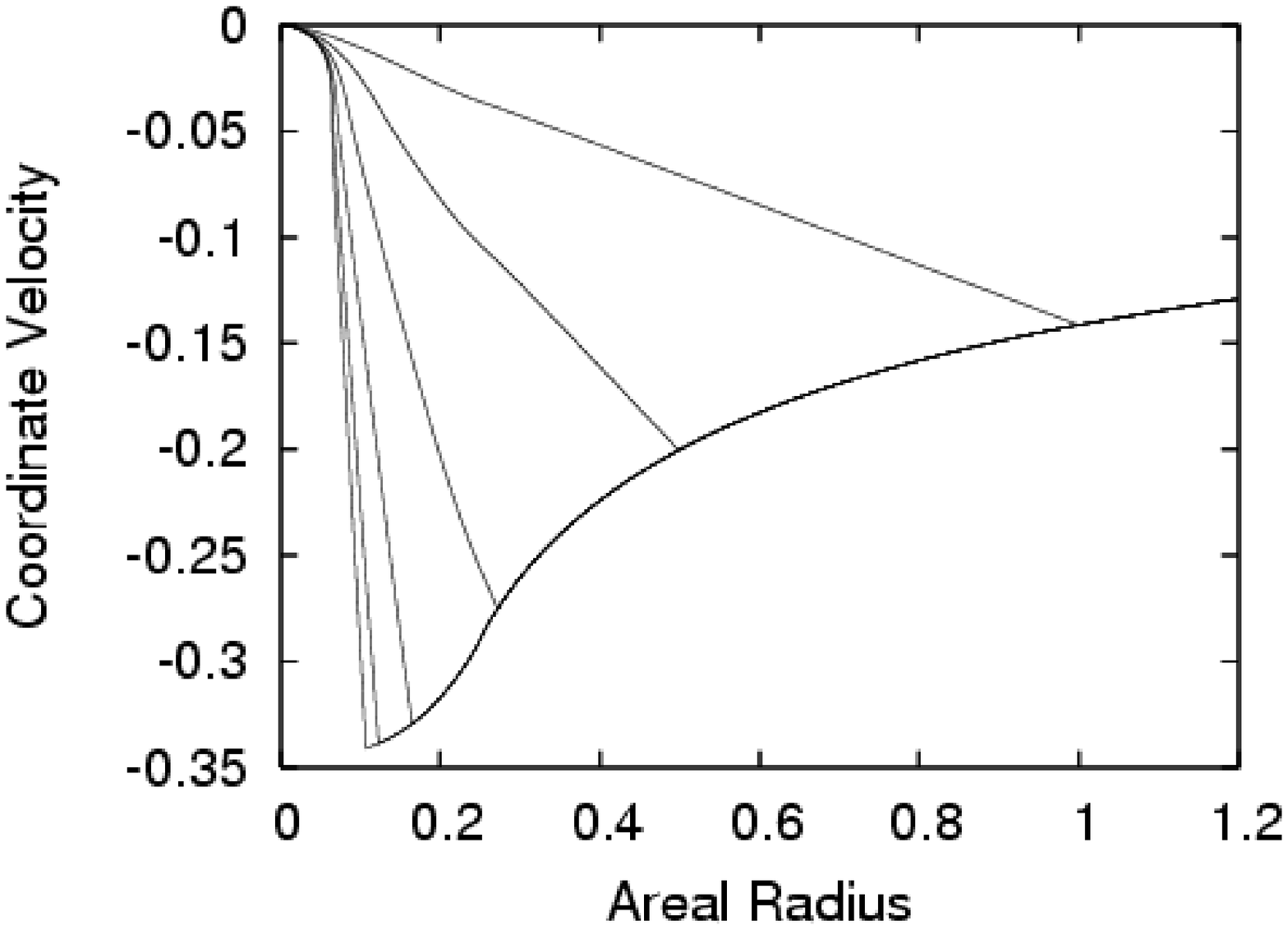}}\\
\subfigure[Quasi-local mass]
{\includegraphics[scale=.4]{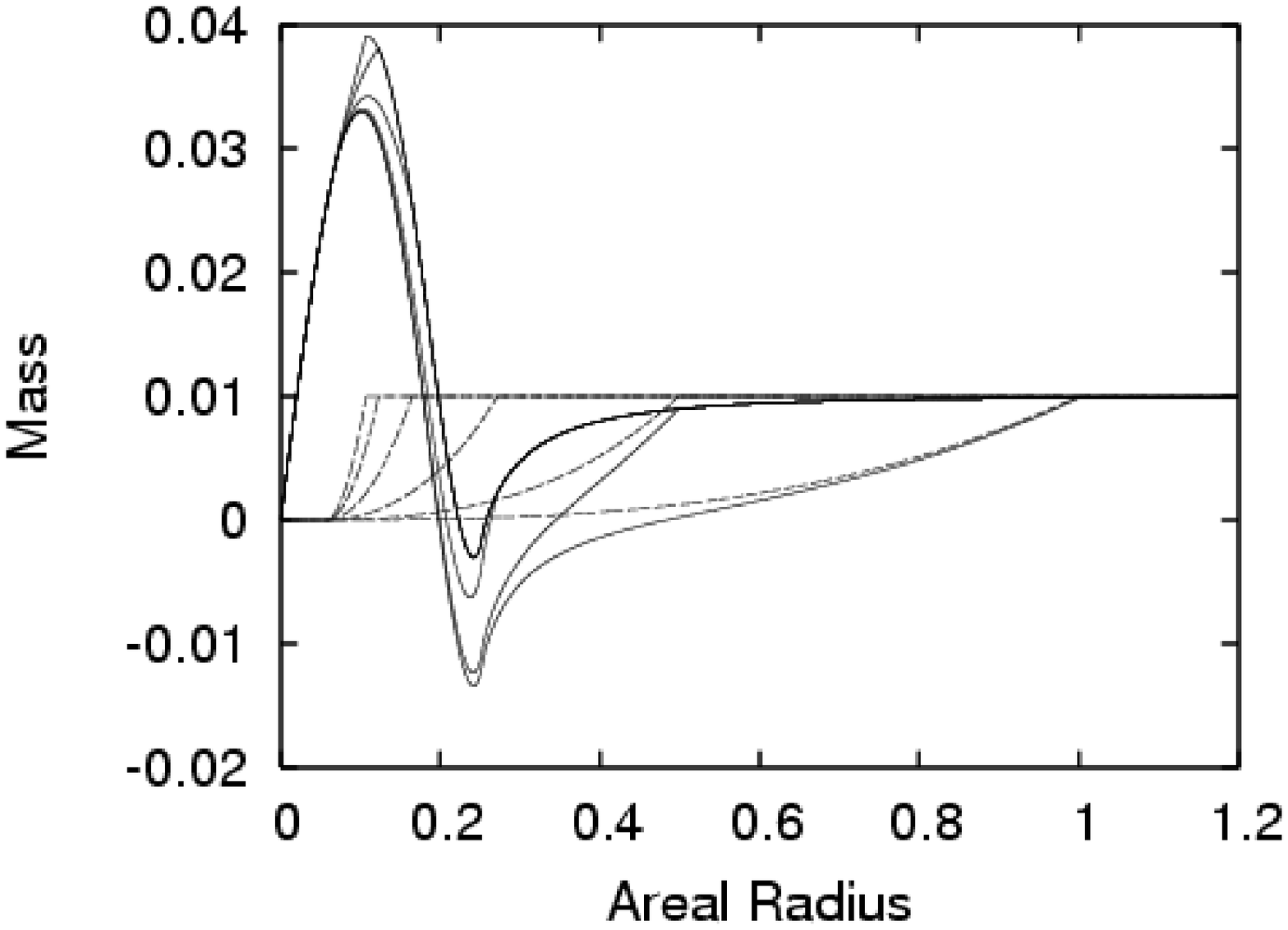}}&
\subfigure[Mass-radius ratio]
{\includegraphics[scale=.4]{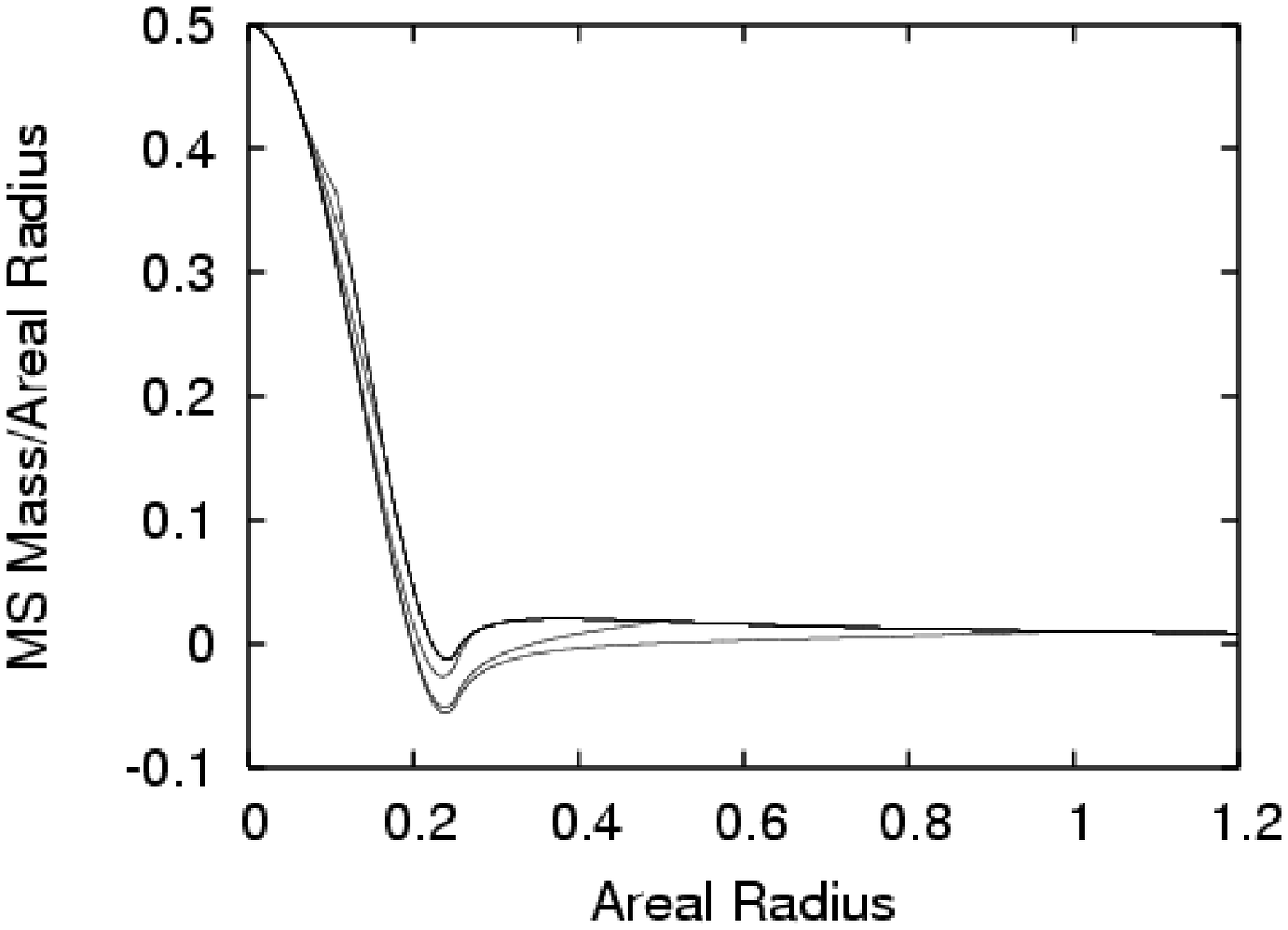}}
\end{tabular}
\end{center}
\caption{\label{osr1_1}
The collapse of a homogeneous dust ball with $R_{\rm s}=1$ and $M=0.01$
in the first version of loop quantum gravity: 
the snapshots at $t=0$,
3.06904628, 4.04642236, 4.39347087, 4.51871685, and 4.56404342
are plotted.
In (a), the solid and dashed lines denote the effective density
and the conserved mass density, respectively.
In (c), the solid and dashed lines denote the Misner-Sharp mass
and the conserved mass, respectively.}
\end{figure}

Fig.~\ref{osr1_1} shows the collapse of an initially homogeneous
ball in the first version of consistent inverse triad corrections from
loop quantum gravity with $R_{\rm s}=1$ and $M=0.01$.  
Besides the
conserved mass density $\rho_{\rm cons}$, we can naturally define the
effective density by
\begin{equation}
\rho_{\rm eff}=\frac{m'}{4\pi G R^{2}R'}.
\end{equation}
This is defined so that when we integrate this with the invariant
3-dimensional volume element on the constant $t$ spacelike 
hypersurface we recover the Misner-Sharp mass. 
This is directly related to the $(t,t)$-component of the
Einstein curvature tensor and not necessarily positive definite. 

Fig.~\ref{osr1_1} (a) shows the evolution of the profiles of both
the conserved mass density (dashed line) 
and the effective density (solid line).  
We can see that 
the cloud becomes inhomogeneous in spite of its initial homogeneity. 
{}From Figs.~\ref{osr1_1} (a) and (b), we can see that the collapse is
strongly slowed down in the central region $R\lesssim 0.05$, while the
collapse continues to take place as in the classical case for the
outer region $R\gtrsim 0.2$.  As a result, as we can see in
Fig.~\ref{osr1_1} (a), the conserved mass density at the central
region remains almost unchanged, while it increases almost
homogeneously in the outer region.  We can also see in the same figure
that the effective density is diverging at the center. As the cloud
surface falls inside $R\simeq  0.2$, the effective density is still
nonzero even outside the cloud surface.  Moreover, for $0.11\lesssim 
R\lesssim 0.24$ the effective density becomes negative, which is not shown in
Fig.~\ref{osr1_1} (a). A spike develops also
in the conserved mass density field at $R\simeq 0.06$ and then
the calculation breaks down soon after $t= 4.56404342$.  The spike in the
conserved mass density field should be identified with a curvature
singularity, which we can identify with the shell-crossing singularity
as it also appears in the classical Lemaitre-Tolman-Bondi solution.
Shell-crossing singularities can be naked but gravitationally
weak~\cite{ShellCrossing,ShellCrossing2}.  This singularity is so weak
in curvature strength that it is generally believed to be extendible
in a distributional sense~\cite{SingAnalysis,CrossingSing}.
Fig.~\ref{osr1_1} (c) shows the evolution of the Misner-Sharp mass
and the conserved mass as a function of $R$.  Although their total
values are both 0.01, their distributions are quite different for
$R\lesssim 1$. 
The Misner-Sharp mass dominates the conserved mass for $R\lesssim 0.2$.
It takes a maximum $\simeq 0.04$ at $R\simeq 0.11$.
The Misner-Sharp mass is a decreasing function of
$R$ for $0.11\lesssim R\lesssim 0.24$, which implies the effective density is
negative there.  Fig.~\ref{osr1_1} (d) shows that the center is 
always marginally trapped. This is actually seen from the definition
of the Misner-Sharp mass $m^{(I)}$ in Eq.~(\ref{mI}).
Except at the center the ratio $m/R$ is less than a half, implying that
the shell-crossing singularity is not covered by a trapping horizon. 

\begin{figure}[htbp]
\begin{center}
\begin{tabular}{cc}
\subfigure[Density profile]{\includegraphics[scale=.4]
{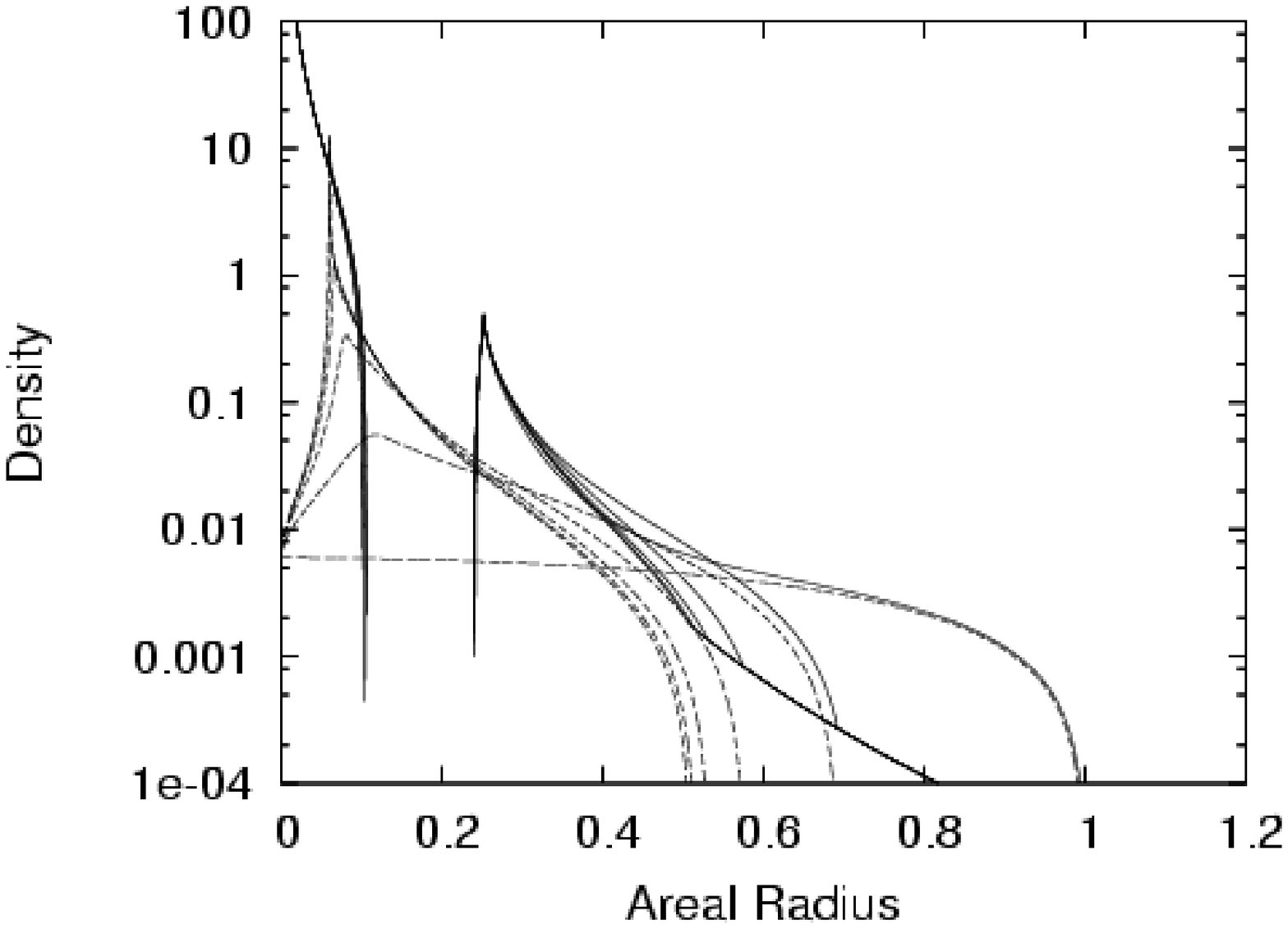}}&
\subfigure[Velocity profile]{\includegraphics[scale=.4]{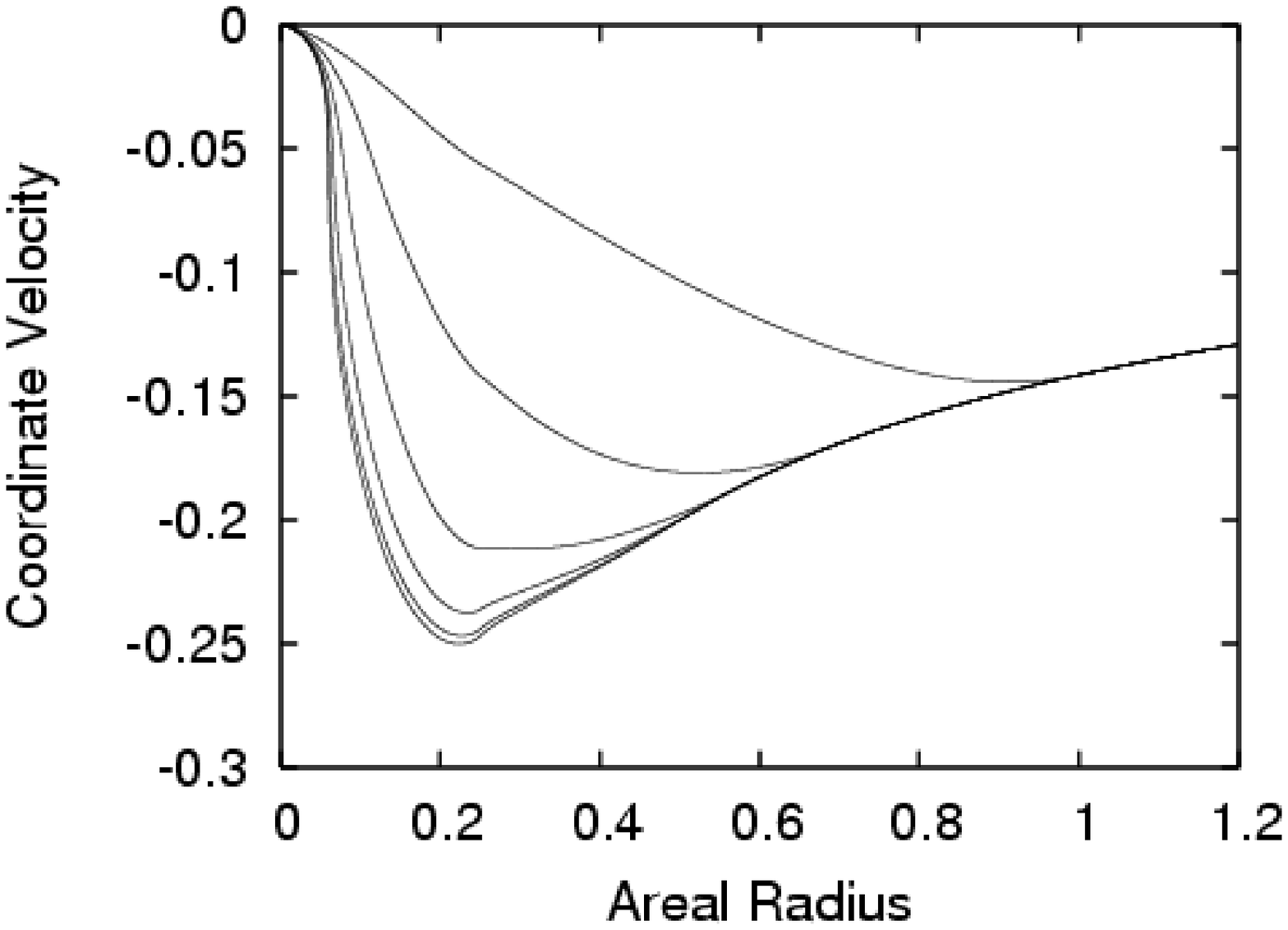}}\\
\subfigure[Quasi-local mass]
{\includegraphics[scale=.4]{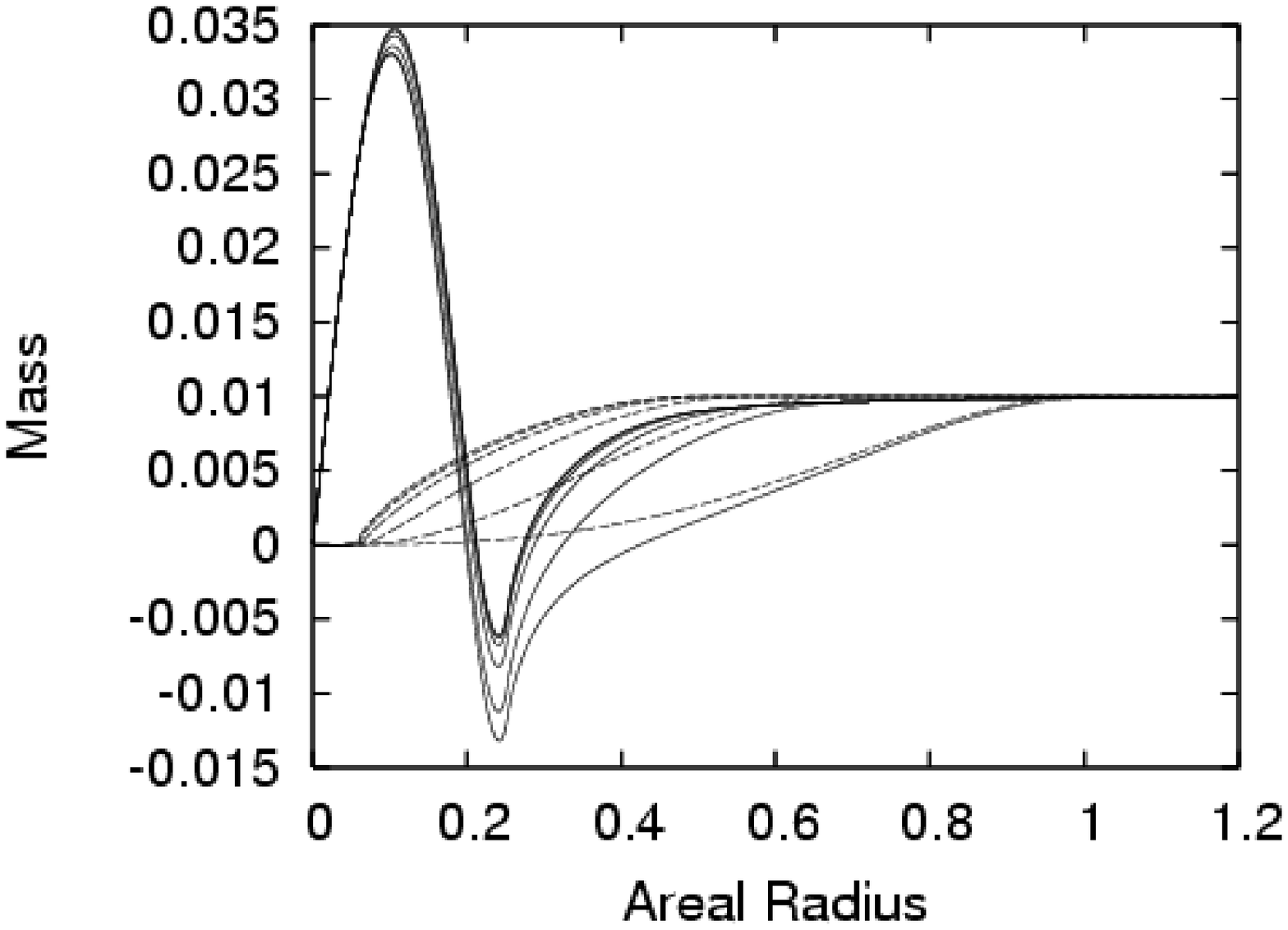}}&
\subfigure[Mass-radius ratio]
{\includegraphics[scale=.4]{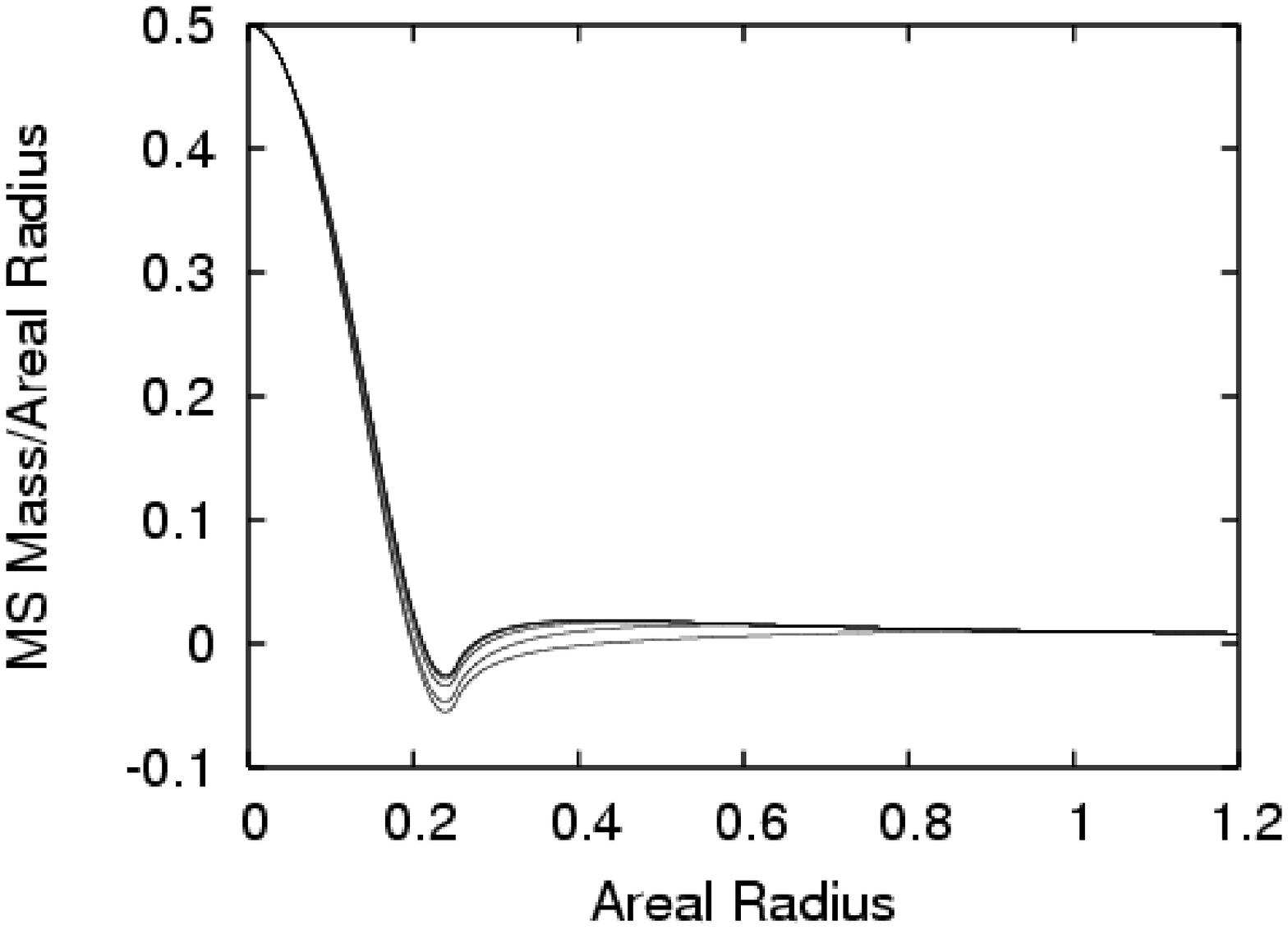}}
\end{tabular}
\end{center}
\caption{\label{fg:ltbr1_1} 
The collapse of an inhomogeneous dust ball with $R_{\rm s}=1$ and $M=0.01$
in the first version of loop quantum gravity:
the snapshots at $t=0$,
2.00754321, 2.65773896, 2.88917737, 2.97274431, and 3.00299176
are plotted.
In (a), the solid and dashed lines denote the effective density
and the conserved mass density, respectively.
In (c), the solid and dashed lines denote the Misner-Sharp mass
and the conserved mass, respectively.
}
\end{figure}

Fig.~\ref{fg:ltbr1_1} shows the collapse of an initially
inhomogeneous ball with $R_{\rm s}=1$ and $M=0.01$ in the first version
of inverse triad corrections in loop quantum gravity. The qualitative
properties are the same as in the initially uniform case.  Also here,
the collapse of the central region $R\lesssim 0.05$ is strongly slowed
down, while the outer dust falls onto the slowly collapsing central
region.  Again, a spike develops both in the effective density field
and the conserved mass density field at $R\simeq 0.06$ and then the
calculation breaks down soon after $t=3.00299176$.  
The maximum value of $m/R$ 
is a half attained at the center 
during this simulation as seen in Fig.~\ref{fg:ltbr1_1}(d).
This means that for this case, no trapping horizon is formed
before the spike or shell-crossing singularity is formed.

Without loop quantum effects, the collapse generically ends in the
formation of a shell-focusing singularity.
Hence, in regard of
singularity formation, the collapse ends in a tamer shell-crossing 
singularity prior to the possible shell-focusing singularity
due to the present loop quantum effects.

\begin{figure}[htbp]
\begin{center}
\begin{tabular}{cc}
\subfigure[Density profile]{\includegraphics[scale=.4]
{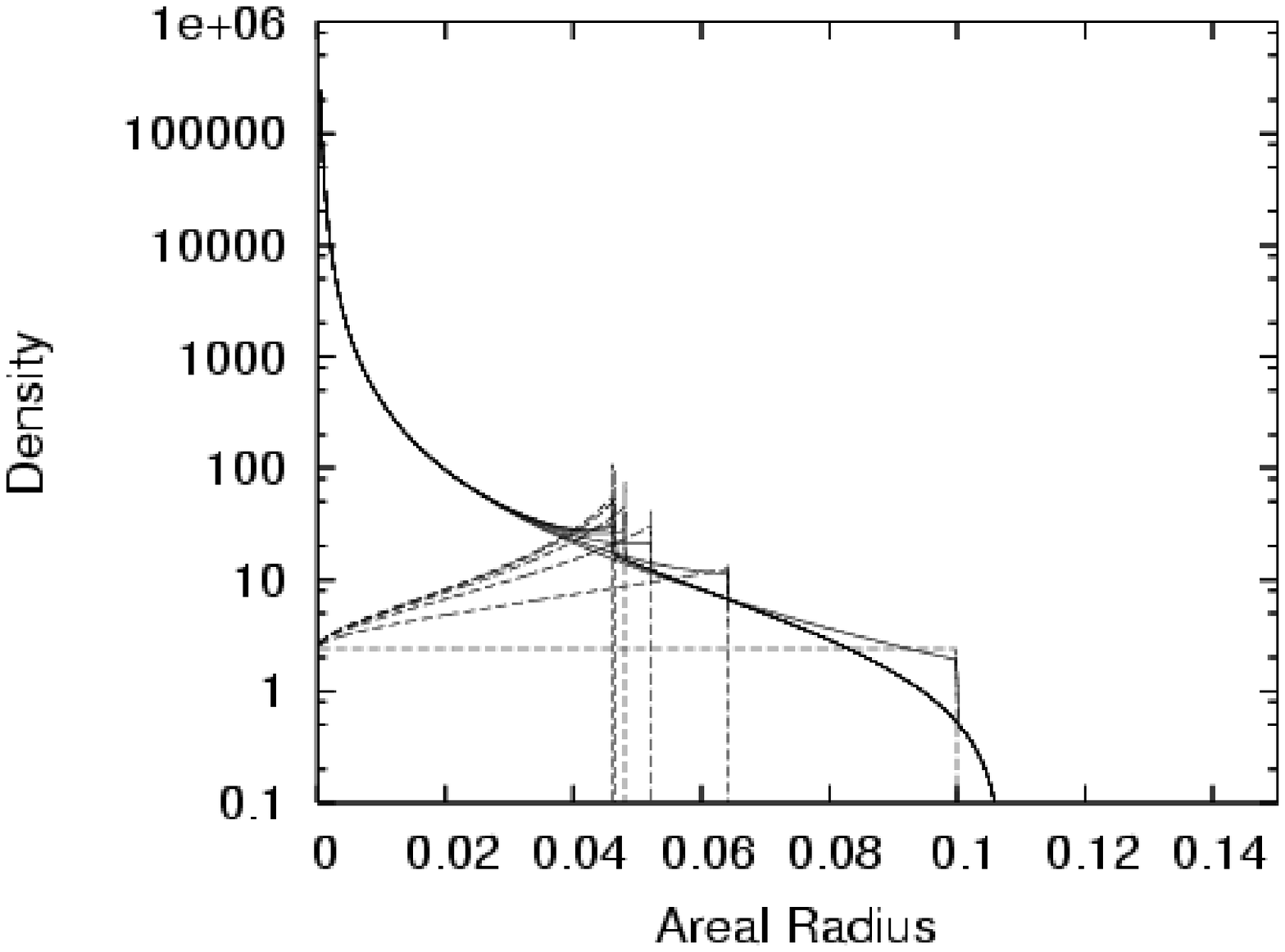}}&
\subfigure[Velocity profile]{\includegraphics[scale=.4]{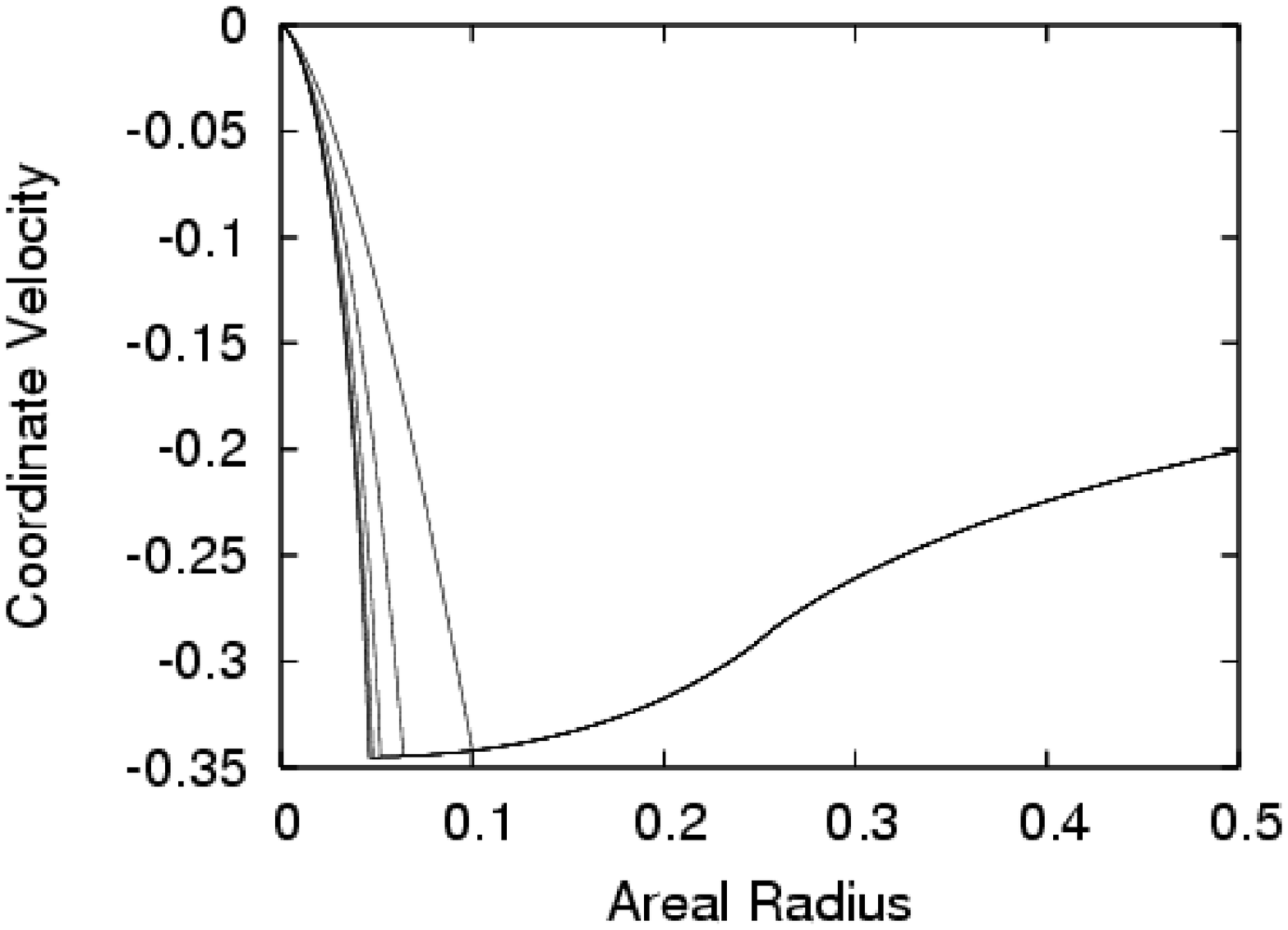}}\\
\subfigure[Quasi-local mass]
{\includegraphics[scale=.4]{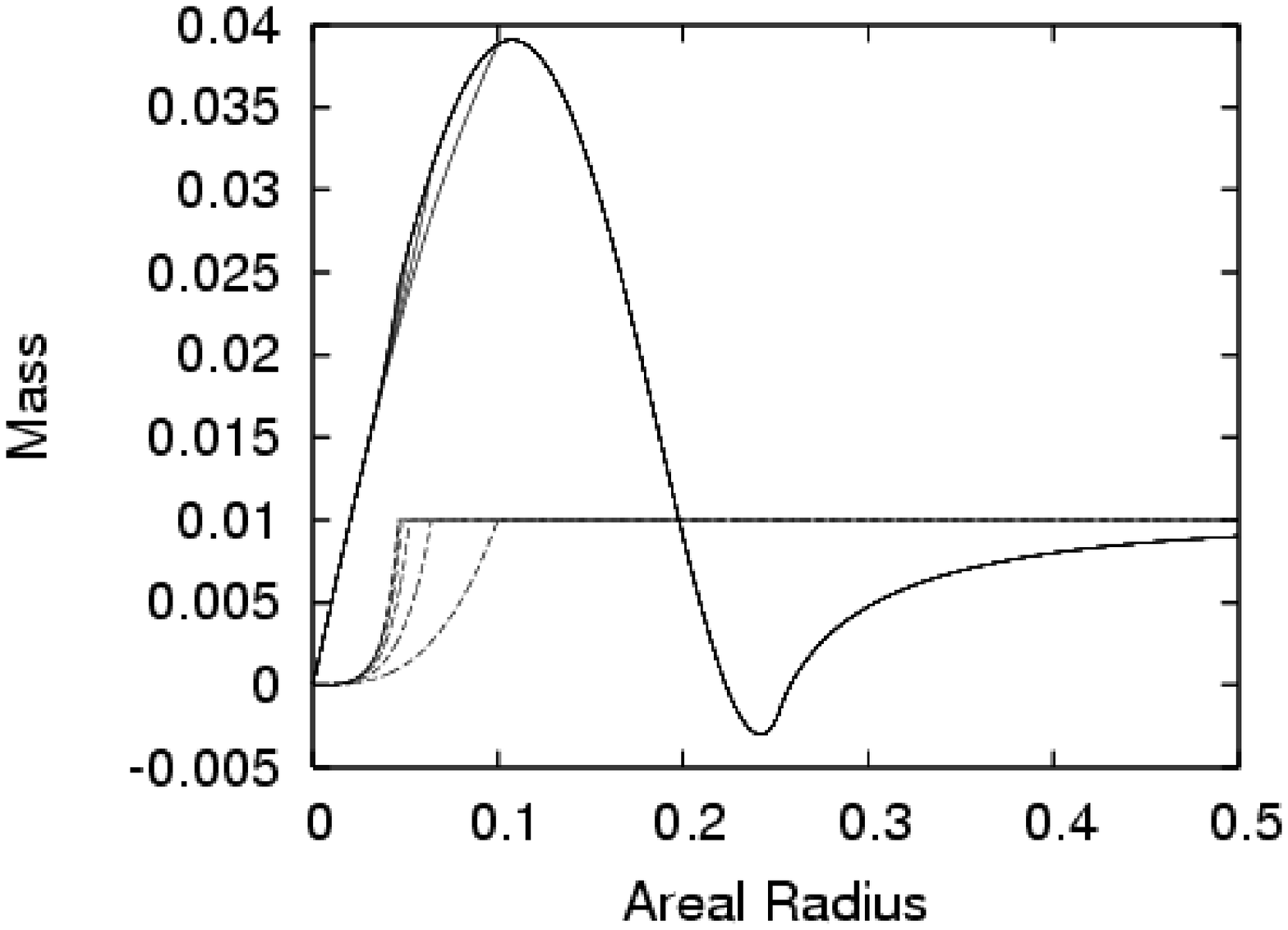}}&
\subfigure[Mass-radius ratio]
{\includegraphics[scale=.4]{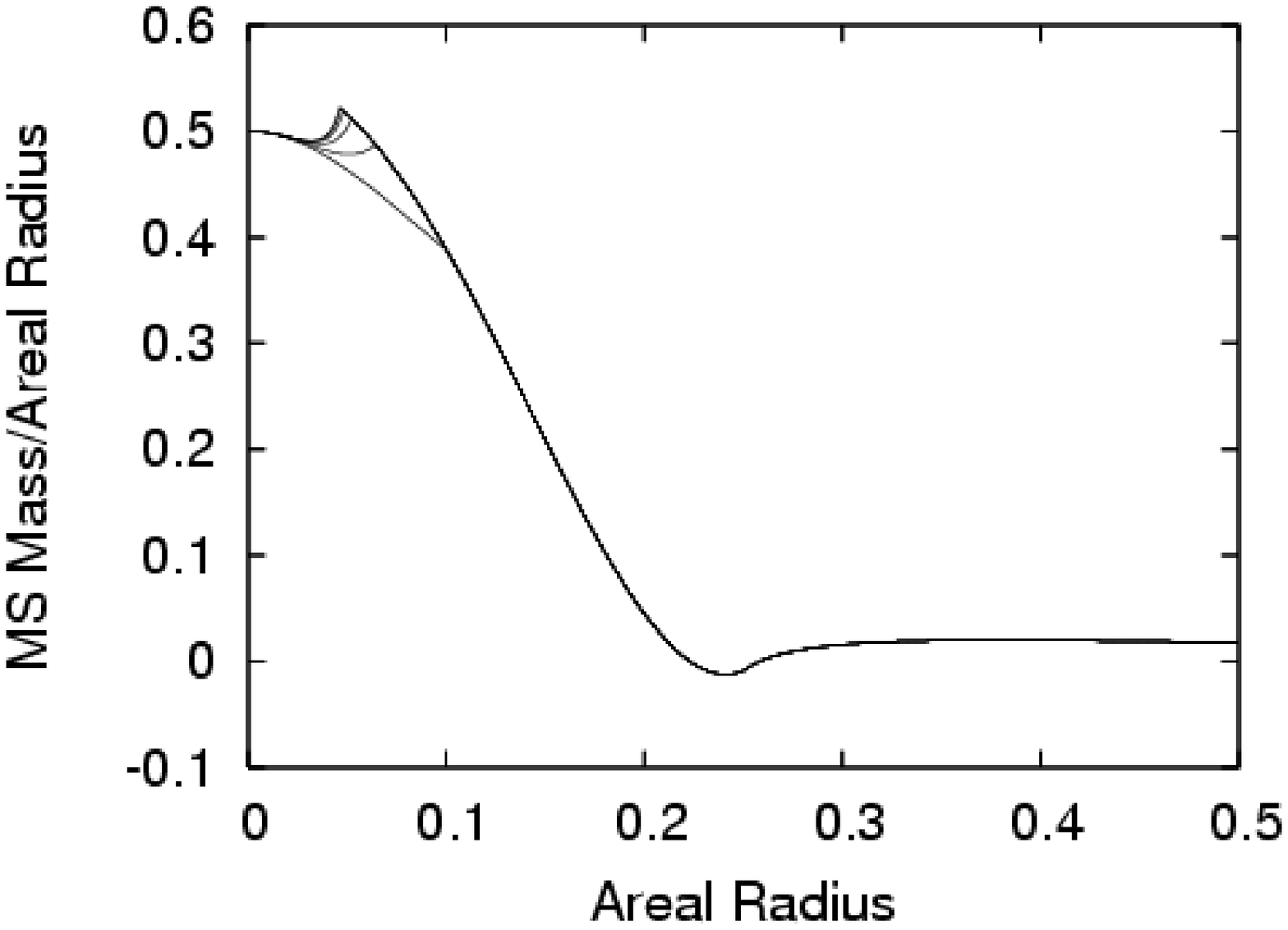}}
\end{tabular}
\end{center}
\caption{\label{fg:osr01_1}
The collapse of a homogeneous dust ball with $R_{\rm s}=0.1$ and $M=0.01$
in the first version of loop quantum gravity: 
the snapshots at $t=0$,
0.105124094,  0.139369943, 0.151074789, 0.155215854, and 
0.156702662 are plotted.
In (a), the solid and dashed lines denote the effective density
and the conserved mass density, respectively.
In (c), the solid and dashed lines denote the Misner-Sharp mass
and the conserved mass, respectively.
}
\end{figure}

If the initial radius of the cloud is smaller than the critical radius
$R_{*}\simeq 0.25$, the situation becomes slightly different.
Fig.~\ref{fg:osr01_1} shows the collapse of a homogeneous dust ball
with $R_{\rm s}=0.1$ and $M=0.01$.  As seen in Fig.~\ref{fg:osr01_1}
(a), the conserved mass density has its maximum at the cloud surface.
Then a spike develops at $R\simeq 0.046$ and the calculation breaks
down soon after $t=0.156702662$.  The effective density outside the
cloud is positive for $R\lesssim 0.11$ but negative for $0.11\lesssim
R\lesssim 0.24$.  The Misner-Sharp mass takes its maximum value at
$R\simeq 0.11$.  Outside the cloud, the profile of the Misner-Sharp
mass and therefore the effective density are almost unchanged compared
to the classical behavior as seen in Fig.~\ref{fg:osr01_1} (c).  In
fact, this feature is also seen for the cases where the initial radius
is larger than the critical radius.  Fig.~\ref{fg:osr01_1} (d) shows
that the maximum value of the ratio $m/R$ greater than a half and this
is attained at $R\simeq 0.046$, which implies that the shell-crossing
singularity is covered by a trapping horizon in this case.

\begin{figure}[htbp]
\begin{center}
\begin{tabular}{cc}
\subfigure[Density profile]{\includegraphics[scale=.4]
{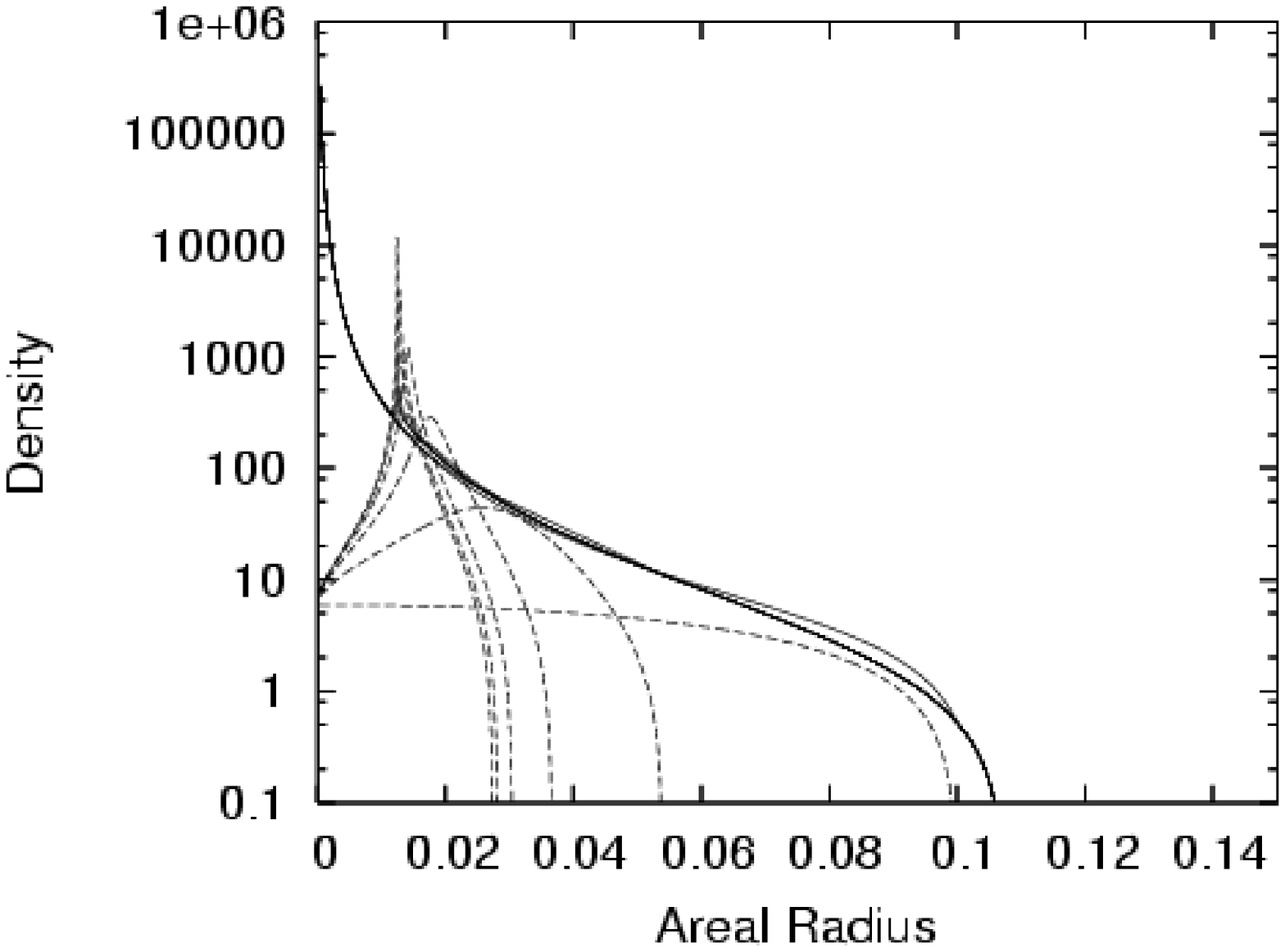}}&
\subfigure[Velocity profile]{\includegraphics[scale=.4]{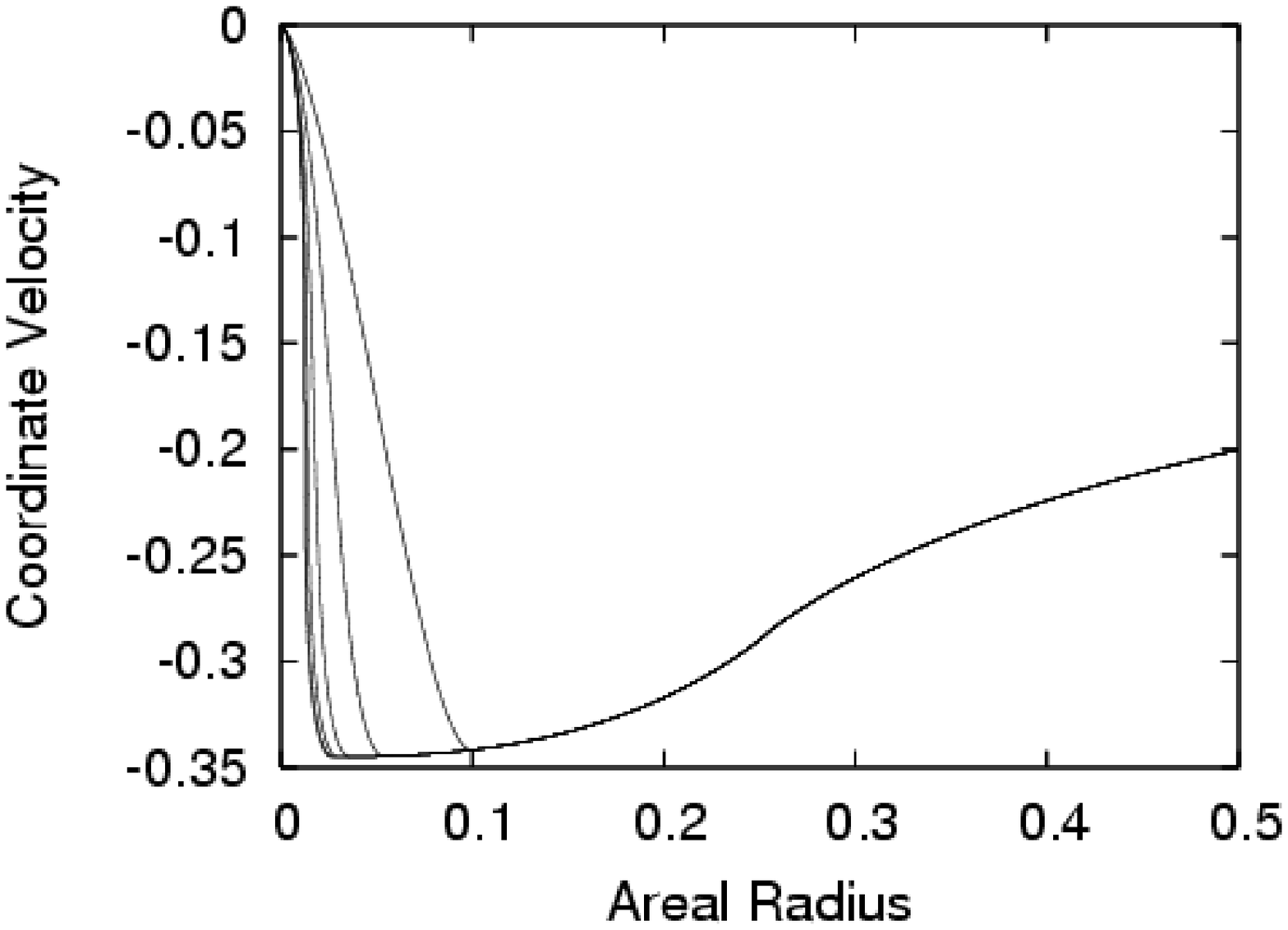}}\\
\subfigure[Quasi-local mass]
{\includegraphics[scale=.4]{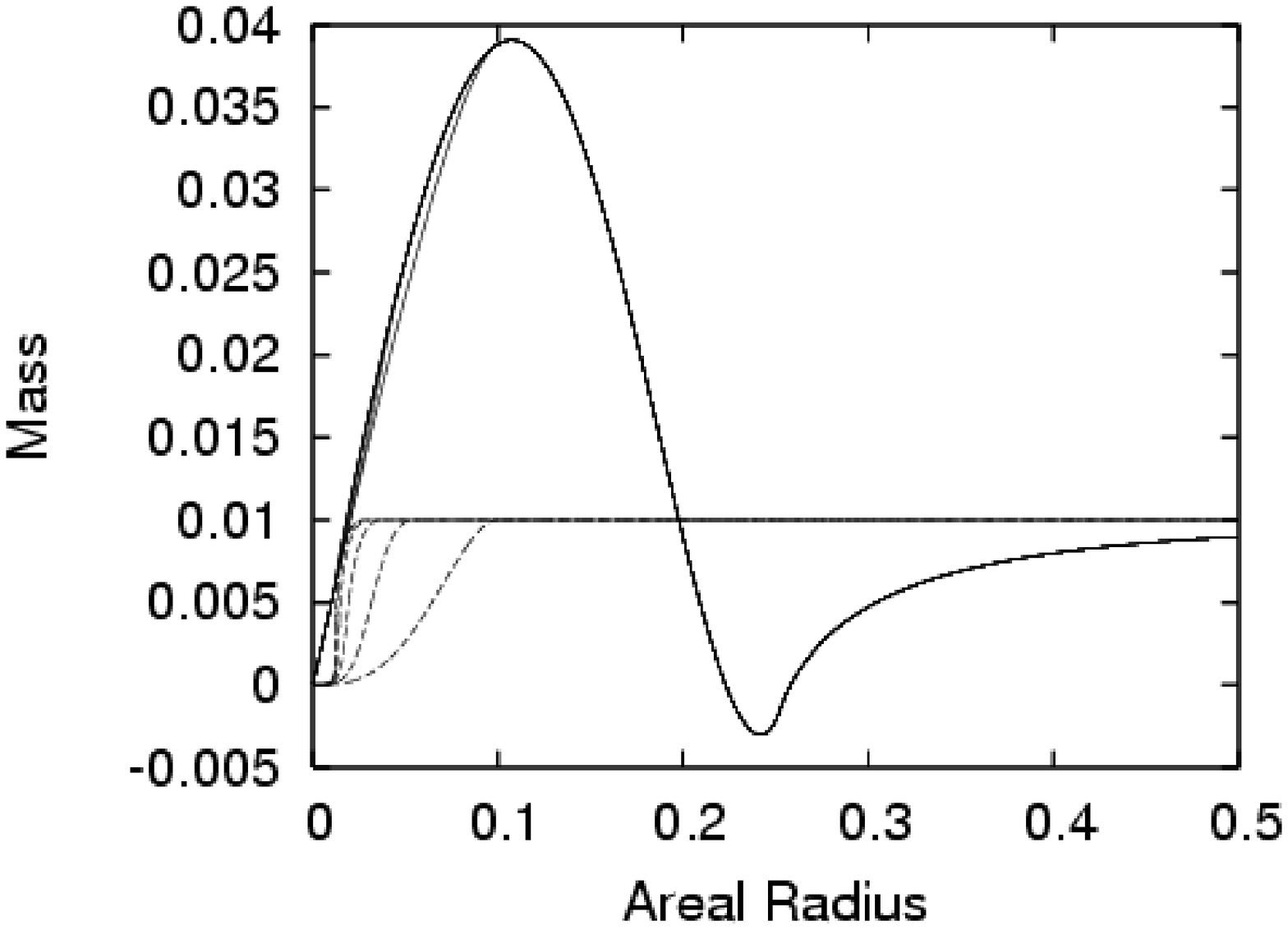}}&
\subfigure[Mass-radius ratio]
{\includegraphics[scale=.4]{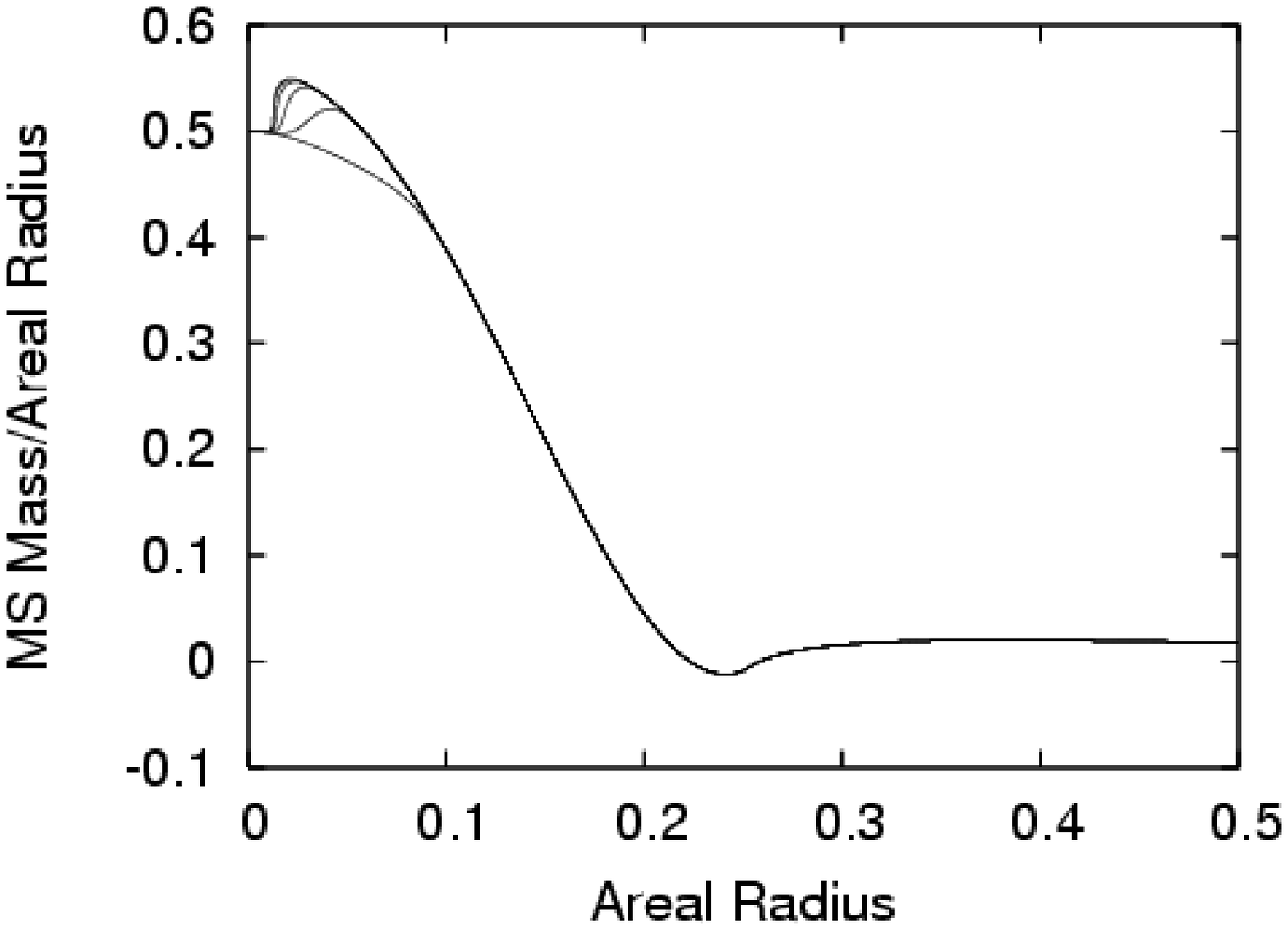}}
\end{tabular}
\end{center}
\caption{\label{fg:ltbr01_1}
The collapse of an inhomogeneous dust ball $R_{\rm s}=0.1$ and 
$M=0.01$ in the first version of loop quantum gravity:
the snapshots at $t=0$,
0.134900072, 0.18447396, 0.202429874, 0.208934676, and 
0.211291203
are plotted.
In (a), the solid and dashed lines denote the effective density
and the conserved mass density, respectively.
In (c), the solid and dashed lines denote the Misner-Sharp mass
and the conserved mass, respectively.
}
\end{figure}

For the collapse of an initially inhomogeneous ball, 
which is shown in Fig.~\ref{fg:ltbr01_1}, the qualitative
feature is almost the same. However, the central region where the
collapse is strongly slowed down is much smaller than in the case
where the initial radius is larger than the critical radius. As a
result, a spike develops at the radius $R\simeq  0.012$.  
The shell-crossing singularity is covered by a trapping horizon
in this case.

\subsubsection{Inverse triad corrections: Second version}

\begin{figure}[htbp]
\begin{center}
\begin{tabular}{cc}
\subfigure[Density profile]{\includegraphics[scale=.4]
{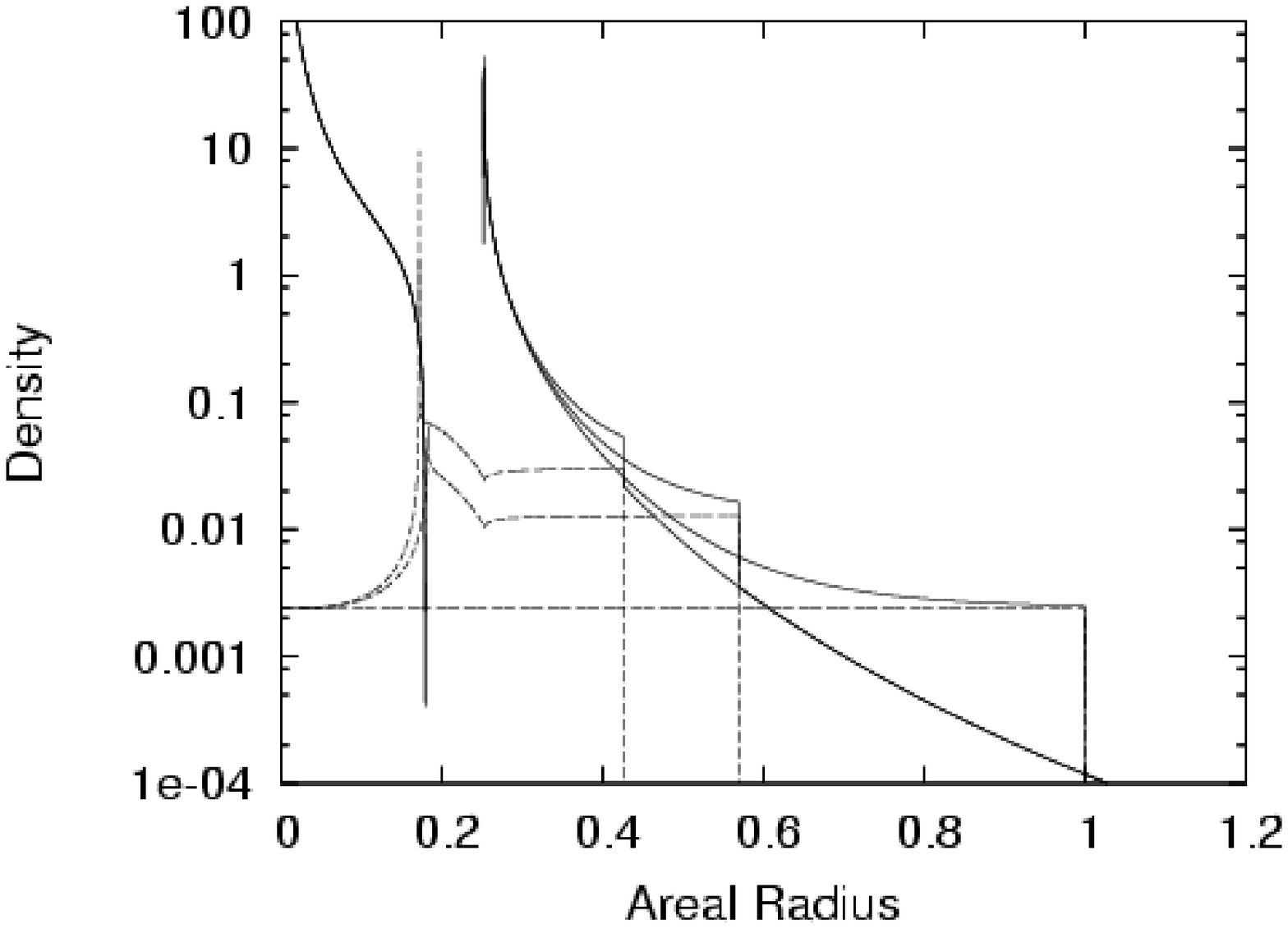}}&
\subfigure[Velocity profile]{\includegraphics[scale=.4]{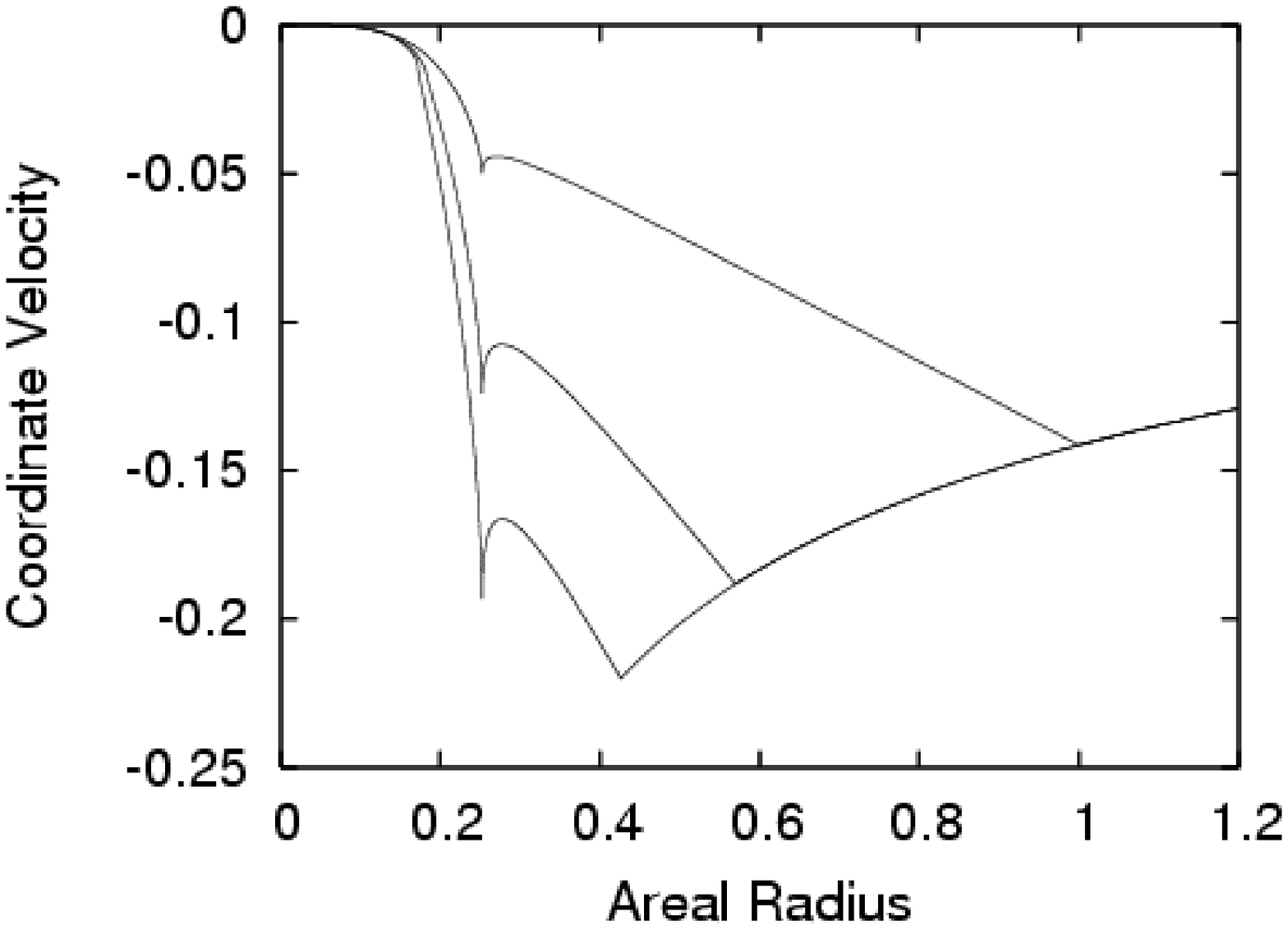}}\\
\subfigure[Quasi-local mass]
{\includegraphics[scale=.4]{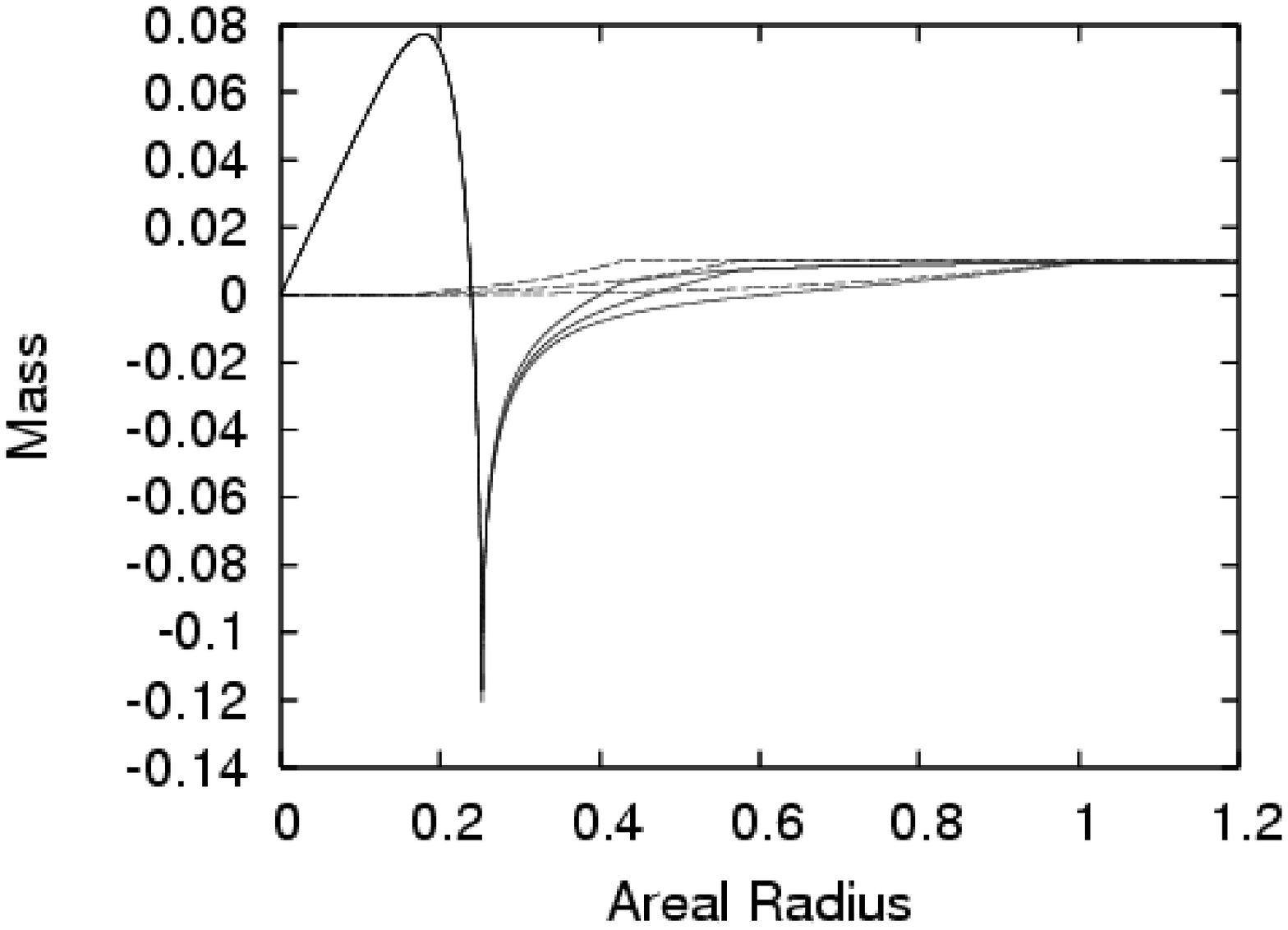}}&
\subfigure[Mass-radius ratio]
{\includegraphics[scale=.4]{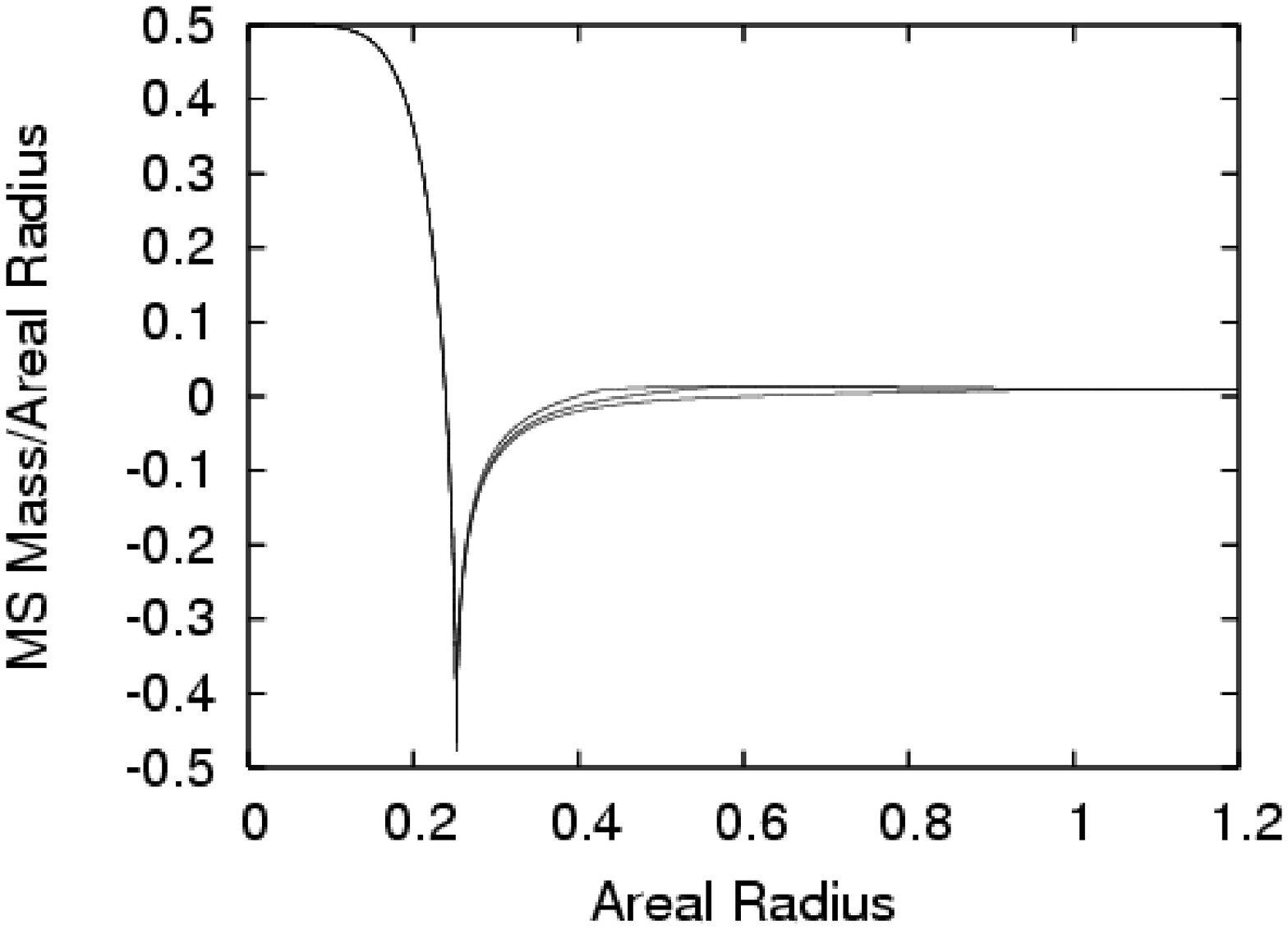}}
\end{tabular}
\end{center}
\caption{\label{fg:osr1_2}
The collapse of a homogeneous dust ball with $R_{\rm s}=1$ and $M=0.01$
in the second version of loop quantum gravity:
the snapshots at $t=0$,
1.43300373, 2.34719999, 2.95414244, 3.36400543, and 3.39105418
are plotted.
In (a), the solid and dashed lines denote the effective density
and the conserved mass density, respectively.
In (c), the solid and dashed lines denote the Misner-Sharp mass
and the conserved mass, respectively.
}
\end{figure}

Fig.~\ref{fg:osr1_2} shows the collapse of an initially homogeneous
ball with $R_{\rm s}=1$ and $M=0.01$ 
in the second version of consistent inverse triad corrections in
loop quantum gravity.  We can easily see that in spite of the very
different formulation, the key feature that the collapse of the
central region is strongly slowed down is the same as in the first
version.  The radius of this almost stopped central region is $\simeq 
0.2$, which is much larger than in the first version.

In the conserved mass density profile a spike develops at $R\simeq
0.19$, $t=3.39105418$.  There is a dip in the profile of the conserved
mass density at $R\simeq 0.25$, which corresponds to the critical
radius $R=R_{*}$.  For $R\gtrsim R_{*}$, the collapse proceeds as it
does classically. As in the first version, the effective density is
diverging at the center.  If we go further outside, it decreases very
rapidly to negative values and then turns to increase to positive
values.  The velocity profile shows two minimum values at the cloud
surface and at $R\simeq R_{*}$ as seen in Fig.~\ref{fg:osr1_2} (b).
The Misner-Sharp mass dominates the conserved mass for $R\lesssim 0.2$
and it has a maximum about 0.08 at $R\simeq 0.18$ as seen in
Fig.~\ref{fg:osr1_2} (c).  The center is always marginally trapped and
the surrounding region is untrapped.  Hence, the shell-crossing
singularity is not covered by a trapping horizon.

\begin{figure}[htbp]
\begin{center}
\begin{tabular}{cc}
\subfigure[Density profile]{\includegraphics[scale=.4]
{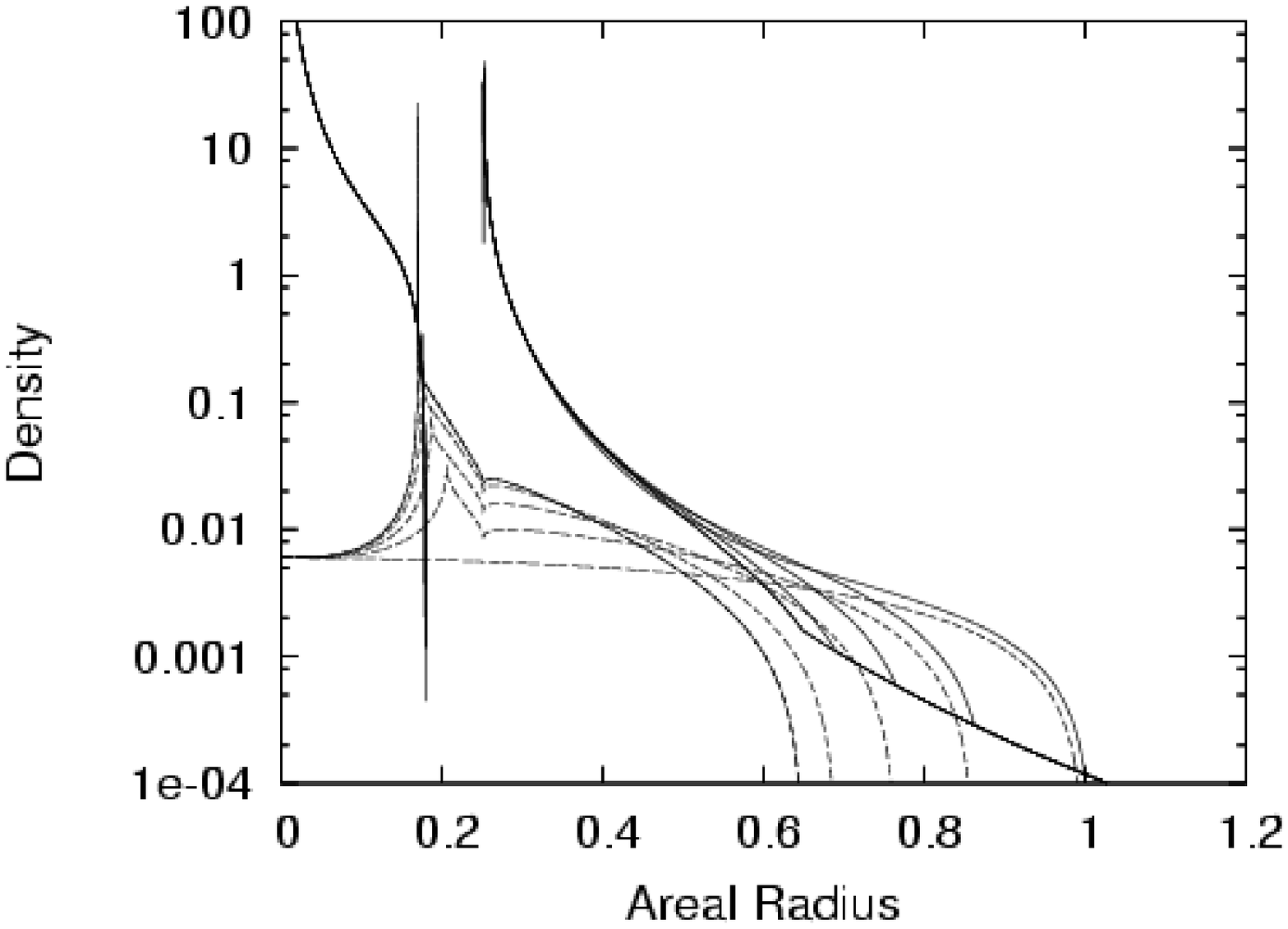}}&
\subfigure[Velocity profile]{\includegraphics[scale=.4]{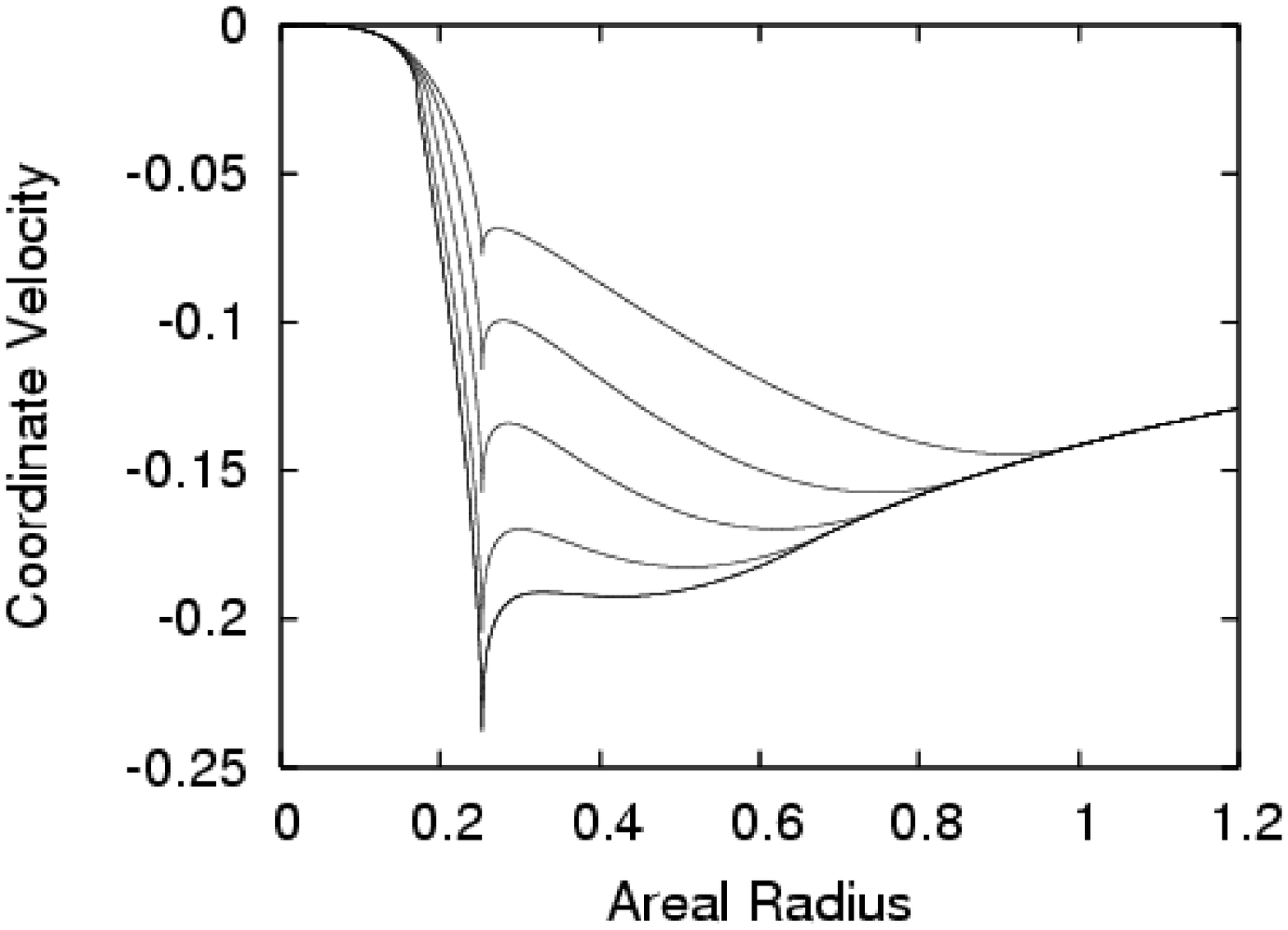}}\\
\subfigure[Quasi-local mass]
{\includegraphics[scale=.4]{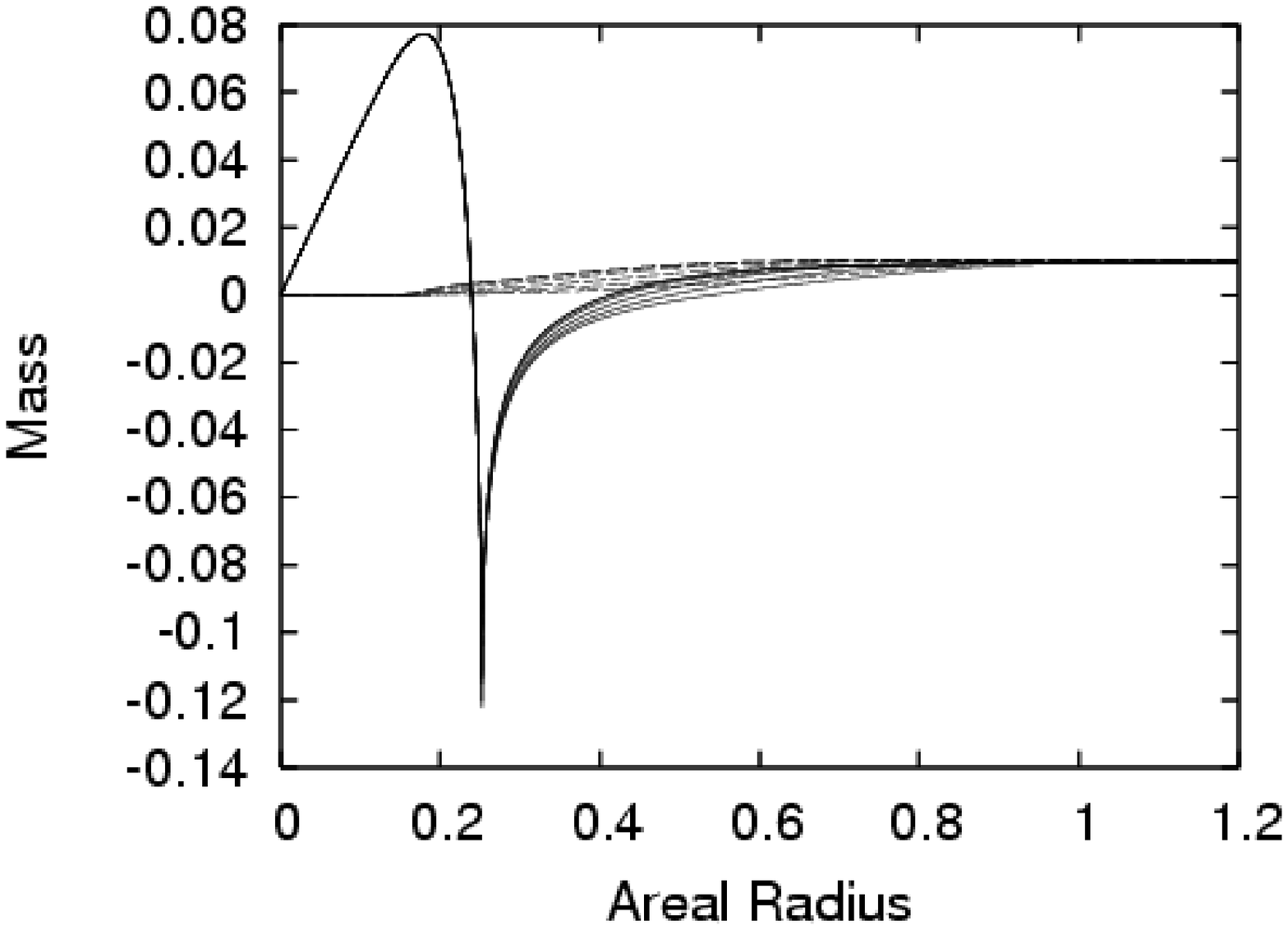}}&
\subfigure[Mass-radius ratio]
{\includegraphics[scale=.4]{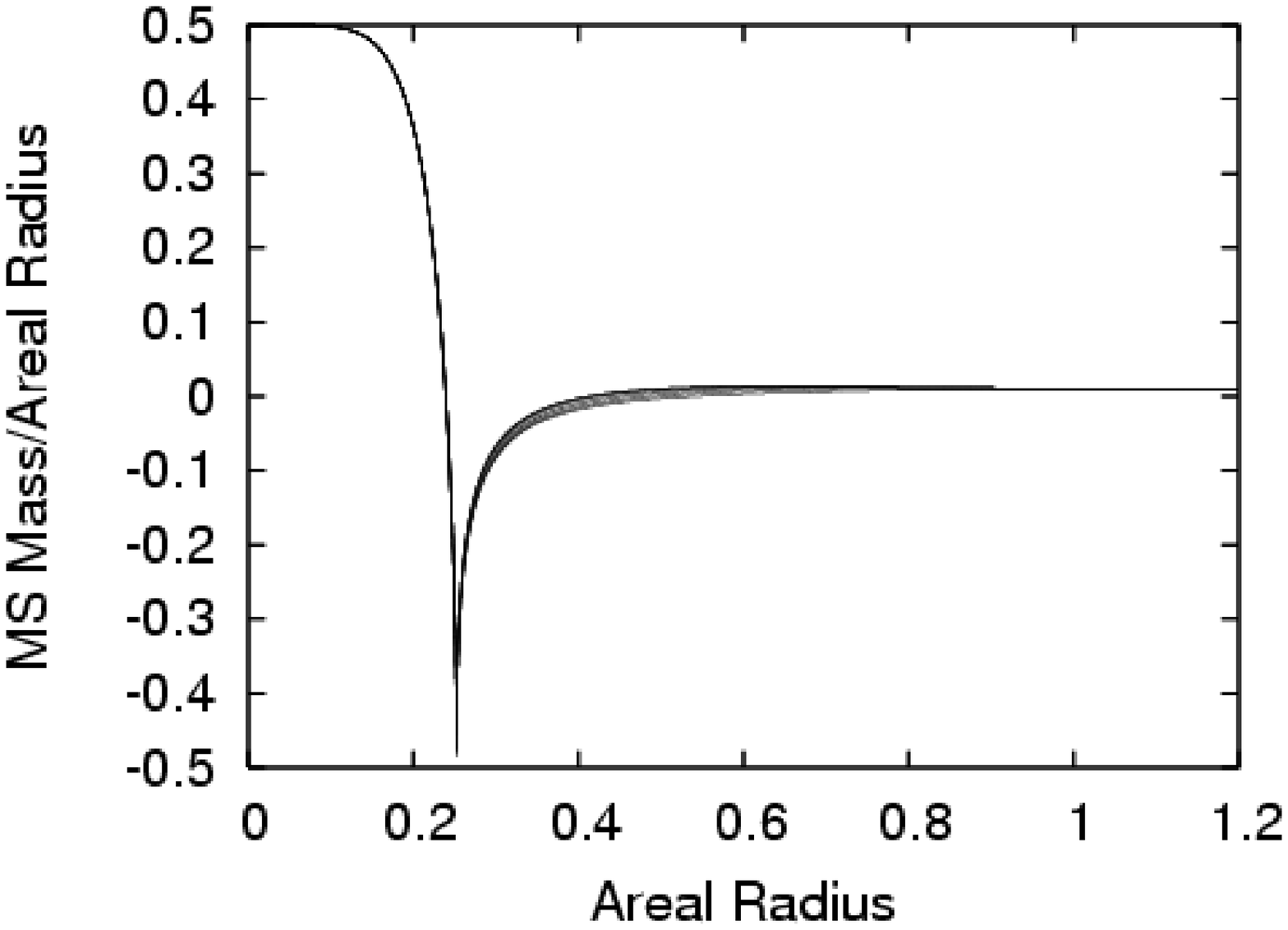}}
\end{tabular}
\end{center}
\caption{\label{fg:ltbr1_2} The collapse of an 
inhomogeneous dust ball with $R_{\rm s}=1$ and $M=0.01$
in the second version of loop quantum gravity:
the snapshots at $t=0$,
0.937882973, 1.5570413, 2.00350064, 2.2452289, and 
2.2463422 are plotted.
In (a), the solid and dashed lines denote the effective density
and the conserved mass density, respectively.
In (c), the solid and dashed lines denote the Misner-Sharp mass
and the conserved mass, respectively.
}
\end{figure}

Fig.~\ref{fg:ltbr1_2} shows the evolution of an initially
inhomogeneous dust ball with $R_{\rm s}=1$ and $M=0.01$ in the second
version of loop quantum gravity.  The qualitative features are common
with the homogeneous case except that the velocity profile has only
one minimum during the late stage of collapse.  A spike develops in the
density fields at $R\simeq  0.19$, $t=2.2452289$.  
The spike or shell-crossing singularity is not covered by 
a trapping horizon.

\begin{figure}[htbp]
\begin{center}
\begin{tabular}{cc}
\subfigure[Density profile]{\includegraphics[scale=.4]
{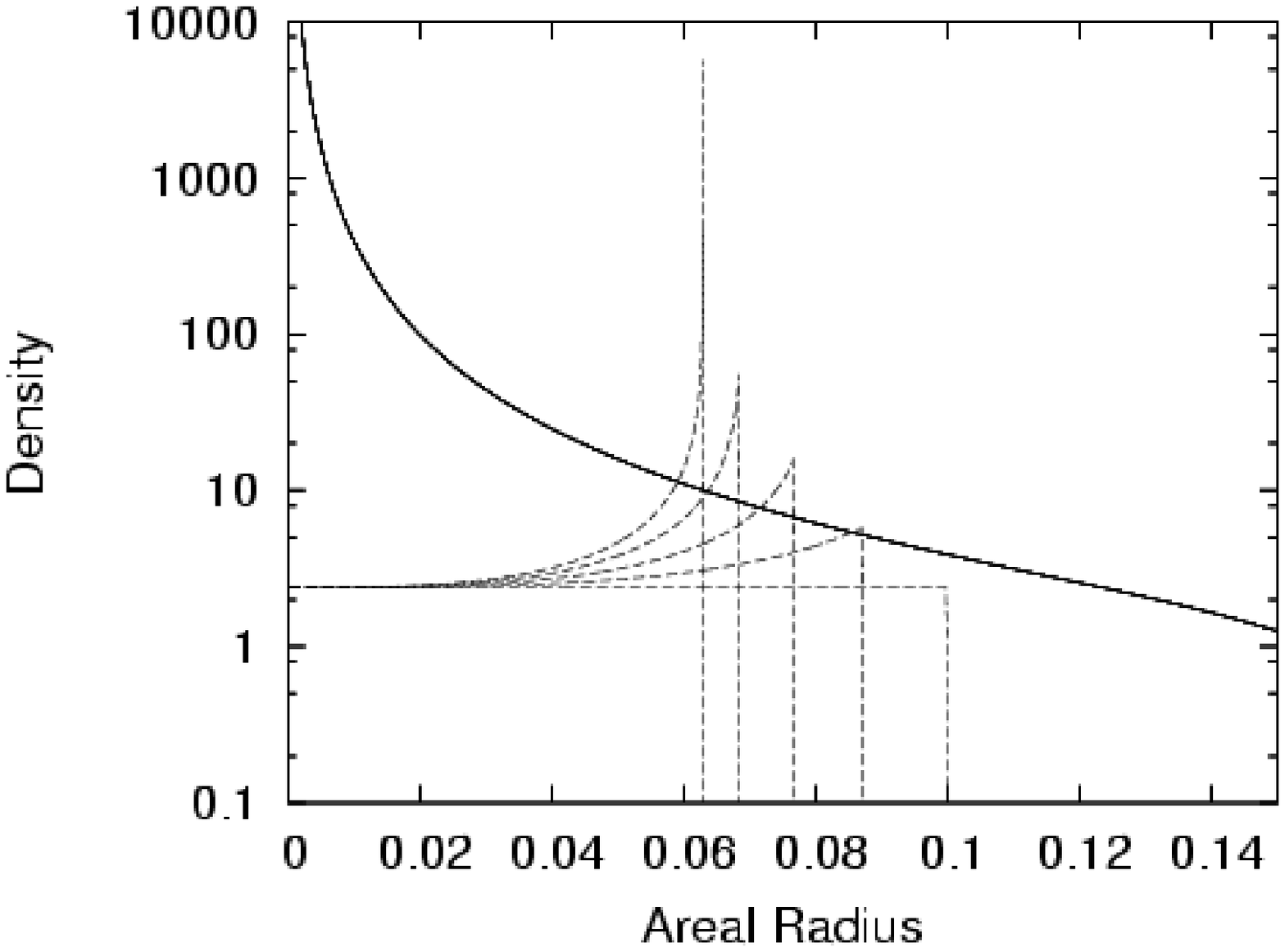}}&
\subfigure[Velocity profile]{\includegraphics[scale=.4]{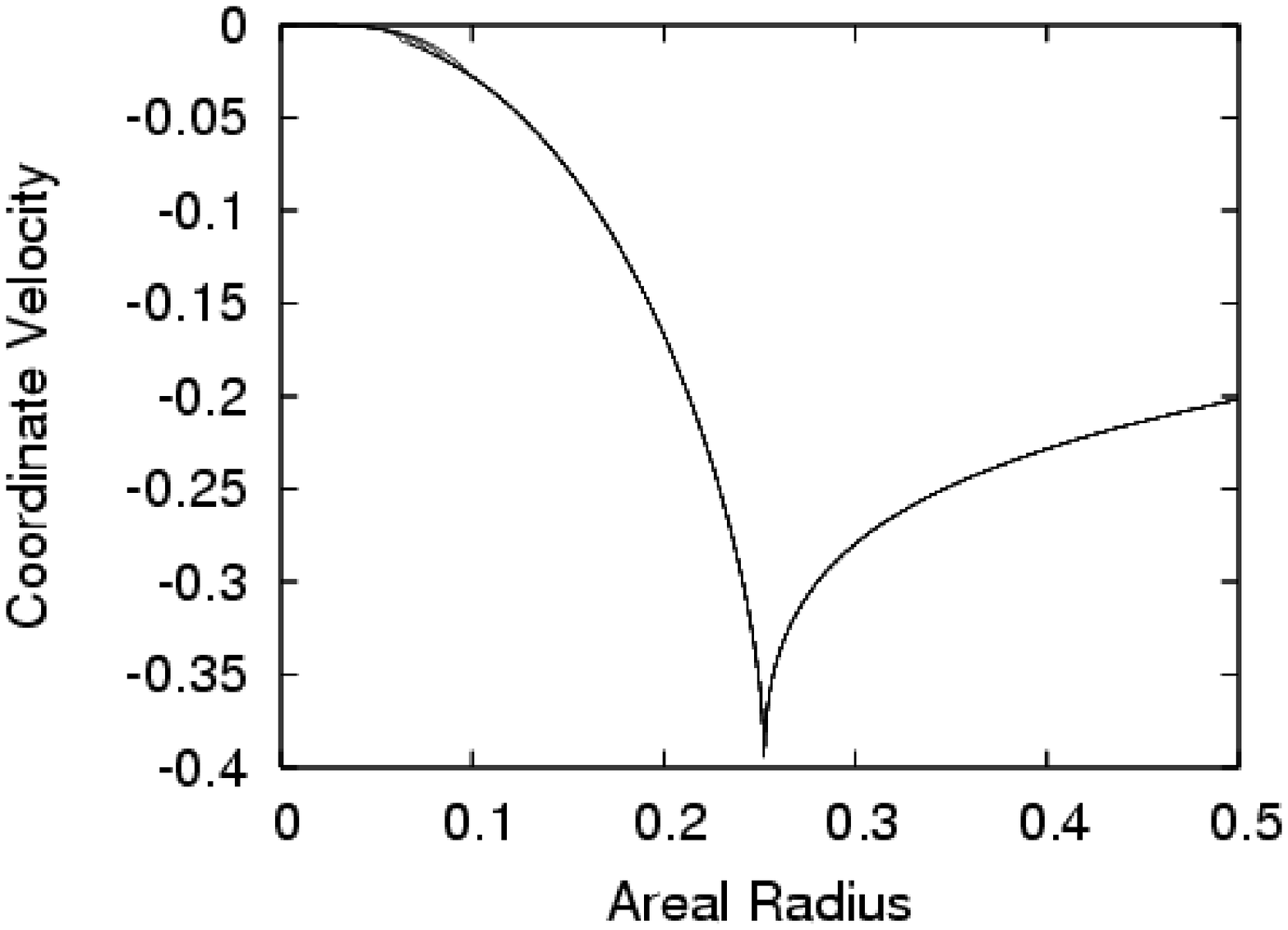}}\\
\subfigure[Quasi-local mass]
{\includegraphics[scale=.4]{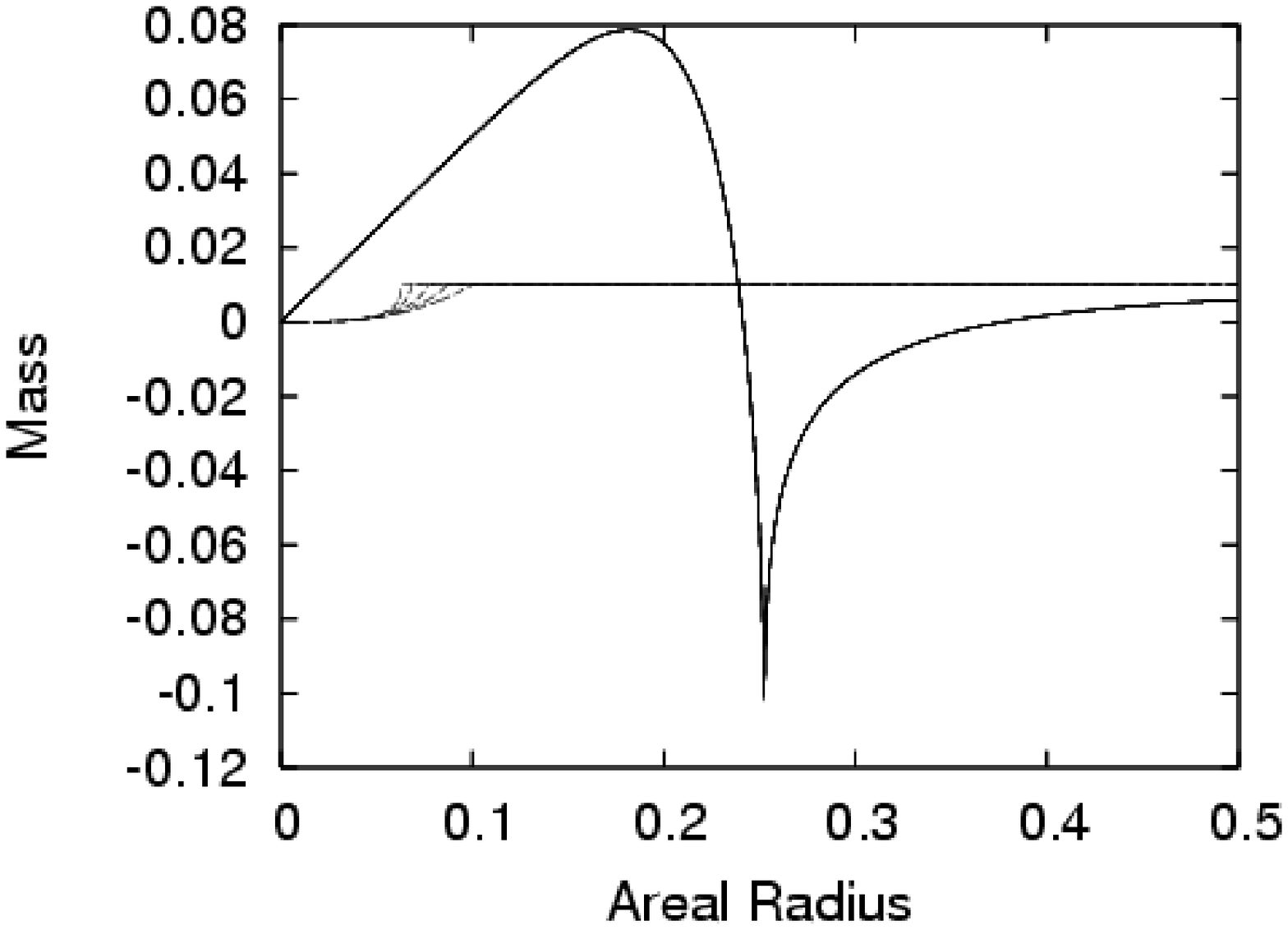}}&
\subfigure[Mass-radius ratio]
{\includegraphics[scale=.4]{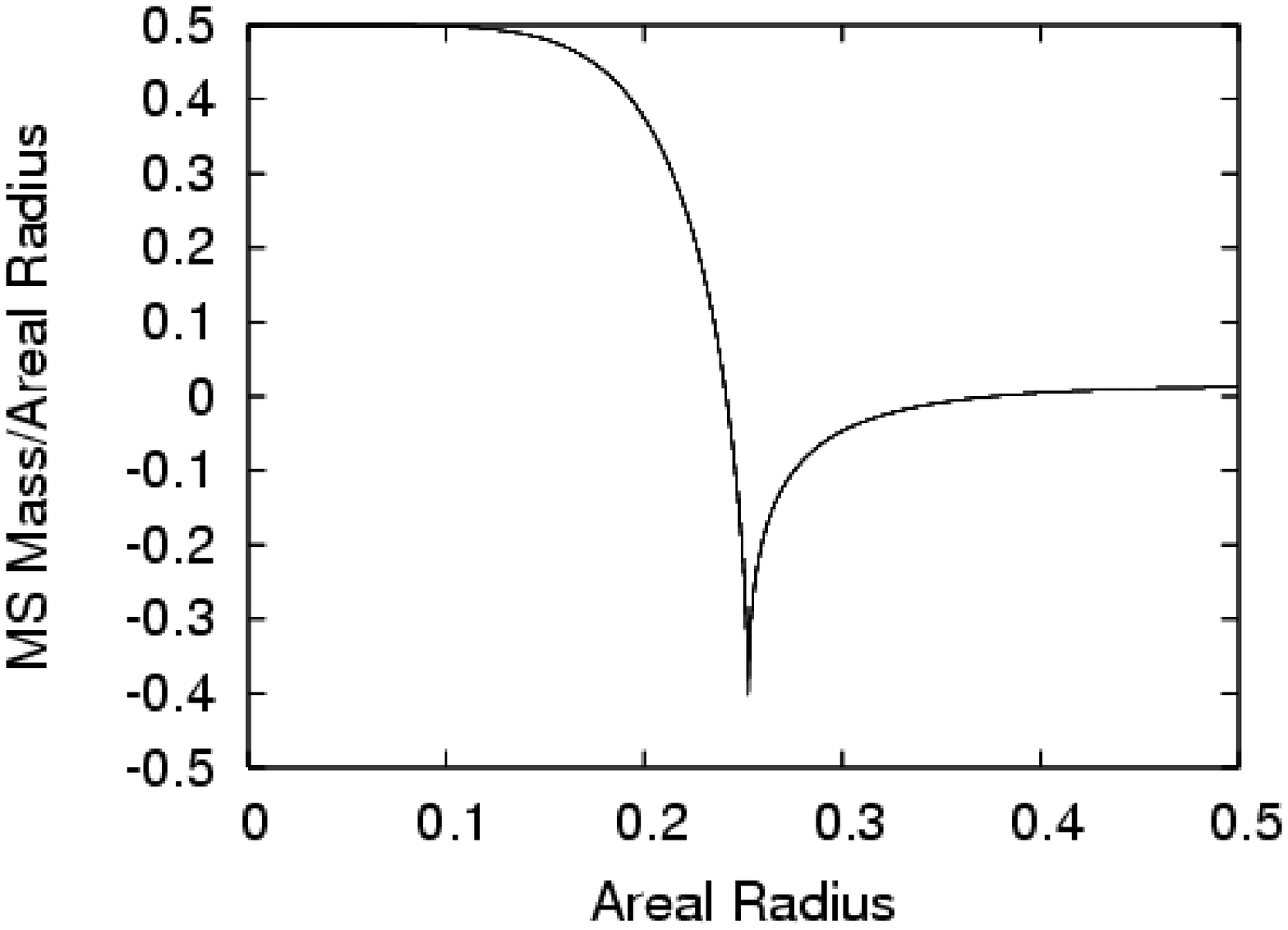}}
\end{tabular}
\end{center}
\caption{\label{fg:osr01_2}
The collapse of a homogeneous dust ball with $R_{\rm s}=0.1$ and $M=0.01$
in the second version formulation of loop quantum gravity: 
the snapshots at $t=0$,
0.550926063, 1.1729359, 1.83622205, and 2.39522102
are plotted.
In (a), the solid and dashed lines denote the effective density
and the conserved mass density, respectively.
In (c), the solid and dashed lines denote the Misner-Sharp mass
and the conserved mass, respectively.
}
\end{figure}

Fig.~\ref{fg:osr01_2} shows the evolution of an initially
homogeneous dust ball with $R_{\rm s}=0.1$ and $M=0.01$.  In this case,
the collapse proceeds very slowly in comparison to the classical
evolution.  It should be noted that in classical general relativity,
the collapse ends in singularity formation at $t=\sqrt{2}/3\times
(0.1^{3}/0.01)^{1/2}\simeq 0.1490712\cdots$.  
In the second version of loop quantum
gravity, the velocity field is kept very small within the whole cloud
as seen in Fig.~\ref{fg:osr01_2} (b).  The profile of the conserved
mass density shows a spike developing at the cloud surface at $R\simeq 
0.062$ as seen in Fig.~\ref{fg:osr01_2} (a). The simulation breaks down
due to this spike soon after $t=2.39522102$.  The profile of the
effective density is very different from that of the conserved mass
density. The effective density becomes negative for $0.18\lesssim R
\lesssim 0.25 $.
This is seen better in Fig.~\ref{fg:osr01_2} (c).
The maximum value of the
Misner-Sharp mass is about 0.08. The maximum
value of the ratio $m/R$ is a half, which is attained at the center.
Hence, the spike or shell-crossing singularity is not covered by 
a trapping horizon.

\begin{figure}[htbp]
\begin{center}
\begin{tabular}{cc}
\subfigure[Density profile]{\includegraphics[scale=.4]
{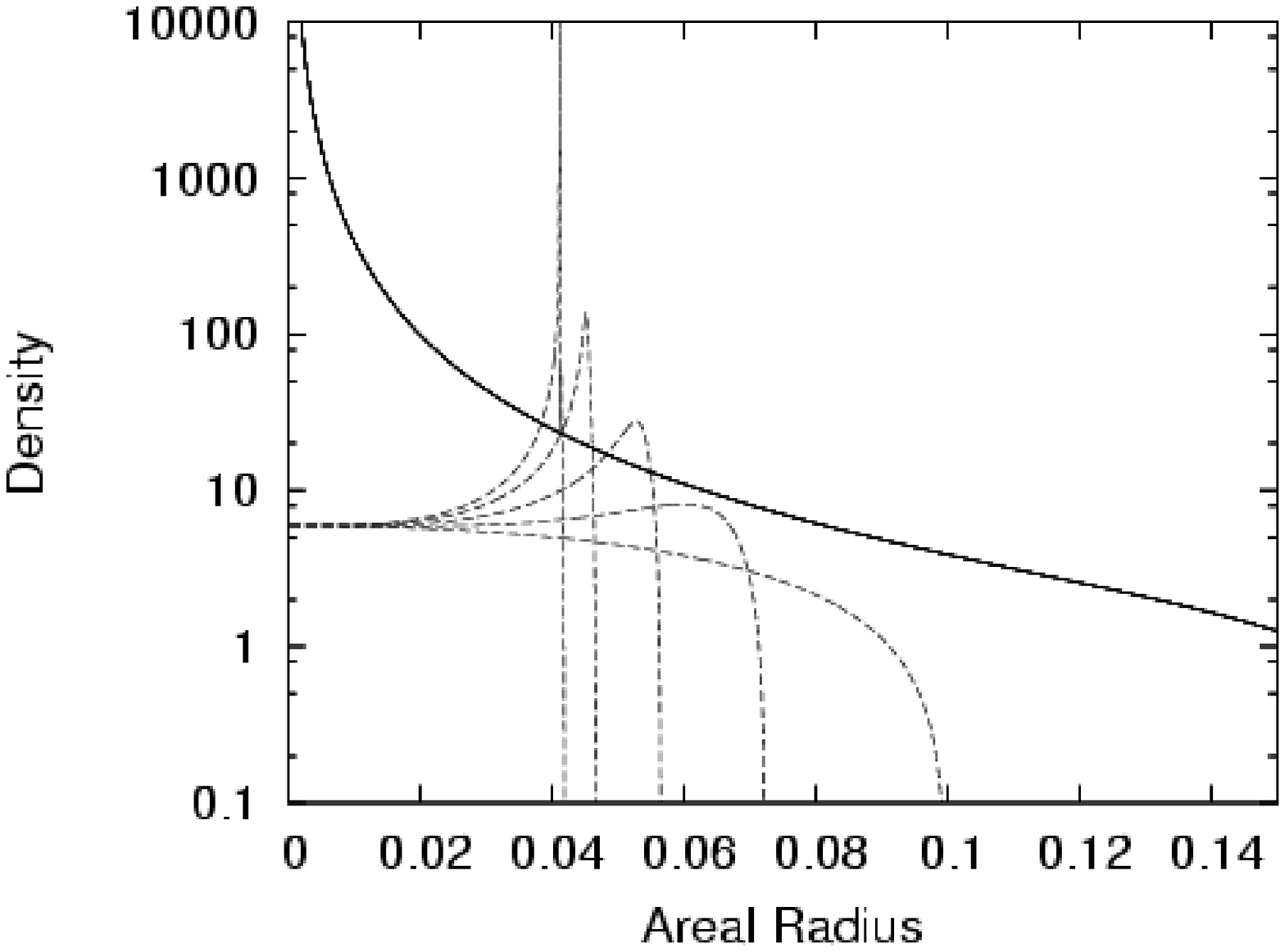}}&
\subfigure[Velocity profile]{\includegraphics[scale=.4]{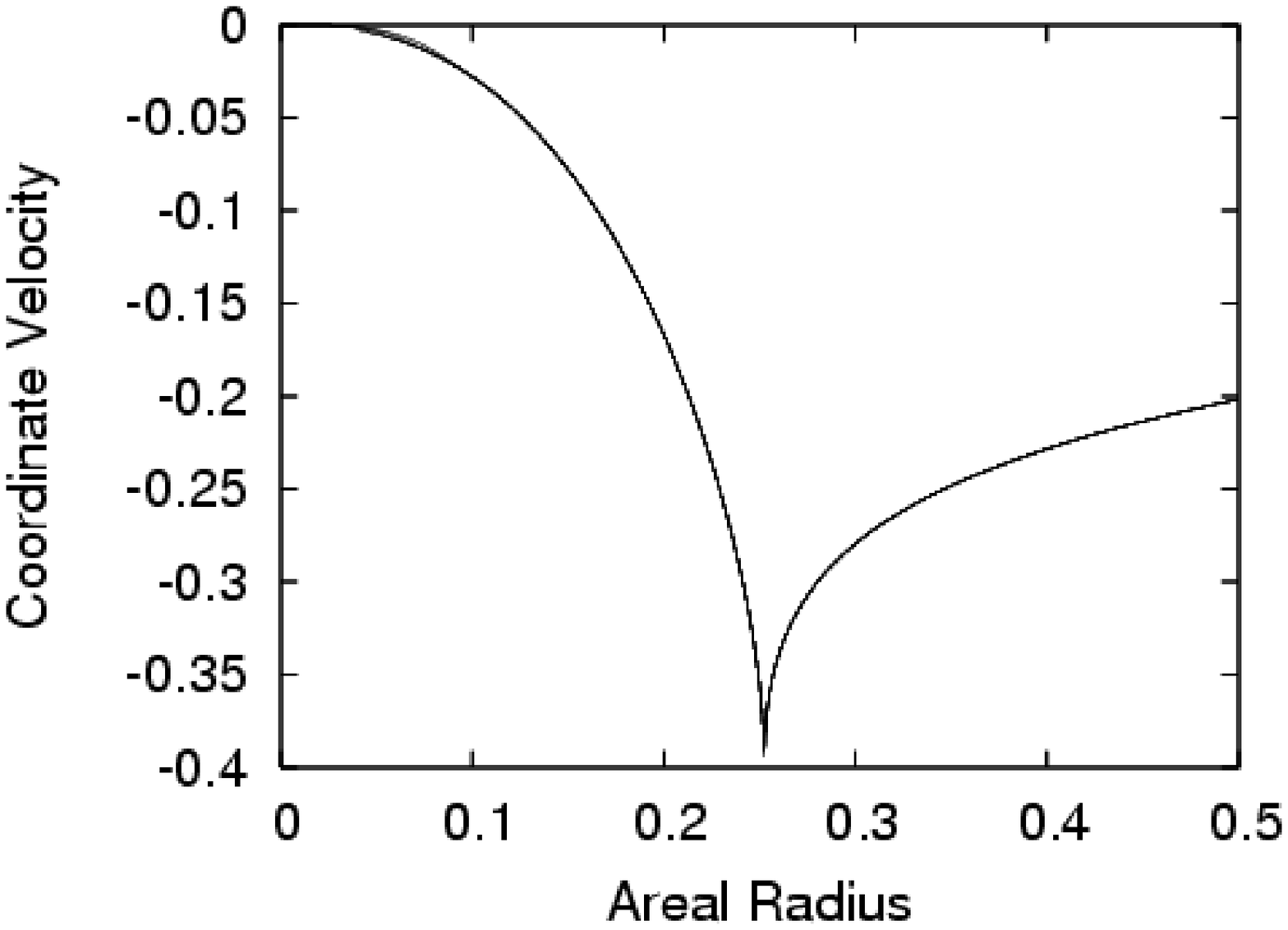}}\\
\subfigure[Quasi-local mass]
{\includegraphics[scale=.4]{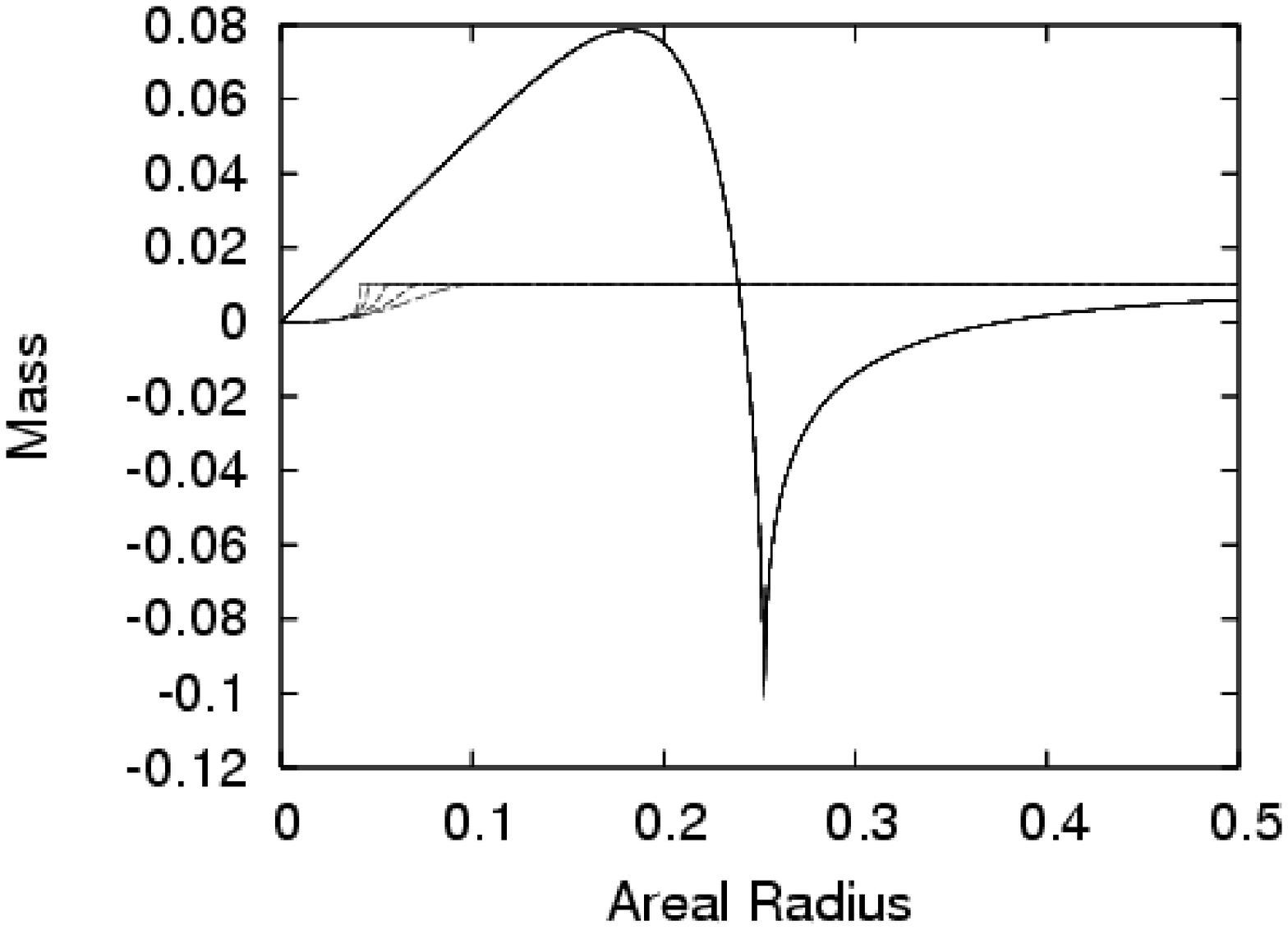}}&
\subfigure[Mass-radius ratio]
{\includegraphics[scale=.4]{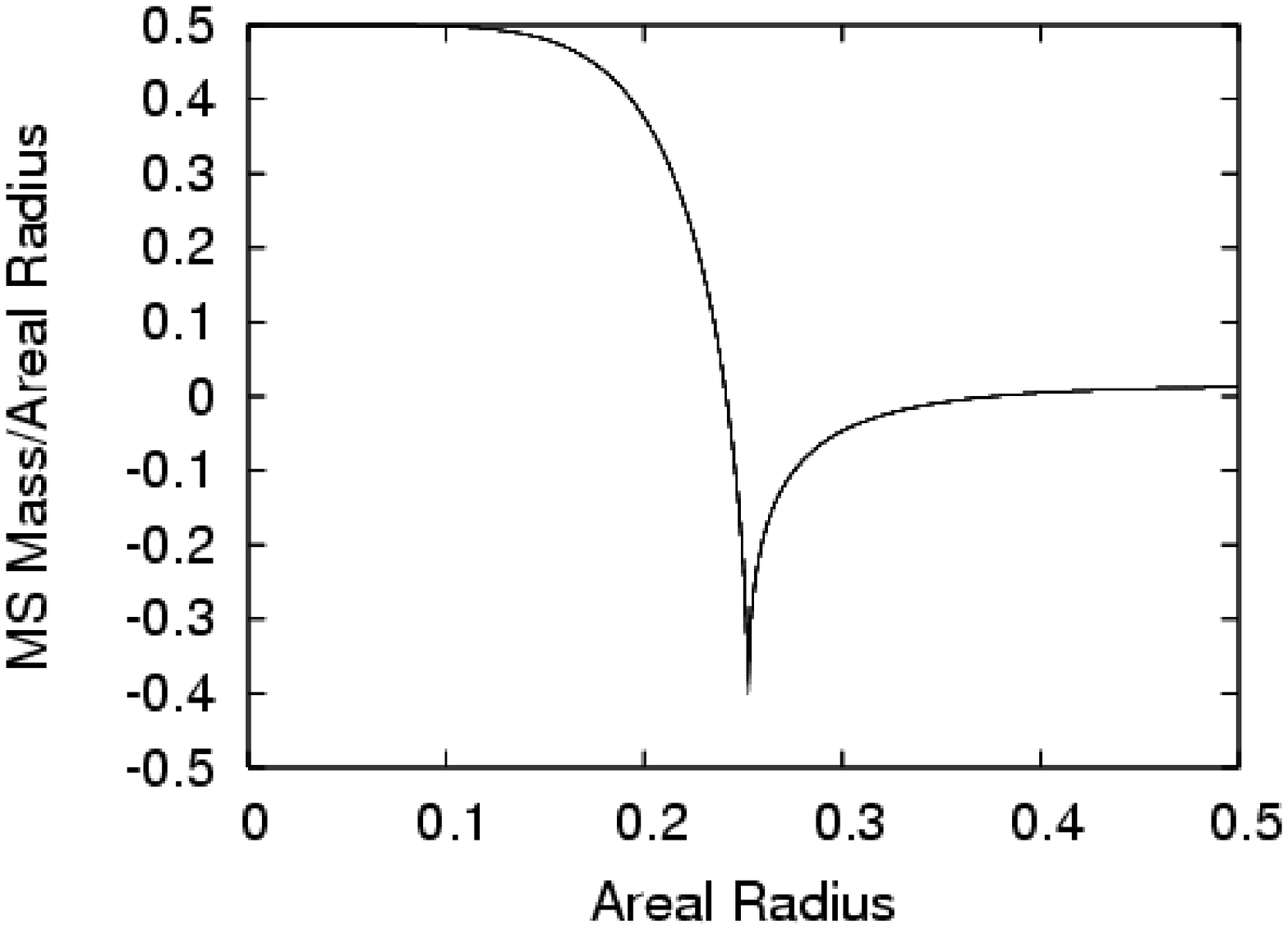}}
\end{tabular}
\end{center}
\caption{\label{fg:ltbr01_2}The collapse of an inhomogeneous 
dust ball with $R_{\rm s}=0.1$ and $M=0.01$ in the 
second version formulation of loop quantum gravity: the 
snapshots at $t=0$,
1.50057314, 3.23870859, 5.12101428, 6.4440209
are plotted.
In (a), the solid and dashed lines denote the effective density
and the conserved mass density, respectively.
In (c), the solid and dashed lines denote the Misner-Sharp mass
and the conserved mass, respectively.
}
\end{figure}

Finally, Fig.~\ref{fg:ltbr01_2} shows the collapse of an initially
inhomogeneous dust cloud with $R_{\rm s}=0.1$ and $M=0.01$ in the second
version of loop quantum gravity.  Also in this case, the evolution is
strongly slowed down.  As seen in Fig.~\ref{fg:ltbr01_2} a spike
develops in the density fields inside the cloud at $R\simeq  0.04$,
$t=6.4440209$.  The evolution outside the cloud is identical to that in
the homogeneous case.  The Misner-Sharp mass profile and therefore the
effective density profile are almost constant in time. The maximum value of
the Misner-Sharp mass is about 0.08 attained at $R\simeq  0.18$.  
The maximum value of the ratio $m/R$ is a half attained at the 
center, which implies the spike or shell-crossing singularity
is not covered by a trapping horizon.

\subsection{Summary of the numerical analysis}

In summary, the numerical results show that the loop quantum effects
significantly slow down the collapse of the central region which is as
large as $\sqrt{\gamma}l_{\rm P}$, while in the outer region these
effects are not important.  (There are however quantitative
differences between the different versions of consistent equations. In
the second version the collapse of the central region is more strongly
slowed down and the radius of this central ``core'' is larger than in
the first version.)  As a result, the density spike develops at around
the radius $\sqrt{\gamma}l_{\rm P}$, which results in the formation of
a shell-crossing singularity.  Another key feature is that the center
is always marginally trapped due to the inverse triad correction to
the metric.  This suggests that the shell-focusing singularity
appearing at the center might be covered by a horizon.

Since the collapse of an inhomogeneous dust cloud in classical general
relativity generically ends in shell-focusing naked singularities,
this means that although the singularity formed in the generic
spherical dust collapse in the present formulation of loop quantum
gravity is still locally naked, the shell-crossing singularity appears
prior to the shell-focusing singularity and hence the curvature
strength of locally naked singularities formed in the gravitational
collapse is weakened due to the loop quantum effects.  In a naive
sense, this seems to favor the cosmic censorship hypothesis in
effective loop quantum gravity because shell-crossing singularities
are generally believed to be extendible in a distributional sense.
Moreover, since the shell-crossing singularity will appear near the
center, it is likely to be trapped due to the inverse triad correction
to the metric.  We have seen that this is the case for some collapse
models in the first version.

However, one should be careful with any definite conclusion because
the present work can only provide insights into some of the quantum
effects in gravitational collapse and spacetime singularities. (For
instance, as mentioned earlier we have not included holonomy effects
in the numerical analysis because that would require a more detailed
analysis of the parameter space.) The slow-down of the collapse
indicates that repulsive forces of quantum gravity are indeed at work,
similar to those that resolve big bang singularities in homogeneous
models. But in our inhomogeneous context this does not appear
sufficient to provide a uniform bound on energy densities. The more
complicated nature of the problem is indicated by the formation of
shell-crossing singularities which may be a consequence of the still
present spherical symmetry and the dust idealization used for matter.

\section{Conclusions}

We have analyzed the existence of LTB-type models in the framework of
loop quantum gravity, starting with an implementation of the
corresponding class of spatial metrics at the kinematical Hilbert
space level. We first discussed the spherically symmetric setting and
in particular noted the role of lattice refinements in
Sec.~\ref{s:HolRef}. The discussion turned out to be much cleaner than
in purely homogeneous settings, especially regarding the scale
dependence of quantum equations. As an immediate consequence, not all
refinement models used so far for anisotropic homogeneous models can
be embedded in spherically symmetric models. In particular,
non-trivial refinement schemes which would give rise to equidistant
difference equations (where point holonomies are of the form
$\exp(i\mu(p^I)c_I)$ with only the variable $p^I$ conjugate to a
connection component $c_I$ entering) do not seem realizable in
spherical symmetry. It is thus important to improve the analysis of
non-equidistant difference equations, for instance along the lines of
\cite{RefinedNumeric,RefinementNumeric}. Despite some restrictions on
the refinement scheme, a whole class of refinements varying with the
spatial size remains allowed. All these schemes have the correct
scaling behavior under coordinate changes, even though further
reduction to homogeneous models may suggest improper scalings for some
of these models. This shows that it is only the reduction to
homogeneity which may suggest improper scalings because
scaling-dependent parameters arising in the reduction are
overlooked. The restriction of refinement models based solely on their
rescaling behavior in homogeneous models is thus too strict.

We then turned to the LTB conditions and showed that the classical
conditions translate easily to the kinematical quantum level, which
allows further studies of quantum reduction mechanisms of loop quantum
gravity along the lines of \cite{SymmRed}. However, the constraint
algebra makes a discussion difficult at the dynamical level where the
consistency issue of the operator algebra of constraints together with
LTB conditions would have to be analyzed. At this stage, further
progress is possible based on an effective treatment of correction
terms. While we have not derived complete effective equations which
would contain all relevant correction terms at once, the inclusion of
individual correction terms of certain types can already be used to
see how quantum effects can be realized consistently.

Indeed, we have found consistent formulations of the LTB reduction for
different types of corrections: the constraint equations
(\ref{ConsHam1}), (\ref{ConsHam2}) and (\ref{ConsHam3}) with the
evolution equations (\ref{evolution eq}), (\ref{EOM2}) and
(\ref{evolution eq for kphi correction}), respectively. Moreover, the
effects can be combined in consistent equations summarized in
Sec.~\ref{s:ConsDisc}. The consistency of an anomaly-free formulation
relates corrections in different terms of the Hamiltonian to each
other, but also to required corrections of the classical LTB
conditions. Here, our general understanding of an LTB-type reduction
is that metric coefficients in (\ref{CanonMetric}) are related to each
other by a proportionality $L\propto R'$. Classically, the factor is
one, but consistent quantum corrections require a non-trivial
dependence on $R$ or $\dot{R}$ which would be important in some
regimes. Thus, while we have the same type of reduction of degrees of
freedom, explicit dynamical implications for the metric may change. In
particular, the relation between $L$ and $R$ affects, for instance,
the position and behavior of horizons in addition to what a change in
the dynamical behavior of $R$ would imply. Also the behavior near
classical singularities changes due to correction factors in the
metric which may even diverge right at a classical singularity. A
complete space-time analysis of the resulting effective metrics
remains to be done.

The change in the form of the metric had unexpected implications:
Further reductions which are possible classically, such as
Friedmann--Robertson--Walker solutions, no longer exist, although on
large scales there are approximate such solutions. This is quite
unexpected and suggests that caution is necessary regarding the
dynamical realization of homogeneous models. Moreover, in combination
with corrections to the classical equations of motion, we have seen
new terms in effective densities which can become negative even where
mass densities remain positive. Also numerical simulations suggest
that repulsive forces of quantum gravity are active on small
scales. All this can affect the horizon behavior as well as
singularities if negative energies and corresponding repulsive forces
become strong enough. However, our analysis of central singularities,
which may be spacelike or non-spacelike, did not reveal any indication
that they would be prevented completely. While spherically symmetric
loop quantum gravity is singularity-free at a fundamental level of
difference equations \cite{SphSymmSing} (see also
\cite{EinsteinRosenQuant,Hybrid} for Gowdy models), the development of
intuitive geometrical pictures requires non-singular equations for an
effective geometry.  We have not analyzed full effective equations and
did not yet consider all possible correction terms; it may be that
some of the corrections which are more complicated to include can
avoid singularities more generally. Nevertheless, the difficulties in
avoiding inhomogeneous singularities, in contrast to the fact that
several spacelike singularities have been shown to be prevented based
on homogeneous models of loop quantum gravity, suggest that general
singularities present a qualitatively different issue compared to what
has been realized so far.  From our numerical simulations we
might speculate a possible resolution mechanism more subtle than the
analog of a cosmological bounce: Strong curvature singularities seem
to be replaced by weak singularities which may be extendable by
distributional solutions or with more realistic matter models.

\section*{Acknowledgements}

We are grateful to T.P.~Singh for several discussions and for valuable
comments on the manuscript.  M.B.\ thanks T.P.~Singh for hospitality at
TIFR, where some of this work was done. He was supported in part by
NSF grant PHY0653127. T.H. thanks K. Nakao for helpful comment. He was
supported by the Grant-in-Aid for Scientific Research Fund of the
Ministry of Education, Culture, Sports, Science and Technology, Japan
(Young Scientists (B) 18740144). R.T. thanks Ghanshyam Date for useful discussions and hospitality at IMSc, where some of this work was done.

\end{document}